\documentclass[11pt,a4paper]{article}

\usepackage[margin=1in]{geometry}
\usepackage[T1]{fontenc}
\usepackage{lmodern,microtype}
\usepackage{amsmath,amssymb,amsthm,bm,nicefrac,centernot}
\usepackage{graphicx,adjustbox,rotating}
\usepackage{booktabs,multirow,threeparttable,longtable}
\usepackage{float,placeins,subcaption}
\usepackage{tikz}
\usetikzlibrary{positioning,shapes.geometric,arrows.meta}
\usepackage{authblk}
\usepackage[round,authoryear]{natbib}
\usepackage[hidelinks]{hyperref}

\newtheorem{assumption}{Assumption}

\newtheorem{proposition}{Proposition}
\newcommand{\ind}{\perp\!\!\!\!\perp}

\title{\LARGE\bfseries Accounting for Attrition in\\Respondent-Driven Sampling Cohort Studies}
\author[1]{Vanessa McNealis\thanks{Correspondence \href{mailto:vanessa.mcnealis@glasgow.ac.uk}{\nolinkurl{vanessa.mcnealis@glasgow.ac.uk}}}}
\author[2]{Erica E. M. Moodie}
\author[3]{Alain Fourmigue}
\author[3]{Milada Dvorakova}
\author[2]{Joseph Cox}
\affil[1]{School of Mathematics and Statistics, University of Glasgow, Glasgow, United Kingdom}
\affil[2]{Department of Epidemiology, Biostatistics and Occupational Health, McGill University, Montréal, Canada}
\affil[3]{Research Institute of the McGill University Health Centre, Montréal, Canada}
\date{}

\begin{document}
\maketitle

\begin{abstract}
Respondent-driven sampling (RDS) is widely used to recruit populations lacking conventional sampling frames. Because RDS oversamples well-connected individuals, weights are commonly applied to improve representativeness. Although methods for cross-sectional RDS analyses are well developed, guidance for longitudinal studies is limited, despite additional selection bias from missed visits and dropout. We develop a framework that jointly accounts for unequal inclusion probabilities at recruitment and attrition during follow-up. The approach combines baseline RDS weights with inverse-probability-of-non-response weights estimated from observed follow-up histories, producing visit-specific weights for estimating prevalences, means, and regression parameters. We clarify when the RDS design should inform attendance models rather than being addressed only during outcome analysis. Using simulations of RDS recruitment, we assess prevalence and standardized prevalence differences under varying recruitment and attrition mechanisms, focusing on bias, variance, and confidence interval coverage. Weighting the response model by baseline RDS weights was important when an unmeasured determinant of degree also predicted response, even when it did not predict the outcome. We illustrate the framework using data from the Montreal site of Engage, a multisite RDS cohort of sexual minority men in three Canadian cities, from 2017–2023. Follow-up involved intermittent missed visits, attrition, and terminal dropout that was more common among participants with smaller networks. Our findings provide practical guidance for combining design-based and attrition adjustments to support valid inference in longitudinal RDS studies.
\end{abstract}


\section{Introduction}\label{sec1}

\subsection{Motivation and problem setting}

The World Health Organization (WHO) estimated a global incidence of 218 million cases of three curable bacterial sexually transmitted infections (bSTIs) each year, including 128 million cases of \textit{Chlamydia trachomatis} (CT), 82 million cases of \textit{Neisseria gonorrhoeae} (NG), and 7 million cases of syphilis \citep{who2021globalprogress}. One population group that is especially vulnerable to bSTIs is gay, bisexual, and other men who have sex with men (GBM), among whom some of the largest increases in reported bSTIs have been observed. Globally, the prevalence of bSTIs, particularly CT, NG, and syphilis, is especially high in this population. For instance, the global pooled prevalence of syphilis among GBM from 2000 to 2020 was 7.5\%, markedly higher than the 2020 estimate of 0.56\% among men in the general population \citep{who2021globalprogress}. Despite being among the populations most at risk, it is harder to survey GBM populations notably because of social stigma and difficulties in creating a representative sampling frame. Prevalence estimates of these infections in GBM populations are typically obtained from
clinics specialising in sexual health and care of sexually transmitted infections, however these data may overrepresent persons who engage in at-risk
sexual behaviours or have better access to health services \citep{lambert2026incidence}.

A common sampling technique used to survey populations for which there is no sampling frame such as GBM is \textit{respondent-driven sampling} (RDS), a link-tracing sampling method which leverages connections among individuals. The RDS process begins with a convenience sample made up of initial \textit{seed} participants who, after being interviewed, receive a fixed number of \textit{coupons} to distribute among peers belonging to the population of interest. Newly recruited participants are then interviewed and invited to recruit their own peers. This process is repeated through several waves until a sufficient number of participants are recruited in the sample. Through many waves of recruitment, the process samples farther from the initial recruits, which should ensure greater representativeness and  generalizability of the sample. 

RDS cohorts are affected by complex sampling design, characterized by unequal selection probabilities and network dependence. Longitudinal studies also commonly face panel attrition, which arises when a participant goes missing for one or more study visits. In longitudinal RDS cohorts, missingness may occur through intermittent missed visits, after which participants later return, or through terminal dropout, after which participants are no longer observed. Thus, the set of participants observed at a given follow-up visit may differ from both the original sample and the sets observed at earlier visits. Attrition rarely occurs completely at random and therefore induces selection bias, especially when the probability of being observed depends on the outcome or on factors associated with the outcome.

RDS weights are widely used for estimating cross-sectional population prevalences, but their use in longitudinal analyses is less well articulated. Inverse probability of observation (IPO) weights are well established in the survey and missing data, where the weights of non-respondents are effectively redistributed among respondents with similar observed characteristics. Such weights are commonly constructed by fitting regression models for the probability that a sampled participant is observed at a particular visit. However, there is limited guidance on how RDS design weights and IPO weights should be combined when estimating means, prevalences, or regression model parameters. In particular, it is unclear whether and how the RDS weights should be incorporated when modelling the non-response mechanism, or how the resulting weights should be used in the final analysis. The non-response model is particularly important in RDS studies, where network characteristics (e.g., network degree) may affect both recruitment and subsequent non-response. This gap hinders valid inference in RDS cohorts with non-response, which are common in HIV and STI epidemiological studies.

\subsection{Related work on combining design, propensity score and non-response weights}

Longitudinal survey analyses may combine design weights with inverse-probability-of-treatment (IPT) or IPO weights \citep{thompson2015using}. Whether the models used to construct the latter weights should account for the sampling design remains debated. For IPT weighting, \cite{brumback2010testing} recommend estimating treatment probabilities using survey weights when the model is intended to describe the target population. \cite{ridgeway2015propensity} show that this is generally necessary for consistent estimation of population parameters unless the variables governing selection are included in the treatment model. In the survey non-response literature, however, weighting a response-propensity model has not consistently improved estimation. Incorporating design information as model covariates or through calibration may also be effective \citep{little2003weighting, kott2012one, chen2015incorporating}. A more detailed literature review is included in Web Appendix A.

\subsection{Contributions and organization}

%Although related literatures have considered whether survey weights should be incorporated when estimating propensity score or non-response models, no published work has, to our knowledge, provided a clear methodological framework for simultaneously incorporating RDS design weights and inverse probability of observation weights in cross-sectional or longitudinal analyses from RDS samples. This gap hinders valid inference in RDS cohorts with non-response, which are common in HIV and STI epidemiology. In this paper, we propose a weighting approach that accounts jointly for unequal inclusion probabilities induced by RDS and selection arising from non-response. We examine the conditions required for the resulting estimators to be valid and investigate the consequences of omitting key variables in the non-response or outcome models.

To our knowledge, no published work has provided a clear methodological framework for simultaneously incorporating RDS weights and IPO weights in cross-sectional or longitudinal analyses from RDS samples. In this paper, we propose a weighting approach that accounts jointly for unequal inclusion probabilities induced by RDS and selection arising from non-response. We examine the conditions required for the resulting estimators to be valid and investigate the consequences of omitting key variables in the non-response or outcome models. We also examine when baseline RDS weights should be incorporated in the estimation of IPO weights.

The remainder of the paper is organized as follows. In Section 2, we set up our notation and assumptions and describe a weighting methodology to simultaneously correct for the design and selection bias due to non-response in longitudinal RDS studies. The validity of the proposed methodology is investigated in simulations presented in Section 3. In Section 4, we analyse the Engage data collected in Montreal to estimate population prevalences of bSTIs and the relationship between chemsex-related substance use and bSTI prevalence, providing reliable parameter estimates and comparing alternative approaches to uncertainty quantification. We conclude in Section 5 with a discussion of the approach and future considerations.

\section{Proposed attrition adjustment in RDS cohort studies}\label{sec3}

\subsection{RDS sampling and baseline weights}Suppose that a target population consists of \(N\) individuals connected by an
undirected social network \(G=(\mathcal{N},\mathcal{E})\), where
\(\mathcal{N}=\{1,\ldots,N\}\). Let \(d_i\) denote the number of neighbours of
individual \(i\). Assume that for nodes in $\mathcal{N}$, we observe the population data $\mathcal{DP}= \{\bm{O}_i; i = 1, \ldots N\}$. Similar to \cite{yauck2021general}, we assume that $\mathcal{DP}$ is a random realization of a random sample from a distribution $\mathcal{F}_0$. 

An RDS process recruits participants along network edges,
without recruiting any individual more than once, and yields a sample
\(\mathcal{N}_T\subseteq\mathcal{N}\) of size \(n\) and an observed recruitment
forest \(G_T=(\mathcal{N}_T,\mathcal{E}_T)\). We let \(S_i=1\) if individual
\(i\) is included in the RDS sample, and \(S_i=0\) otherwise. Following the RDS sampling process, we observe the tree data $\mathcal{DT}= \{\bm{O}_i ; i \in \mathcal{N}_T\}$, which is a subset of $\mathcal{DP}$.

The inclusion probabilities \(\pi_i\) are unknown in an RDS study. We
consider two baseline RDS weights, \(w_i^{\mathrm{RDS-II}} =w_i^{\mathrm{VH}}=1/\widehat{\pi}_i^{\mathrm{VH}}\)
and \(w_i^{\mathrm{SS}}=1/\widehat{\pi}_i^{\mathrm{SS}}\). For the RDS-II weights,
\(\widehat{\pi}_i^{\mathrm{VH}}=nd_i/\sum_{j\in\mathcal{N}_T}d_j\), reflecting
sampling probabilities proportional to degree. The successive-sampling
(SS) weights use estimated inclusion probabilities
\(\widehat{\pi}_i^{\mathrm{SS}}\equiv\widehat{\pi}_i^{\mathrm{SS}}(d_i)\) under a
without-replacement approximation and require \(N\) or an estimate of it.
We write \(w_i^{\mathrm{RDS}}\) for either choice of baseline weight.

\subsection{Cohort study framework and identification assumptions}\label{subsec31}

Suppose that an RDS sample $G_T=(\mathcal{N}_T,\mathcal{E}_T)$ is drawn from the target population $\mathcal{N}$, and sampled individuals are followed for $K$ planned visits. For each $i\in\mathcal{N}$, let
$\bm{O}_i=(\bm{Y}_i,\bm{L}_i,\bm{t}_i)$ denote the data that would arise under the cohort protocol, where $\bm{Y}_i=(Y_{i1},\ldots,Y_{iK})$, $\bm{L}_i=(\bm{L}_{i1},\ldots,\bm{L}_{iK})$, and $\bm{t}_i=(t_{i1},\ldots,t_{iK})$ contain the outcomes, covariates, and planned visit times. These quantities are defined in the population, but observed only for sampled individuals $S_i=1$.

Within the RDS sample, we let \(\delta_{ik}\) denote an observation indicator at visit $k$, where \(\delta_{ik} = 1\) if individual \(i\)'s outcome is recorded at visit \(k\). We distinguish intermittent non-attendance (INA), after which a participant may return, from terminal loss to follow-up (LTFU), after which no further visits are observed. Let $A_{ik}=1$ indicate attendance at visit \(k\) conditional on remaining under follow-up, and let $D_{ik}=1$ indicate retention at that visit. Since $A_{ik}=0$ by convention after terminal dropout, $\delta_{ik}=A_{ik}\times D_{ik}$. The observed counterparts $Y_{ik}^{\mathrm{obs}}$ and $\bm{L}_{ik}^{\mathrm{obs}}$ equal their full-data values when $\delta_{ik}=1$ and may be missing otherwise. Figure~\ref{fig:po-framework} illustrates this framework for \(K=1\).

Finally, \(\mathcal{H}_{ik}\) denotes the history observed immediately before visit \(k\). It may include baseline covariates, previous response indicators, observed outcomes and covariates, visit number, calendar time, and time since the last attended visit. In particular, it does not require the outcome or covariates from a missed visit.

We assume that terminal dropout and intermittent non-attendance are missing at random (MAR) conditional on the observed participant history. Specifically, for terminal dropout, we assume \begin{equation}P(D_{ik}=1 \mid D_{i,k-1}=1, \bar{\bm Y}_i, \bar{\bm L}_i, \bar {\bm A}_{i, k-1}, \bar {\bm D}_{i,k-1})
=
P(D_{ik}=1 \mid D_{i,k-1}=1, \mathcal{H}^{D}_{ik}),
\label{eqn:dropout}
\end{equation}
where $\bar {\bm A}_{i, k-1}$ and $\bar {\bm D}_{i,k-1}$ represent retention and attendance histories up to visit $k-1$, respectively, $ \bar{\bm Y}_i, \bar{\bm L}_i, \bar {\bm A}_i, \bar {\bm D}_i$, represent full, potentially unobserved or future, outcomes, and covariates respectively for participant $i$, and \(\mathcal{H}^{D}_{ik}\) denotes the observed history used to model remaining under follow-up at visit \(k\). For intermittent non-attendance, we assume
\begin{equation}P(A_{ik}=1 \mid D_{ik}=1, \bar{\bm Y}_i, \bar{\bm L}_i, \bar {\bm A}_{i, k-1}, \bar {\bm D}_{i, k-1})
=
P(A_{ik}=1 \mid D_{ik}=1, \mathcal{H}^{A}_{ik}),
\label{eqn:attendance}
\end{equation}
where \(\mathcal{H}^{A}_{ik}\) denotes the observed history used to model attendance at visit \(k\). In other words, conditional on the observed history, neither the probability of remaining under follow-up nor the probability of attending a scheduled visit depends on unobserved current or future outcomes or covariates.

We further require positivity. For all observed histories with positive probability, we assume
$P(D_{ik}=1 \mid D_{i,k-1}=1, \mathcal{H}^{D}_{ik})>0$
and
$P(A_{ik}=1 \mid D_{ik}=1, \mathcal{H}^{A}_{ik})>0.$
That is, among participants with observed histories that occur in the RDS sample, there must be a nonzero probability of remaining under follow-up and, conditional on remaining under follow-up, a nonzero probability of attending the scheduled visit.

Since \(\delta_{ik}=A_{ik} \times D_{ik}\) and $\delta_{ik}= 1 \iff A_{ik} \text{ and } D_{ik} = 1$, one also may write\begin{equation}
    P(\delta_{ik}=1 \mid D_{i,k-1}=1, \mathcal{H}_{ik})
=
P(D_{ik}=1 \mid D_{i,k-1}=1,\mathcal{H}^{D}_{ik}) \times 
P(A_{ik}=1 \mid D_{ik}=1,\mathcal{H}^{A}_{ik}).
\label{eq:probresponse}
\end{equation}
That is, the response probability for a participant at visit $k$ corresponds to the product of the probability of remaining under follow-up at visit $k$ and the probability of attending visit $k$. This suggests an alternative weighting strategy for attrition adjustment that targets the overall probability of response instead of its individual components. This will be discussed in more detail in the next section.

Our interest lies in identifying a mean outcome, prevalence, a regression parameter, or an association estimand in the original population if there had been no INA nor LTFU among the participants. In addition to a correction for the sampling design through RDS weights, corrections are required for selection biases that might ensue with INA or LTFU. We can make these adjustments through inverse probability weighting, which is based on assigning a visit-specific weight \(w_{ik}\) or time-specific weight \(w_{it}\), depending on the estimand of interest, to each individual-visit or individual-time so that they account in the analysis not only for themselves, but also for those who were not included in the analysis, i.e., those who were not sampled, those who missed a visit, or those who were lost to follow-up.

For exposure–outcome contrasts, additional assumptions are required beyond those concerning response. Let $E_{ik}$ denote the exposure of interest at visit $k$, let $Y_{ik}(e)$ denote the potential outcome under exposure level $e$, and let $\bm{V}_{ik}$ denote a set of measured pre-exposure covariates included in $\bm{L}_{ik}$. A causal interpretation of the standardized contrast requires consistency, no unmeasured confounding (NUC), and exposure positivity:
$ Y_{ik}=Y_{ik}(E_{ik}), \ Y_{ik}(E_{ik}) \ind E_{ik}  \ \rvert \ \bm{V}_{ik},$
and $ 0 < P(E_{ik} = e \ \rvert \ \bm{V}_{ik} = v) < 1$
for every visit $k$ and all relevant \(e\) and covariate values \(v\) with positive probability. These assumptions are not required for estimation of marginal outcome means or prevalences. When they are not considered plausible, standardized exposure contrasts should instead be interpreted as adjusted population-level associations.

\newcommand{\scenarioDAG}[1]{%
\begin{tikzpicture}[
  >=stealth, thick,
  var/.style={draw,circle,minimum size=6mm,inner sep=0pt},
  latent/.style={draw,circle,dashed,fill=gray!15,
                 minimum size=6mm,inner sep=0pt},
  selected/.style={draw,circle,fill=red!20,
                   minimum size=6mm,inner sep=0pt}]
  \node[var] (d) at (0,2) {$d$};
  \ifnum#1=1
    \node[var] (W) at (1,1) {$W$};
    \node[latent] (Z) at (1,2) {$Z$};
  \else
    \node[latent] (W) at (1,1) {$W$};
    \node[var] (Z) at (1,2) {$Z$};
  \fi
  \node[selected] (S) at (0,1) {$S$};
  \node[var] (Y) at (2,2) {$Y$};
  \node[var] (delta) at (0,0) {$\delta$};
  \node[var] (E) at (2,0) {$E$};

  \draw[->] (Z) -- (Y);
  \draw[->] (W) -- (delta);
  \draw[->] (W) -- (d);
  \draw[->] (Z) -- (d);
  \draw[->] (E) -- (Y);
  \draw[->] (E) -- (delta);
  \draw[->] (d) -- (S);
  \draw[dotted,bend right=40] (d) to (delta);
  \draw[dotted,bend left=40] (d) to (Y);

  \ifnum#1=1
    \draw[->] (W) -- (Y);
    \draw[->] (W) -- (E);
  \else\ifnum#1=2
    \draw[->] (Z) -- (delta);
  \else
    \draw[->] (W) -- (Y);
    \draw[->] (W) -- (E);
    \draw[->,bend left=5] (Z) to (E);
    \draw[->,bend right=5] (Z) to (delta);
  \fi\fi
\end{tikzpicture}%
}

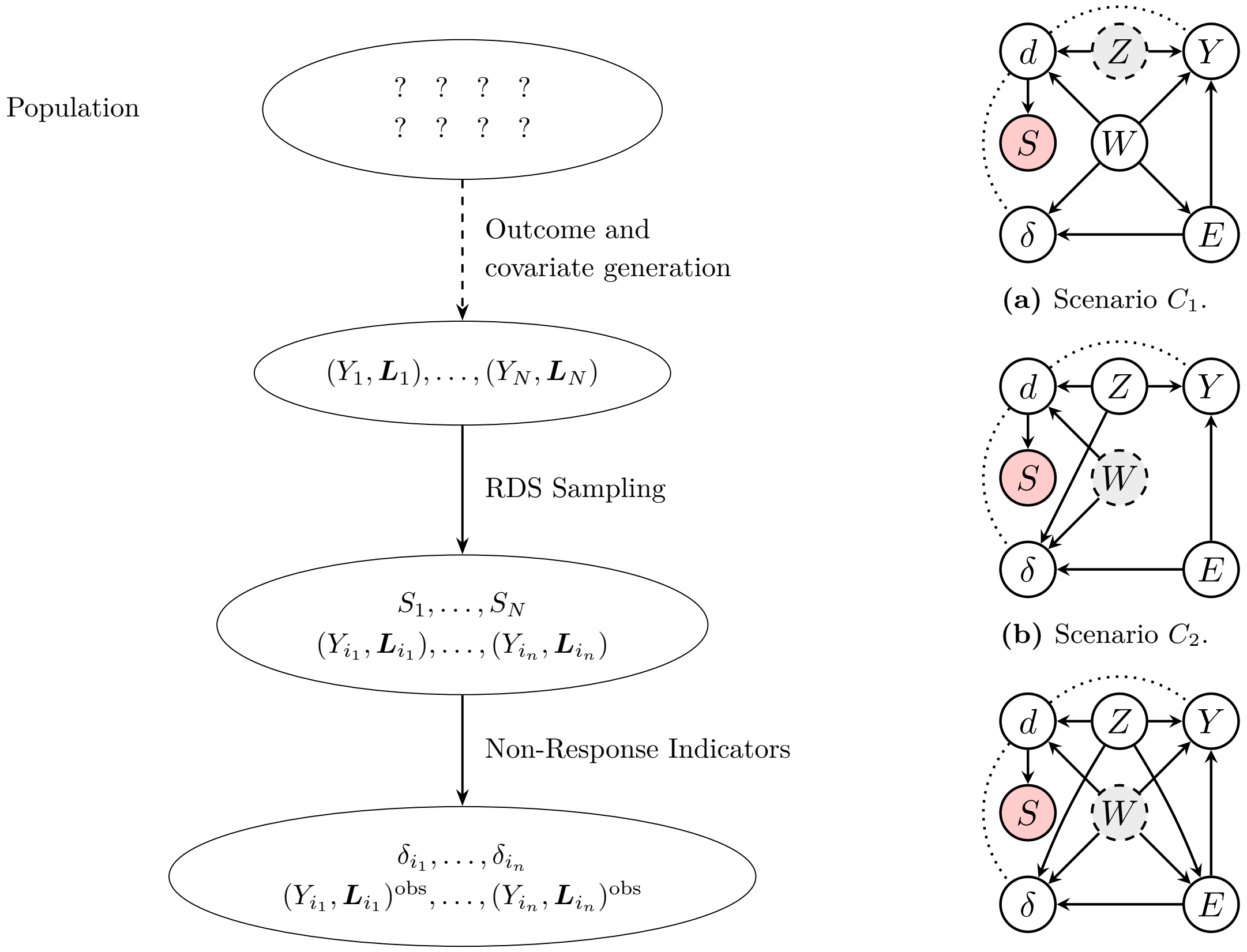
\begin{figure}[!htbp]
\centering
\begin{minipage}[c]{0.58\linewidth}
  \centering
  \begin{adjustbox}{width=\linewidth}
  \begin{tikzpicture}[
    scale=1.5,
    every node/.style={font=\small},
    oval/.style={draw,ellipse,align=center,
                 minimum width=5cm,minimum height=1.3cm},
    labelleft/.style={anchor=east},
    arrow/.style={-{Stealth[length=6pt]},thick},
    dashedarrow/.style={-{Stealth[length=6pt]},thick,dashed}]
    \node[oval] (popR) at (0,2.6) {
      \begin{tabular}{c}
        ?\quad ?\quad ?\quad ?\\[1pt]
        ?\quad ?\quad ?\quad ?
      \end{tabular}};
    \node[labelleft] at (-2.6,2.6) {Population};

    \node[oval] (sampPotR) at (0,0.4)
      {$(Y_1,\bm{L}_1),\ldots,(Y_N,\bm{L}_N)$};

    \node[oval] (sampleS) at (0,-1.7) {
      \begin{tabular}{c}
        $S_1,\ldots,S_N$\\[1pt]
        $(Y_{i_1},\bm{L}_{i_1}),\ldots,
         (Y_{i_n},\bm{L}_{i_n})$
      \end{tabular}};

    \node[oval] (sampleR) at (0,-3.8) {
      \begin{tabular}{c}
        $\delta_{i_1},\ldots,\delta_{i_n}$\\[1pt]
        $(Y_{i_1},\bm{L}_{i_1})^{\mathrm{obs}},\ldots,
         (Y_{i_n},\bm{L}_{i_n})^{\mathrm{obs}}$
      \end{tabular}};

    \draw[dashedarrow] (popR.south) --
      node[right=4pt,align=left]
      {Outcome and\\covariate generation} (sampPotR.north);
    \draw[arrow] (sampPotR.south) --
      node[right=4pt,align=left]
      {RDS Sampling} (sampleS.north);
    \draw[arrow] (sampleS.south) --
      node[right=4pt] {Non-Response Indicators} (sampleR.north);
  \end{tikzpicture}
  \end{adjustbox}
\end{minipage}\hfill
\begin{minipage}[c]{0.40\linewidth}
  \centering
  \begin{subfigure}{\linewidth}
    \centering
    \resizebox{0.48\linewidth}{!}{\scenarioDAG{1}}
    \caption{Scenario $C_1$.}\label{fig:dag-c1}
  \end{subfigure}\par\smallskip
  \begin{subfigure}{\linewidth}
    \centering
    \resizebox{0.48\linewidth}{!}{\scenarioDAG{2}}
    \caption{Scenario $C_2$.}\label{fig:dag-c2}
  \end{subfigure}\par\smallskip
  \begin{subfigure}{\linewidth}
    \centering
    \resizebox{0.48\linewidth}{!}{\scenarioDAG{3}}
    \caption{Scenario $C_3$.}\label{fig:dag-c3}
  \end{subfigure}
\end{minipage}
\caption{\textit{Left}: Graphical representation of the assumed data generating sequence, starting with the full participant data in the population followed by selection in the RDS sample and missing data indicators. Here, we assume a point-treatment study with only one follow-up visit. \textit{Right}: Directed acyclic graph representations for the data generating mechanisms considered in the three simulation study scenarios, where nodes represent the degree $d$, the sample inclusion indicator $S$, the covariates $Z$ and $W$, the exposure $E$, the outcome $Y$, and the unit response indicator $\delta$. Note that scenario $C_4$ also follows causal graph (b).}
\label{fig:po-framework}\label{fig:dags}
\end{figure}

\subsection{Proposed weighting procedures} \label{subsec32}

The proposed adjustments combine RDS-II weights calculated at baseline with inverse probability weights for INA or for LTFU, depending on the estimand considered. The goal is to make the final sample resemble the sample we would obtain under random (not respondent-driven) sampling with no attrition. Participants who were retained throughout the study despite a small probability of being recruited at baseline and/or attendance receive more weight. 

We first consider visit-specific weights, which can be used for estimating visit-specific means and prevalences or regression model parameters, such as GEE parameters. Under the decomposition in (\ref{eq:probresponse}), the response weight can be written as the product of an inverse-probability-of-dropout weight and an inverse-probability-of-attendance weight. Given a participant $i$ and a visit $k$, the survey and attrition adjustment weights can be expressed as the product of three terms: \begin{equation}
  w_{ik} = w_i^{RDS} \times w_{ik}^D \times w_{ik}^A  = \pi_i^{-1}\times (p^D_{ik})^{-1} \times (p^A_{ik})^{-1}, 
  \label{eqn:weightproduct}
\end{equation}
where $p^D_{ik} = 1/w_{ik}^D$ is given by $
p^D_{ik} =  P(D_{i1} = 1, D_{i2} = 1, \ldots, D_{ik} = 1 \ | \ \mathcal{H}_{ik}^D)= \prod_{j=1}^k P(D_{ij} = 1 \ | D_{i(j-1)} =1, \  \mathcal{H}_{ij}^D),$ with $D_{i0} = 1$ by convention, and $ p^A_{ik} = 1/w_{ik}^A $ is given by $p^A_{ik} =  P(A_{ik}=1 \mid D_{ik}=1,\mathcal{H}^{A}_{ik}). $ Finally, $w_i^{RDS} = \pi_i^{-1}$ is a time-invariant RDS weight (either $w_i^{VH}$ or $w_i^{SS}$) described in the RDS preliminaries in Web Appendix B. The first product term in (\ref{eqn:weightproduct}) acts as a survey weight to represent the unstandardized population. The second term is an inverse-probability-of-retention weight that standardizes the population in order to adjust for LTFU and the third term is an inverse-probability-of-attendance weight that standardizes the population in order to adjust for INA. This decomposition can be particularly useful if it is thought that the drivers of LTFU $\mathcal{H}_{ik}^D$ are very different from the drivers of INA $\mathcal{H}_{ik}^A$. Alternatively, one could also consider weights based on the overall response probability at visit $k$, given by
$ p_{ik}^{O*} = P(\delta_{ik} =1 \ | D_{i,k-1}=1, \mathcal{H}_{ik}),$
where $\mathcal{H}_{ik}$ contains the participant history up to visit $k$ that is relevant to the overall response mechanism. The corresponding IPO weight is
$w^{R}_{ik} = 1/p_{ik}^{O*}$ and the combined weight would be $w_{ik} = w_i^{RDS} \times w^{R}_{ik} $.

The weights $w_{ik}$ are then incorporated in the same way as conventional RDS weights in cross-sectional analyses, whether it be the estimation of a mean or prevalence or an associational analysis at visit $k$. For instance, for a given follow-up visit $k$, the modified RDS-II or SS estimator of the mean (or prevalence, in the case of a binary variable) of $Y_k$ that accounts for both the RDS design and attrition is given by $ \tilde{\mu}_{k} = \left[\sum_{i=1}^N  \delta_{ik} S_i Y^{\text{obs}}_{ik} {w_{ik}}\right]/ \left[{\sum_{i=1}^N \delta_{ik} S_i {w_{ik}}}\right].$
In a regression analysis using GEE or generalised linear mixed-effects models, the adjustments for the survey design and attrition are made by incorporating the weights $w_{ik}$ in the model equations, such as in the GEE estimating equation in Web Appendix B.

\subsection{Estimation of the response weights}

In practice, the individual retention/attendance/observation propensities \(p^D_{ik}\), \(p^A_{ik}\), and \(p^{O*}_{ik}\) are unknown and must be estimated. Following the inverse-probability weighting approach used for longitudinal survey data and marginal structural models, we estimate these probabilities using pooled logistic regression models (unless there are substantive reasons to suggest separate mechanisms govern the models at each time point) \citep{robins2000marginal, brumback2010testing}. For terminal retention at visit $j$, $j = 1, \ldots K$, we fit the model 
$\mathrm{logit}\{P(D_{ij}=1 \mid D_{i,j-1}=1,\mathcal{H}^D_{ij})\}
=
\bm{\alpha}_{Dj}^{\top}\bm{h}^D_{ij}$,
among person-visits for which \(D_{i,j-1}=1\), where
\(\bm{h}^D_{ij}\) is a vector of functions of the observed history
\(\mathcal{H}^D_{ij}\), and \(\bm{\alpha}_{Dj}\) is a vector of regression coefficients. Because the target response probabilities are defined in the population rather than within the realised RDS sample, similar to \cite{brumback2010testing}, we estimate \(\bm{\alpha}_{Dj}\) using the baseline RDS weights (either $w_i^{VH}$ or $w_i^{SS}$). For the pooled logistic retention model, \(\widehat{\bm{\alpha}}_{Dj}\) solves
\begin{equation}
\sum_{i\in \mathcal{N}}
w_i^{RDS}S_i
\sum_{j=1}^K
D_{i,j-1}
\bm{h}^D_{ij}
\left[
D_{ij}
-
\operatorname{expit}\{\bm{\alpha}_{Dj}^{\top}\bm{h}^D_{ij}\}
\right]
=
0.
\label{eqn:weedropout}
\end{equation}
These weighted estimating equations fit the dropout model in a
pseudo-population in which the RDS sample has been reweighted to represent the
baseline target population.
This yields fitted probabilities $\widehat{p}^D_{ij}=\operatorname{expit}(\widehat{\bm{\alpha}}_{Dj}^{\top}\bm{h}^D_{ij})$,
and therefore $\widehat{p}^D_{ik}=\prod_{j=1}^k \widehat{p}^D_{ij}$.

Similarly, for attendance at visit $k$, we fit $\mathrm{logit}\{P(A_{ik}=1 \mid D_{ik}=1,\mathcal{H}^A_{ik})\}
=
\bm{\alpha}_{Ak}^{\top}\bm{h}^A_{ik}$
among person-visits for which \(D_{ik}=1\), where
\(\bm{h}^A_{ik}\) is a vector of functions of the observed history
\(\mathcal{H}^A_{ik}\), and \(\bm{\alpha}_{Ak}\) is a vector of regression coefficients. The attendance model is estimated using
\begin{equation}
\sum_{i\in \mathcal{N}} w_i^{RDS}S_i \sum_{k=1}^K D_{ik} \bm{h}^A_{ik}
\left[ A_{ik} - \operatorname{expit}\{\bm{\alpha}_A^{\top}\bm{h}^A_{ik}\} \right] =0.
\label{eqn:weeattendance}
\end{equation}
This yields $\widehat{p}^A_{ik}
=
\operatorname{expit}(\widehat{\bm{\alpha}}_{Ak}^{\top}\bm{h}^A_{ik})$ and the final estimated weight is therefore
\begin{equation}
\widehat{w}_{ik}
=
w_i^{RDS}
\times
\frac{1}{\prod_{j=1}^k \widehat{p}^D_{ij}}
\times
\frac{1}{\widehat{p}^A_{ik}}.
\label{eqn:finalweight}
\end{equation}

The models for \(D_{ij}\) and \(A_{ik}\) should include predictors $\mathcal{H}_{ik}^D$ and $\mathcal{H}_{ik}^A$ that make
the corresponding sequential MAR assumptions plausible. %These may include baseline covariates, previous outcomes, previous attendance indicators, time since the last attended visit, visit number, calendar time, and other features of the observed follow-up history. 

If the drivers of LTFU, $\mathcal{H}_{ik}^D$, and INA,  $\mathcal{H}_{ik}^A$, are believed to be the same, then overall response probability $\delta_{ik}$ can be modelled directly by solving similar weighted estimating equations to (\ref{eqn:weedropout}) and (\ref{eqn:weeattendance}) among participants who were still enrolled at visit $k-1$.

The distinction between population-standardized and sample-conditional response probabilities is formalized in the following proposition, with a derivation provided in Web Appendix C. Without loss of generality, we consider the estimation of the overall non-response ($\delta$) probability in a single follow-up visit setting.

\begin{proposition}
\label{prop:response_weighting}
Consider an RDS cohort with a single follow-up visit $(K=1)$ and
suppose that $Y_i \ind \delta_i \mid \mathcal{H}_i.$. Define $p_i^{O*}=P(\delta_i=1\mid\mathcal{H}_i)$ as response mechanism standardized to the population distribution of relevant characteristics under hypothetical recruitment. Then, the combined RDS and inverse-probability-of-observation weighted
estimator is consistent for the superpopulation mean $\mu_0=\mathbb{E}[Y]$,
under the stated positivity conditions.

In contrast, using the sample-conditional response probability $p_i^{O}=P(\delta_i=1\mid S_i=1,\mathcal{H}_i)$
does not generally yield consistency. A sufficient additional
condition is $S_i \ind \delta_i \mid Y_i,\mathcal{H}_i.$
\end{proposition}
 Consequently, when sampling and non-response remain associated after
conditioning on the observed history, the response model should
target $p_i^{O*}$, for example by estimating it using the baseline
RDS design weights. This is because if some design information which is also linked to non-response is omitted or only partially captured by the model covariates, then $\forall k$ $P(\delta_{ik} \ | D_{i,k-1}=1, \mathcal{H}_{ik}) \neq P(\delta_{ik} \ | \  S_i=1, D_{i,k-1}=1, \mathcal{H}_{ik})$ and analyses may result in bias. Otherwise, if all design-related variables that affect dropout and attendance are included, weighted and unweighted fits of the retention, attendance, and overall non-response models may be similar. Estimating these models with RDS weights provides a principled way to align the dropout and attendance weights with the target population represented by the original RDS sample. These guidelines are corroborated by empirical results presented in the following section, where we compare performance of IPO weighting estimators based on $p^O_{ik} = P(\delta_{ik} \ | \ \mathcal{H}_{ik}, S_i=1)$, labeled inverse probability of sample response (IPSR), versus IPO weighting estimators based on the population observation probability $p^{O*}_{ik} = P(\delta_{ik} \ | \ \mathcal{H}_{ik})$, inverse probability of population response (IPPR). %The IPOW estimators use estimators of  $p^{O*}_{ik}$, $\hat{p}^{O*}_{ik}$, that are constructed from RDS-weighted pooled logistic models.

\section{Simulations}\label{sec4}
We conducted simulation studies with $R=1000$ replicates to assess the accuracy of the prevalence and regression parameter estimators under different modelling scenarios. The target estimands were: (i) the population prevalence of a binary outcome $Y$, and (ii) the marginal association between an exposure $E$ and the outcome $Y$, expressed as a marginal prevalence difference. For instance, in typical RDS epidemiological study, $Y$ may correspond to an active syphilis infection, while $E$ could represent recent crystal methamphetamine use. 

We considered a point-treatment study with one follow-up visit, with the possibility of attrition at follow-up. This scenario is similar to the one considered in Proposition \ref{prop:response_weighting}, and allows us to only consider the overall non-response indicator $\delta$. In the following subsection, we detail the generation of the population network $G$, baseline covariates $Z$ and $W$, outcome $Y$, exposure $E$, sample inclusion indicator $S$, and non-response indicator at follow-up $\delta$ for each Monte Carlo replicate, where we distinguish four main data generating mechanisms denoted $C_1$--$C_4$. The scenarios $C_1$, $C_2$, and $C_3$ describe different scenarios of dependence between the outcomes, covariates, non-response indicators, and participants' degrees, as shown in the directed acyclic graphs in Figure \ref{fig:dags}. The scenario $C_4$ uses the same diagram as $C_2$ but additionally incorporates network autocorrelation in the outcomes through a SAR error model. This reflects the practical reality that participants' network degrees are often related to their characteristics and creates settings in which RDS weighting becomes important, as supported by previous empirical studies \citep{avery2019unweighted, yauck2021general}.

The primary simulation results presented here focus on scenario $C_2$, which was designed to isolate the main methodological question considered in this paper. This scenario corresponds precisely to the setting considered in Proposition \ref{prop:response_weighting}, in which incorporating the sampling design into the non-response model is necessary for valid inference. %Although restricting to the RDS sample can associate the latent $W$ with $Y$ through their shared dependence on $d$ and $Z$, according to Figure \ref{fig:dags}(b), conditioning on the observed $Z$ blocks this path. Thus MAR holds given observed $E$ and $Z$. NUC also holds because $E$ has no causal parents in $C_2$. However, $W$ affects both degree $d$ and non-response $\delta$, thereby inducing an association between inclusion in the RDS sample and subsequent non-response. 

\subsection{Data generating mechanisms}

For each simulation scenario, we first generated a finite population network of size $N=1000$ with a fixed degree sequence and assigned the baseline covariates $Z$ and $W$. This network and these covariates were held fixed across Monte Carlo replicates. Within each replicate, we generated the remaining individual-level variables $E$, $Y$, followed by RDS sampling, response indicator at follow-up $\delta$, andthe application of the estimators. Thus, the simulation results assess estimator performance conditional on a fixed population network, while allowing the outcome and attrition mechanisms to vary across replicates.

\subsubsection{Population network and baseline covariates.} We generated a population network from an initial set of $N$ individuals. Each individual had a binary baseline covariate \(Z_i\) representing age group and a standard normal covariate \(W_i\). In scenarios \(C_1\)--\(C_3\), we set \(P(Z_i=1)=0.67\), where \(Z_i=1\) denotes age 30 years or older. Target degrees were drawn from a negative binomial distribution with a mean depending on both \(Z_i\) and \(W_i\), and bounded between 1 and 150. We generated the network using the configuration model, then removed self-loops, duplicate edges, and any resulting isolated individuals. Further details of the network generation are given in Web Appendix E. Web Figure E1 displays the overall observed degree distribution in $G$, as well as the degree distributions stratified by $Z$. As expected, individuals with $Z=1$ tend to have higher network degree on average.

\subsubsection{Generation of outcome and exposure}

In scenario $C_2$, the outcome was generated according to the model
\begin{equation*}
\mathrm{logit}[P(Y_i=1 \ \rvert \ E_i, Z_i]) = \eta_0 + \eta_1 E_i + \eta_2 Z_i + \eta_{12} E_i Z_i,  \tag{M2} \label{eqn:M2}
\end{equation*}
with coefficients $\eta_0 =-5$, $\eta_1 = 1.5$, $\eta_2 = 2$, and $\eta_{12} = 0.6$. As opposed to other scenarios, $E$ had no causal parents in scenario $C_2$, with $P(E_i = 1) = 0.47$. A detailed description of the generation of exposures and outcomes for scenarios $C_1$--$C_4$ can be found in Web Appendix E.

\subsubsection{RDS sampling and generation of non-response indicator at follow-up}

From each simulated population, we drew an RDS sample using the population network $G$ according to the procedure described in Web Appendix E.
%From each simulated population, we drew an RDS sample using the population network $G$. Seeds were sampled with probability proportional to degree. Unless otherwise stated, each simulated RDS study used $s=10$ seeds and each participant received up to $c=3$ coupons. The number of successful recruits generated by each participant was sampled uniformly from $\{0,1,\ldots,c\}$, subject to the availability of unrecruited neighbours and the remaining target sample size. Sampling proceeded without replacement along edges of the population network until the target sample size was reached or no further recruitment was possible. We considered sampling fractions $f \in \{ 0.10, 0.20, 0.30\}$, corresponding to target sample sizes $\lfloor fN \rfloor$. 
Then, a response indicator at follow-up was generated among recruited participants as
\begin{equation*}
    \mathrm{logit}[P(\delta_i=0 \ \rvert \ E_i, Z_i, W_i)] = \alpha_0 + \alpha_1 E_i + \alpha_2 Z_i + \alpha_3 W_i + \alpha_{23} Z_i W_i,\tag{R3} \label{eqn:R3}
\end{equation*}
with coefficients $\alpha_0 = -3$, $\alpha_1 =  0.5$, $\alpha_2 = \alpha_3=  2$, and $\alpha_{23} = 1$. A detailed description of the generation of non-response for scenarios $C_1$--$C_4$ can be found in Web Appendix E. 

\subsection{Fitting procedures and performance metrics}

For each simulated RDS sample, we computed RDS-II and successive-sampling (SS) weights using \texttt{vh.weights} and \texttt{gile.ss.weights}, respectively, from the R package RDS \citep{handcock2024rds}. Degree was assumed to be measured without error and computation of the SS weights also used the known population size \(N\).

We compared four weighting methods: unweighted (UW), RDS-only (RDS), and combined RDS and inverse-probability-of-observation methods (IPSR+RDS and IPPR+RDS). Their respective analysis weights were $ \omega_i \in \left\{ 1,\; w_i^{\mathrm{RDS}},\; w_i^{\mathrm{RDS}}(\hat p_i^O)^{-1},\; w_i^{\mathrm{RDS}}(\hat p_i^{O*})^{-1}\right\},$
where \(w_i^{\mathrm{RDS}}\) denotes either the RDS-II or SS weight. The IPSR method uses the sample-conditional response probability
\(\hat p_i^O=\widehat P(\delta_i=1\mid S_i=1,\mathcal H_i)\).
The IPPR method instead estimates the population response probability
\(\hat p_i^{O*}=\widehat P(\delta_i=1\mid\mathcal H_i)\)
by fitting the response model using baseline RDS weights, as motivated by Proposition~\ref{prop:response_weighting}. In scenario \(C_2\), both methods used the working model
\begin{equation*}
\mathrm{logit}\{P(\delta_i=0\mid E_i,Z_i)\}
=\omega_0+\omega_1E_i+\omega_2Z_i+\omega_{12}E_iZ_i.
\tag{R2}\label{eqn:R2}
\end{equation*}

For each choice of \(\omega_i\), we estimated the follow-up prevalence and the marginal prevalence difference using $$\hat\mu_\omega =\frac{\sum_{i\in\mathcal N_T}\delta_i\omega_iY_i}{\sum_{i\in\mathcal N_T}\delta_i\omega_i}$$ and
\begin{align}
\hat\Delta_\omega
&=\frac{\sum_{i\in\mathcal N_T}\delta_i\omega_i
\{\hat m_\omega(1,\bm{V}_i)-\hat m_\omega(0,\bm{V}_i)\}}
{\sum_{i\in\mathcal N_T}\delta_i\omega_i},
\label{eqn:simulation-estimators}
\end{align}
where $\hat m_\omega(e,\bm{V}_i)$ is the predicted outcome probability at exposure level \(e\) from a weighted logistic regression model among respondents, and $\bm{V}_i$ denotes the observed covariates used for outcome adjustment. The target prevalence difference was
\(\Delta=P\{Y(1)=1\}-P\{Y(0)=1\}\).
In scenario \(C_2\), we fitted the correctly specified outcome model~(\ref{eqn:M2}). Fitting procedures for \(C_1\), \(C_3\), and \(C_4\) are detailed in Web Appendix E.

Across \(R=1000\) replicates, we assessed relative bias, Monte Carlo standard deviation, relative root mean squared error (RRMSE), average standard error, and empirical coverage. Relative bias was defined as $ \mathrm{Bias }(\hat{\theta}) = 100 \times R^{-1}\left(\sum_{r=1}^R \hat{\theta}^{(r)} - \theta^{(r)}\right)/\bar{\theta},$ where $\hat{\theta}^{(r)}$ is an estimator computed on the $r$-th RDS study, ${\theta}^{(r)}$ is the corresponding estimand computed on the $r$-th generated finite population, and $\bar{\theta} = R^{-1} \sum_{r=1}^R \theta^{(r)}$. Monte Carlo RRMSE was defined as $\mathrm{RRMSE}_{\mathrm{MC}}(\hat{\theta})= 100 \times{\left[ R^{-1} \sum_{r=1}^R \left\{ \hat{\theta}^{(r)} -{\theta}^{(r)} \right\}^2\right]^{1/2}/\bar{\theta}}.$  We obtained standard errors and 95\% percentile confidence intervals using the neighbourhood and tree bootstraps \citep{yauck2022neighborhood,baraff2016estimating}. In \(C_4\), we also examined sandwich standard errors clustered by seed and recruiter, and the corresponding delta method calculation is given in Web Appendix D.

\subsection{Simulation results}
\label{sec:simulation-results}

The primary simulation results can be found in Figure~\ref{fig:boxplotsC2}, which compare the Monte Carlo distributions of the prevalence and prevalence difference estimators under Scenario $C_2$. Table~\ref{tab:simulation-c2} supplements the graphical results for Scenario $C_2$ by reporting the average estimate, relative bias, Monte Carlo standard deviation, relative root mean squared error (RRMSE), average bootstrap standard error, and empirical coverage probability across sampling fractions. A detailed version of this table that includes estimators based on SS weights can be found in Table F2 in Web Appendix F. Our discussion of results focus on point prevalence and prevalence difference estimation in scenario $C_2$, as well as uncertainty quantification performance across all scenarios. Detailed discussions of results for scenarios $C_1$--$C_4$ along with accompanying results figures and tables can be found in Web Appendix F. %Across the four scenarios, the principal determinant of bias was whether the estimation procedure adequately addressed both sources of selection: unequal recruitment through the RDS design and differential response at follow-up. Increasing the sampling fraction generally reduced the Monte Carlo standard deviation. However, it did not eliminate systematic bias arising from an inadequate response or outcome model.

\subsubsection{Scenario \texorpdfstring{$C_2$}{C2}: an unmeasured
determinant of degree and response}
\label{sec:simulation-results-c2}

Scenario $C_2$ isolates the distinction between response in the RDS sample and in the target population (Figure~\ref{fig:dags}(b)). In this scenario, the latent variable $W$ affects degree and non-response, but not the outcome. Although restricting to the RDS sample can associate the latent $W$ with $Y$ through their shared dependence on $d$ and $Z$, according to Figure \ref{fig:dags}(b), conditioning on the observed $Z$ blocks this path. Thus, MAR holds given observed $E$ and $Z$. NUC also holds because $E$ has no causal parents in $C_2$. However, $W$ affects both degree $d$ and non-response $\delta$, thereby inducing an association between inclusion in the RDS sample and subsequent non-response. Thus, %MAR and NUC hold given the observed variables in models~(\ref{eqn:M2}) and~(\ref{eqn:R2}), while 
$P(\delta=1\mid S=1,E,Z) \neq P(\delta=1\mid E,Z)$ since the distribution of $W$ among recruits differs from its population distribution.

\begin{figure}[!htbp]
\centering
\includegraphics[width=0.90\linewidth,height=0.63\textheight,keepaspectratio]{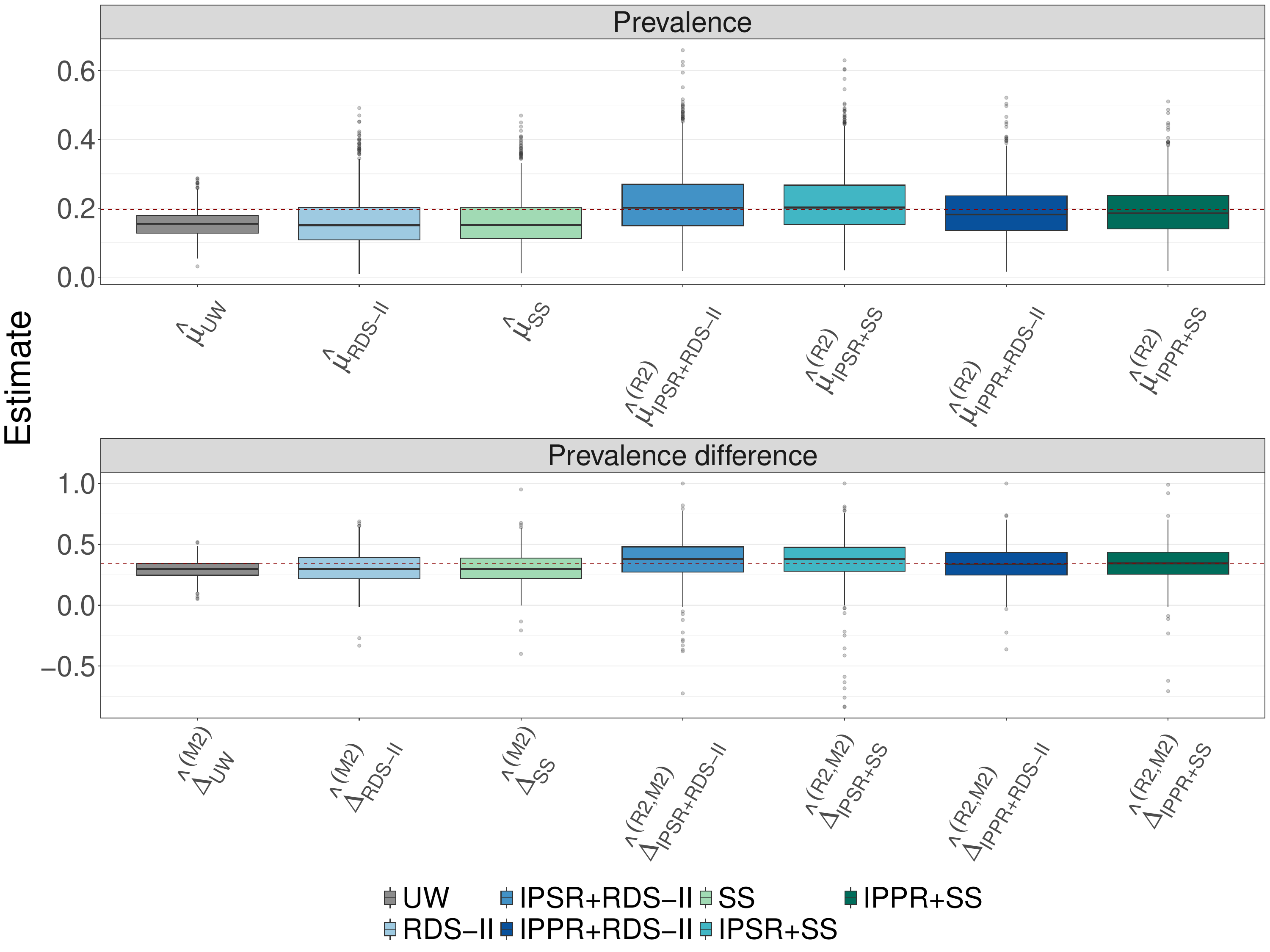}
\caption{Boxplots show the estimated outcome prevalence (top panel) and prevalence difference (bottom panel) under Scenario $C_2$ across 1,000 simulation replicates for the unweighted (UW), RDS, and IPO adjusted estimators (IPSR/IPPR). Adjusted estimators are combined with either RDS-II or SS sampling weights. The sampling fraction displayed is 20\%. Superscripts R2 and M2 denote response-model specification (\ref{eqn:R2}) and marginal standardization model (\ref{eqn:M2}), respectively. The red dashed line indicates the mean population-specific true value of the corresponding estimand.}
\label{fig:boxplotsC2}
\end{figure}

For an average true prevalence of $\mu=0.196$, relative bias was approximately $-20\%$ for unweighted, $-13\%$ to $-16\%$ with RDS weighting alone, and $11\%$ to $16\%$ when the response model was fitted without RDS weights (IPSR). Fitting that model with baseline RDS weights (IPPR) reduced bias to $-2.4\%$ to $2.4\%$ (Figure~\ref{fig:boxplotsC2} and Table~\ref{tab:simulation-c2}). The prevalence difference showed the same pattern: for an average true value of $\Delta=0.345$, relative bias was $-12\%$ to $-15\%$ for unweighted and RDS-only estimators, $7\%$ to $12\%$ for IPSR, and $-2.6\%$ to $-0.4\%$ for IPPR. These results align with Proposition~\ref{prop:response_weighting}, in the sense that weighting only the final outcome analysis did not address the difference between sample and population response probabilities.

The bias reduction came with greater variability. At the $10\%$ sampling fraction, the IPPR prevalence estimators had Monte Carlo standard deviations of approximately $0.100$, versus $0.055$ for unweighted, and did not always have the lowest RRMSE. Variability declined as the sampling fraction increased. RDS-II and SS performed similarly, with SS tending to be slightly less variable at larger sampling fractions.

\subsubsection{Uncertainty estimation}
\label{sec:simulation-results-uncertainty}

Neighbourhood-bootstrap intervals generally had coverage below the nominal $95\%$ level (Table~\ref{tab:simulation-c2}; Web Tables F1--F3). Their average standard errors were usually smaller than the corresponding Monte Carlo standard deviations, suggesting underestimation of variability. Estimator bias also contributed to poor coverage: under $C_2$, coverage for the RDS-only prevalence estimators fell below $73$\% for all sampling fractions considered. For combined RDS- and IPO- weighted estimators, the empirical coverage tended to increase with sample size. The observed undercoverage may reflect the relatively modest sample sizes considered: the population size was $N=1000$, and the largest sample size was $n=300$. Consistent with this interpretation, \citet{yauck2022neighborhood} showed that the neighbourhood-bootstrap variance estimator converged as the sample size increased and reported coverage approaching the nominal 95\% level when $n=1000$.

The tree bootstrap gave larger standard errors and higher coverage. For IPSR and IPPR, coverage ranged from approximately $85\%$ to $100\%$ under $C_1$ and $95\%$ to $99\%$ under $C_2$. It was generally conservative, particularly at larger sampling fractions. Neither bootstrap could restore nominal coverage for systematically biased estimators.

\begin{table}[p]
\caption{Simulation performance for estimation of population prevalence across scenarios and sampling fractions.}
\label{tab:simulation-c2}
\scriptsize
\centering
\setlength{\tabcolsep}{3.5pt}
\begin{threeparttable}
\begin{tabular}{@{}lllrrrrrrrr@{}}
  \toprule
 &  &  &  &  &  &  & NBoot &  NBoot & TBoot & TBoot \\ 
Scenario & $f$ & Estimator & Av. Est. & Rel. bias (\%) & SD & RRMSE (\%) & Av. SE & Cov. (\%) & Av. SE & Cov. (\%) \\ 
  \midrule
  C2 & 10\% & $\widehat{\mu}_{\mathrm{UW}}$ & 0.157 & -19.823 & 0.055 & 33.664 & 0.045 & 76.406 & 0.074 & 95.281 \\ 
  \addlinespace
   &  & $\widehat{\mu}_{\mathrm{RDS-II}}$ & 0.170 & -13.304 & 0.109 & 56.533 & 0.067 & 71.787 & 0.105 & 87.751 \\ 
   \addlinespace
 %  &  & $\widehat{\mu}_{\mathrm{CC+SS}}$ & 0.170 & -13.493 & 0.107 & 55.448 & 0.066 & 72.490 & 0.103 & 88.052 \\ 
   &  & $\widehat{\mu}_{\mathrm{IPSR+RDS-II}}^{(\mathrm{R2})}$ & 0.228 & 15.920 & 0.132 & 68.749 & 0.082 & 82.129 & 0.126 & 95.080 \\ 
   \addlinespace
 %  &  & $\widehat{\mu}_{\mathrm{IPO+SS}}^{(\mathrm{R2})}$ & 0.227 & 15.799 & 0.130 & 67.433 & 0.080 & 82.329 & 0.124 & 95.783 \\ 
   &  & $\widehat{\mu}_{\mathrm{IPPR+RDS-II}}^{(\mathrm{R2})}$ & 0.201 & 2.112 & 0.106 & 53.231 & 0.069 & 82.430 & 0.108 & 95.181 \\ 
 %  &  & $\widehat{\mu}_{\mathrm{IPOW+SS}}^{(\mathrm{R2})}$ & 0.201 & 2.419 & 0.104 & 52.211 & 0.068 & 82.530 & 0.107 & 95.582 \\ 
   \addlinespace
   \addlinespace
  &  & $\widehat{\Delta}_{\mathrm{UW}}^{(\mathrm{M2})}$ & 0.295 & -14.341 & 0.110 & 34.856 & 0.089 & 81.727 & 0.145 & 97.289 \\ 
  \addlinespace
   &  & $\widehat{\Delta}_{\mathrm{RDS-II}}^{(\mathrm{M2})}$ & 0.304 & -11.769 & 0.183 & 54.423 & 0.117 & 79.317 & 0.181 & 94.779 \\ 
   \addlinespace
 %  &  & $\widehat{\Delta}_{\mathrm{CC+SS}}^{(\mathrm{M2})}$ & 0.301 & -12.766 & 0.194 & 57.528 & 0.121 & 80.522 & 0.184 & 95.382 \\ 
   &  & $\widehat{\Delta}_{\mathrm{IPSR+RDS-II}}^{(\mathrm{R2},\mathrm{M2})}$ & 0.380 & 10.247 & 0.224 & 65.791 & 0.141 & 82.028 & 0.212 & 95.080 \\ 
   \addlinespace
%   &  & %$\widehat{\Delta}_{\mathrm{IPO+SS}}^{(\mathrm{R2},\mathrm{M2})}$ & 0.386 & 11.942 & 0.241 & 70.994 & 0.162 & 83.936 & 0.231 & 95.281 \\ 
   &  & $\widehat{\Delta}_{\mathrm{IPPR+RDS-II}}^{(\mathrm{R2},\mathrm{M2})}$ & 0.344 & -0.361 & 0.190 & 55.165 & 0.125 & 81.627 & 0.192 & 95.783 \\ 
   \addlinespace
 %  &  & %$\widehat{\Delta}_{\mathrm{IPOW+SS}}^{(\mathrm{R2},\mathrm{M2})}$ & 0.342 & -0.830 & 0.210 & 60.913 & 0.139 & 83.735 & 0.206 & 95.984 \\ 
   \midrule
   & 20\% & $\widehat{\mu}_{\mathrm{UW}}$ & 0.156 & -20.750 & 0.037 & 27.687 & 0.031 & 69.439 & 0.058 & 96.493 \\ 
   \addlinespace
   &  & $\widehat{\mu}_{\mathrm{RDS-II}}$ & 0.166 & -15.522 & 0.075 & 40.947 & 0.051 & 72.745 & 0.093 & 92.485 \\ 
   \addlinespace
  % &  & $\widehat{\mu}_{\mathrm{CC+SS}}$ & 0.165 & -15.872 & 0.071 & 39.197 & 0.048 & 72.345 & 0.089 & 93.086 \\ 
   &  & $\widehat{\mu}_{\mathrm{IPSR+RDS-II}}^{(\mathrm{R2})}$ & 0.221 & 12.660 & 0.094 & 48.927 & 0.063 & 85.170 & 0.112 & 98.297 \\ 
   \addlinespace
  % &  & $\widehat{\mu}_{\mathrm{IPO+SS}}^{(\mathrm{R2})}$ & 0.221 & 12.395 & 0.089 & 46.442 & 0.059 & 86.673 & 0.108 & 98.497 \\ 
   &  & $\widehat{\mu}_{\mathrm{IPPR+RDS-II}}^{(\mathrm{R2})}$ & 0.192 & -2.051 & 0.075 & 37.691 & 0.052 & 83.267 & 0.094 & 97.695 \\
   \addlinespace
  % &  & $\widehat{\mu}_{\mathrm{IPOW+SS}}^{(\mathrm{R2})}$ & 0.194 & -0.968 & 0.071 & 35.897 & 0.049 & 84.870 & 0.090 & 98.096 \\ 
   \addlinespace
   &  & $\widehat{\Delta}_{\mathrm{UW}}^{(\mathrm{M2})}$ & 0.297 & -13.958 & 0.071 & 24.843 & 0.060 & 79.559 & 0.114 & 98.898 \\ 
   \addlinespace
   &  & $\widehat{\Delta}_{\mathrm{RDS-II}}^{(\mathrm{M2})}$ & 0.302 & -12.447 & 0.132 & 40.298 & 0.090 & 80.261 & 0.159 & 97.495 \\ 
   \addlinespace
  % &  & $\widehat{\Delta}_{\mathrm{CC+SS}}^{(\mathrm{M2})}$ & 0.302 & -12.515 & 0.128 & 39.197 & 0.086 & 80.160 & 0.156 & 97.796 \\ 
   &  & $\widehat{\Delta}_{\mathrm{IPSR+RDS-II}}^{(\mathrm{R2},\mathrm{M2})}$ & 0.376 & 8.916 & 0.176 & 51.720 & 0.111 & 83.768 & 0.189 & 98.397 \\ 
   \addlinespace
 %  &  & %$\widehat{\Delta}_{\mathrm{IPO+SS}}^{(\mathrm{R2},\mathrm{M2})}$ & 0.378 & 9.673 & 0.174 & 51.276 & 0.111 & 84.970 & 0.194 & 98.697 \\ 
   &  & $\widehat{\Delta}_{\mathrm{IPPR+RDS-II}}^{(\mathrm{R2},\mathrm{M2})}$ & 0.338 & -2.052 & 0.142 & 41.214 & 0.097 & 85.471 & 0.169 & 98.397 \\ 
   \addlinespace
   %&  & %$\widehat{\Delta}_{\mathrm{IPOW+SS}}^{(\mathrm{R2},\mathrm{M2})}$ & 0.340 & -1.360 & 0.153 & 44.435 & 0.097 & 86.373 & 0.172 & 98.597 \\ 
   \midrule 
   & 30\% & $\widehat{\mu}_{\mathrm{UW}}$ & 0.157 & -20.039 & 0.030 & 24.723 & 0.025 & 59.200 & 0.050 & 96.600 \\ 
   \addlinespace
   &  & $\widehat{\mu}_{\mathrm{RDS-II}}$ & 0.166 & -15.274 & 0.060 & 34.121 & 0.043 & 72.000 & 0.085 & 94.600 \\
   \addlinespace
  % &  & $\widehat{\mu}_{\mathrm{CC+SS}}$ & 0.165 & -15.825 & 0.055 & 31.978 & 0.038 & 71.400 & 0.078 & 94.400 \\ 
   &  & $\widehat{\mu}_{\mathrm{IPSR+RDS-II}}^{(\mathrm{R2})}$ & 0.219 & 11.544 & 0.075 & 39.507 & 0.053 & 88.700 & 0.102 & 99.100 \\ 
   \addlinespace
  % &  & $\widehat{\mu}_{\mathrm{IPO+SS}}^{(\mathrm{R2})}$ & 0.218 & 11.079 & 0.068 & 36.024 & 0.047 & 88.500 & 0.095 & 99.200 \\ 
   &  & $\widehat{\mu}_{\mathrm{IPPR+RDS-II}}^{(\mathrm{R2})}$ & 0.192 & -2.353 & 0.061 & 30.884 & 0.044 & 85.500 & 0.085 & 98.700 \\ 
   \addlinespace
  % &  & $\widehat{\mu}_{\mathrm{IPOW+SS}}^{(\mathrm{R2})}$ & 0.195 & -0.575 & 0.057 & 28.292 & 0.039 & 87.400 & 0.079 & 99.300 \\ 
   \addlinespace
    &  & $\widehat{\Delta}_{\mathrm{UW}}^{(\mathrm{M2})}$ & 0.295 & -14.562 & 0.057 & 22.033 & 0.048 & 73.000 & 0.097 & 99.000 \\ 
    \addlinespace
   &  & $\widehat{\Delta}_{\mathrm{RDS-II}}^{(\mathrm{M2})}$ & 0.302 & -12.403 & 0.109 & 33.916 & 0.076 & 81.100 & 0.144 & 98.500 \\ 
   \addlinespace
  % &  & $\widehat{\Delta}_{\mathrm{CC+SS}}^{(\mathrm{M2})}$ & 0.302 & -12.520 & 0.101 & 31.886 & 0.068 & 80.600 & 0.136 & 98.600 \\ 
   &  & $\widehat{\Delta}_{\mathrm{IPSR+RDS-II}}^{(\mathrm{R2},\mathrm{M2})}$ & 0.373 & 8.242 & 0.133 & 39.529 & 0.094 & 84.600 & 0.171 & 99.400 \\ 
   \addlinespace
  % &  & %$\widehat{\Delta}_{\mathrm{IPO+SS}}^{(\mathrm{R2},\mathrm{M2})}$ & 0.369 & 6.977 & 0.138 & 40.629 & 0.084 & 84.400 & 0.165 & 99.400 \\ 
   &  & $\widehat{\Delta}_{\mathrm{IPPR+RDS-II}}^{(\mathrm{R2},\mathrm{M2})}$ & 0.336 & -2.577 & 0.115 & 33.295 & 0.082 & 85.600 & 0.153 & 99.200 \\ 
 %  &  & %$\widehat{\Delta}_{\mathrm{IPOW+SS}}^{(\mathrm{R2},\mathrm{M2})}$ & 0.341 & -1.198 & 0.118 & 34.081 & 0.076 & 85.600 & 0.148 & 99.200 \\ 
\bottomrule
\end{tabular}
\begin{tablenotes}
      \small
     \item[] The average true prevalence is $\mu = 0.196$, and the true 
prevalence difference is $\Delta = 0.345$. Abbreviations: Av. Est., 
average estimate; Rel. Bias, relative bias; SD, Monte Carlo standard 
deviation; RRMSE, relative root mean squared error; Av. Boot. SE, average 
bootstrap standard error; Cov., empirical coverage probability; NBoot, neighbourhood bootstrap; TBoot, tree bootstrap.
    \end{tablenotes}
\end{threeparttable}
\end{table}

\section{Case study}\label{sec5}
We applied the proposed methods to data from the Montreal site of Engage, a multisite Canadian study of gay, bisexual, and other men who have sex with men. Engage was designed to investigate individual-, social-, and community-level determinants of HIV and sexually transmitted infections and related behaviours among GBM \citep{hart2021prevalence, moore2021characteristics, cox2021use}. In this case study, we estimated the prevalence of three bSTIs and prevalence differences according to recent use of substances commonly associated with chemsex.

%We applied our method to an RDS dataset collected through the Engage study, a national cross-sectional study conducted in Montreal, Toronto, and Vancouver in Canada, with the goal of determining the individual-, social-, and community-level factors that impact transmission of HIV and sexually transmitted infections among GBM. The main goal of the Engage study is to identify individual, social, and community-level factors that influence HIV and STI transmission and related behaviours among GBM. In this example, we focus on data collected in Montreal, and aim to estimate prevalences of bSTI. 

\subsection{Participant recruitment and data collection}
\begin{figure}[!htbp]
\centering
\includegraphics[width=\linewidth,height=0.58\textheight,keepaspectratio]{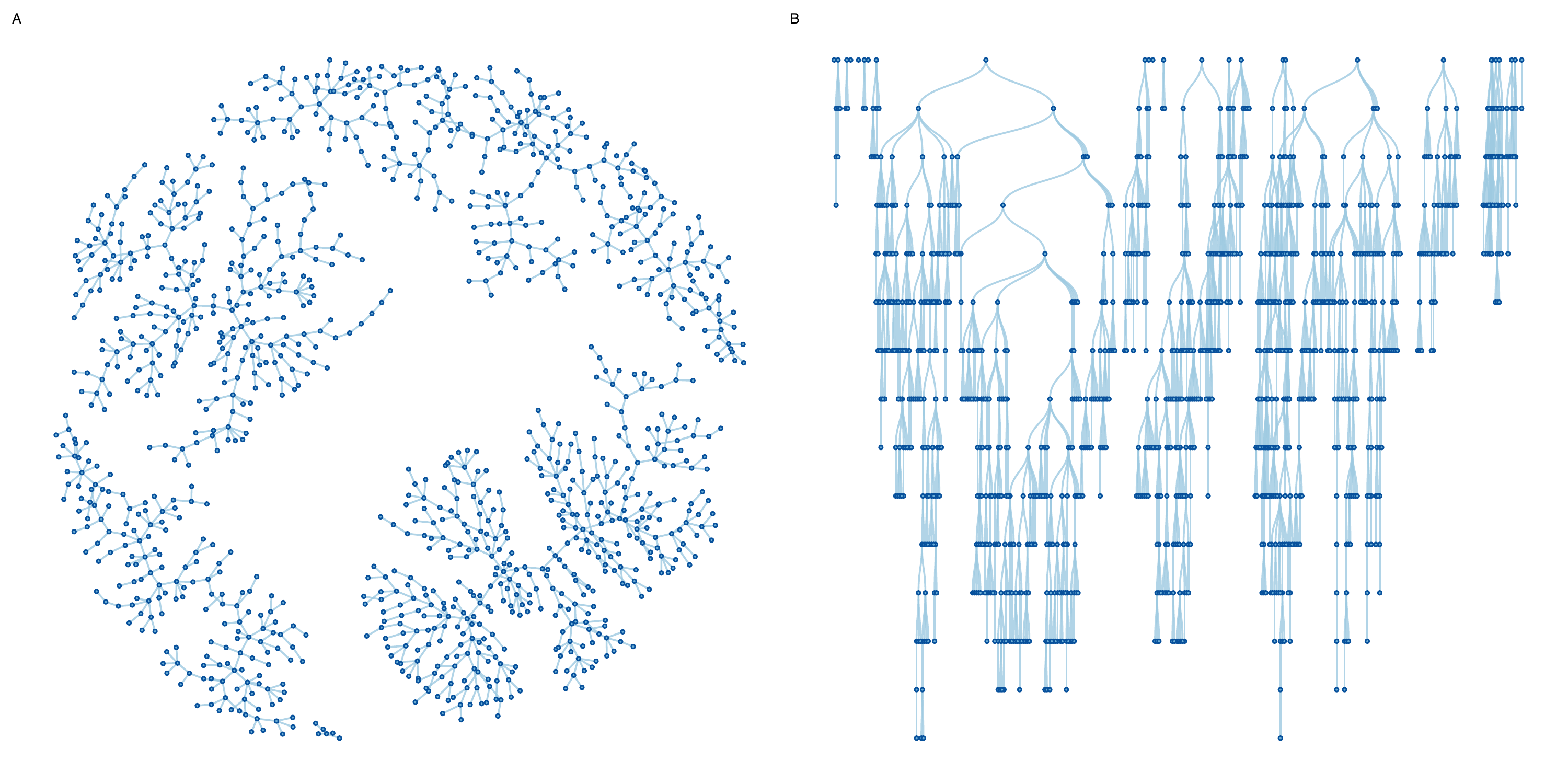}
\caption{Representations of the observed RDS recruitment graph among $n=1179$ gay, bisexual and other men who have sex with men (GBM) in the Engage study, Montreal, 2018. The left-hand side uses a Fruchterman-Reingold layout, whereas the right-hand side uses a tree layout that reveals the forest structure of the recruitment graph. }
\label{fig:recruitment}
\end{figure}

Participants were recruited using respondent-driven sampling (RDS). Starting with a convenience sample of 27 seeds, recruitment conducted between February 2017 and June 2018 led to a baseline sample of $n=1179$ GBM in Montreal. Of the 27 seeds, 21 initiated recruitment chains containing at least one recruit. Figure~\ref{fig:recruitment} presents the observed RDS recruitment graph.

\begin{figure}[!htbp]
\centering
\includegraphics[width=0.96\linewidth,height=0.58\textheight,keepaspectratio]{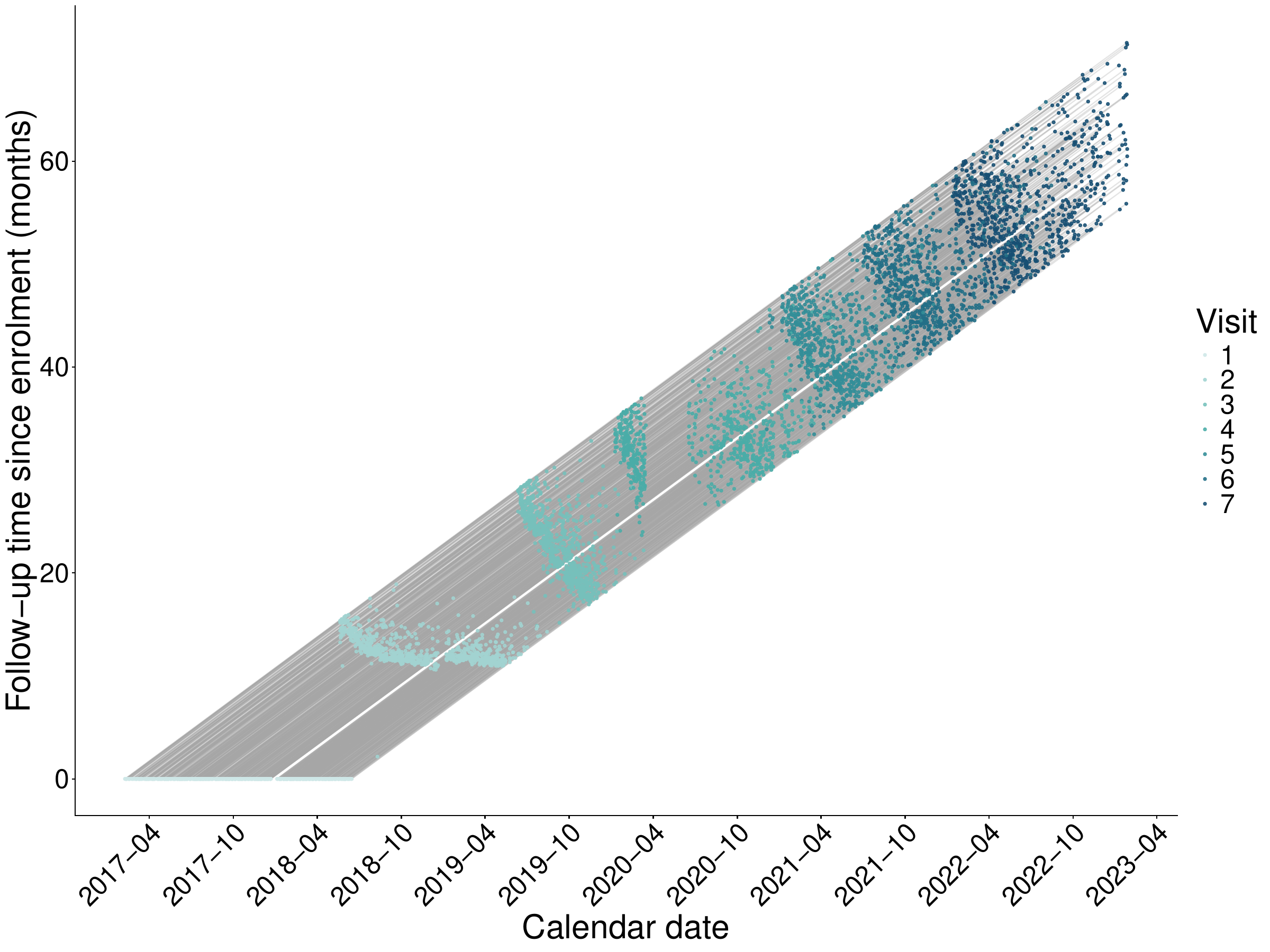}
\caption{Longitudinal follow-up trajectories for all participants in the study cohort. Each grey line represents an individual participant’s trajectory across calendar time, and blue points indicate observed study visits, plotted according to the time elapsed since enrolment. The staggered trajectories reflect rolling enrolment into the cohort and variation in the duration and completeness of follow-up.}
\label{fig:lexis}
\end{figure}

A total of $K=7$ visits including baseline were planned for each participant, with data collection occurring between February 2017 and January 2023. Following the baseline visit, participants were invited to return approximately every 6--12 months. A Lexis diagram representing the longitudinal follow-up trajectories can be visualised in Figure \ref{fig:lexis}. We defined loss to follow-up as a monotone pattern of non-attendance beginning before visit 6, the penultimate planned visit, with no subsequent return. Intermittent non-attendance was defined as missing one or more visits followed by attendance at a later visit. These classifications were not mutually exclusive over the entire follow-up: a participant could return after an intermittently missed visit and subsequently become lost to follow-up. Figure G1 in Web Appendix G shows longitudinal observation patterns for 12 hypothetical participants and contains examples of either LTFU or INA. In total, 371 participants met the definition of LTFU, 252 experienced INA, and 47 experienced INA before subsequently becoming lost to follow-up. A detailed description of the attrition pattern over the course of the study is presented in Table G1 in Web Appendix G. Unweighted and RDS-weighted descriptive statistics are presented in Tables G2 and G3 of Web Appendix G. It is worth noting that participants who met the definition of LTFU tended to have smaller reported network sizes.

\subsection{Outcomes, exposure, and covariates}

At each selected study visit, the outcomes of interest were binary indicators of laboratory-confirmed syphilis, chlamydia, and gonorrhoea. See \cite{hart2021prevalence} and \cite{lambert2026incidence} for more detail regarding the ascertainment of active bSTI infections. Screening for CT and NG was performed using nucleic acid amplification testing or culture, according to provincial laboratory procedures \citep{hart2021prevalence}. For CT and NG, a participant was classified as having an infection if a positive result was obtained from a urine, pharyngeal, or rectal specimen \citep{hart2021prevalence}. 
We did not incorporate self-reported diagnoses occurring during the preceding six months because these represented infections over an interval rather than infection status at the study visit. Hence, resulting estimates should be interpreted as visit-specific prevalences and do not capture infections diagnosed between visits.

The exposure of interest was self-reported use during the preceding six months of any substance commonly associated with chemsex, including crystal methamphetamine, mephedrone, gamma-hydroxybutyrate (GHB), and ketamine \citep{bourne2015illicit,lambert2026incidence}.

Potential confounders were specified a priori based on previous literature and subject-matter knowledge. Baseline covariates included age, Black, Indigenous, and People of Color (BIPOC) indicator, marital status, educational attainment, immigration status, sexual identity, past six months (P6M) number of sexual partners, P6M condomless anal sex, P6M injection drug use, and history of sexual abuse.

\subsection{Analyses}

We fitted pooled logistic models for dropout and attendance described in equations~(\ref{eqn:dropout}) and~(\ref{eqn:attendance}), respectively, both with and without baseline RDS-II weights. The observed history $\mathcal{H}_{ik}$ included the prespecified baseline covariates, baseline STI history, study wave, and the number of previously missed visits. Response weights were combined with RDS-II weights as in equation~(\ref{eqn:finalweight}) and truncated at the 95th percentile.

We estimated crude bSTI prevalences and model-standardized prevalence differences using four methods: unweighted, RDS-II, IPSR+RDS-II, and IPPR+RDS-II. For standardization, we fitted weighted logistic GEE models adjusted for the prespecified confounders, with participant-level clustering and an independent working covariance structure. Prevalences under chemsex-related substance use and no such use were standardized to the empirical covariate distribution. We interpret their difference as an adjusted population-level association.

We report baseline (February 2017--June 2018), visit 4 (January 2020--April 2021), and visit 6 (July 2021--August 2022), whose calendar windows did not overlap. Each participant contributed at most one observation per reported visit, with its corresponding weights. We obtained 95\% confidence intervals using seed-clustered sandwich estimates and neighbourhood- and tree-bootstrap percentile intervals ($B=1000$ each), refitting the response and outcome models and recalculating weights in every bootstrap sample.

\subsection{Results}
The estimated prevalence of each bSTI varied across study waves (Figure \ref{fig:engageprevdiff} and Table G4 in Web Appendix G). Under IPSR+RDS-II, syphilis prevalence was 2.0\% at baseline, 1.5\% at visit 4, and 1.6\% at visit 6. The corresponding estimates were 3.4\%, 4.0\%, and 0.9\% for chlamydia and 6.0\%, 2.5\%, and 3.3\% for gonorrhoea. Thus, gonorrhoea prevalence decreased after baseline before increasing slightly at visit 6. Chlamydia prevalence was lower at visit 6, although visit-specific estimates may depend on assessment timing and do not necessarily reflect trends in incidence. At baseline, RDS-II weighting lowered the estimated prevalence of all three bSTIs relative to the unweighted analysis. Adjustment for attrition (IPSR+RDS-II and IPPR+RDS-II) had its clearest impact on syphilis at visits 4 and 6, where the RDS-II estimate increased from 1.0\% to approximately 1.5--1.6\%, and on chlamydia at visit 4, where it decreased from 5.6\% to approximately 4.0\%. Gonorrhoea prevalence was less sensitive to attrition adjustment.

Across all outcomes, visits, and weighting methods, the estimated standardized prevalence differences were positive, suggesting a higher population prevalence under the hypothetical scenario in which all participants used at least one chemsex-related substance than under the scenario in which none did (Figure \ref{fig:engageprevdiff} and Table G5 in Web Appendix G). The largest differences were observed for gonorrhoea: the IPSR+RDS-II and IPPR+RDS-II estimates were respectively 5.5 and 5.8 percentage points at baseline, 2.8 and 3.0 percentage points at visit 4, and 3.9 percentage points under both methods at visit 6. The corresponding differences for syphilis ranged from 1.6 to 2.3 percentage points. For chlamydia, they were approximately 2.0--2.6 percentage points at baseline and visit 4 but decreased to 0.7 percentage points at visit 6.

The IPSR+RDS-II and IPPR+RDS-II procedures produced similar results, with absolute differences no greater than 0.3 percentage points across all reported prevalence and prevalence difference estimates and generally similar confidence intervals. Thus, although the simulations demonstrated settings in which RDS-weighting the response model was important, doing so had little practical effect in this application. Greater sensitivity arose from the uncertainty-quantification procedure. The neighbourhood-bootstrap intervals were consistently narrower and excluded zero for every prevalence difference estimate, whereas the sandwich intervals included zero for several syphilis and chlamydia contrasts. The tree bootstrap intervals were substantially wider and included zero for every IPSR+RDS-II and IPPR+RDS-II contrast, consistent with the greater variability and conservative coverage observed in the simulations.

\iffalse

\begin{figure}
\includegraphics[scale=0.2]{laboratory_sti_prevalence2.pdf}
\caption{Estimated prevalence of laboratory-confirmed (a) syphilis, (b) chlamydia (CT), and (c) gonorrhoea (NG) at selected study visits, by weighting method. Error bars show 95\% confidence intervals obtained using the neighbourhood bootstrap.}
\label{fig:engageprev}
\end{figure}

\begin{figure}
\includegraphics[scale=0.2]{laboratory_sti_prevalence_difference.pdf}
\caption{Model-standardized bSTI prevalence differences comparing participants reporting chemsex-related substance use with those reporting no such use, standardized to a common covariate distribution, at selected study visits and by weighting method. The error bars represent 95\% neighbourhood bootstrap confidence intervals.}
\label{fig:engageprevdiff}
\end{figure}

\fi

\begin{figure}[!htbp]
\centering
\includegraphics[width=\linewidth,height=0.60\textheight,keepaspectratio]{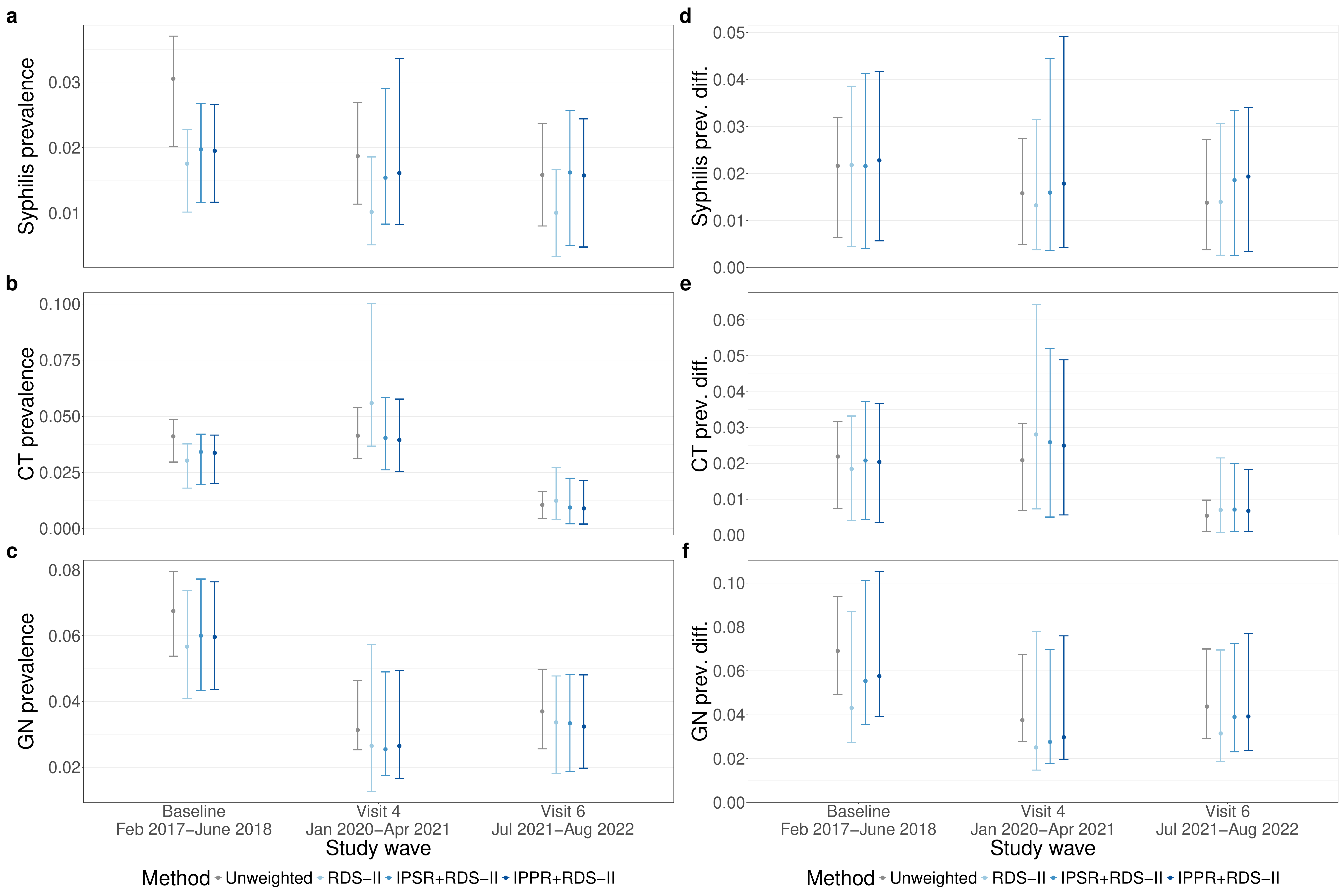}
\caption{\textit{Left}: Estimated prevalence of laboratory-confirmed (a) syphilis, (b) chlamydia (CT), and (c) gonorrhoea (NG) at selected study visits, by weighting method. \textit{Right}: Model-standardized of laboratory-confirmed (d) syphilis, (e) CT, and (f) NG prevalence differences comparing participants reporting chemsex-related substance use with those reporting no such use, standardized to a common covariate distribution, at selected study visits and by weighting method. The error bars represent 95\% neighbourhood bootstrap confidence intervals.}
\label{fig:engageprevdiff}
\end{figure}

\FloatBarrier

\section{Discussion}\label{sec6}

In this work, we propose a weighting framework for longitudinal RDS studies that jointly accounts for unequal inclusion probabilities at recruitment and selection arising during follow-up. The framework distinguishes terminal loss to follow-up from intermittent non-attendance and combines baseline RDS weights with visit-specific IPO weights. We further distinguish population-standardized response mechanism from that in the realised RDS sample. When characteristics related to the sampling design remain associated with response after conditioning on the observed history, fitting the response models using baseline RDS weights provides a principled way to target the population-standardized response mechanism.

The validity of the proposed approach depends on the usual assumptions underlying RDS inference \citep{gile20107}, together with sequential MAR and positivity assumptions for both retention and attendance. In particular, the missingness must be plausibly explainable by the observed participant history. If dropout or attendance depends on unobserved current or future outcomes, even after conditioning on past data, weighting alone may not identify the target estimand. Similarly, prevalence difference estimands require sufficient adjustment for confounding of the exposure--outcome relationship. The procedure also depends on adequate specification of the response and outcome models. Near-violations of positivity may produce highly variable weights, while weight truncation can improve stability but may introduce some bias. More flexible nuisance models and doubly or multiply robust estimators could reduce reliance on correctly specified models.

An additional caveat concerns the population targeted by the longitudinal analysis. The RDS weights and inclusion indicators $S_1,S_2,\ldots,S_N$ are time-invariant and reflect the network structure and population characteristics \textit{at recruitment}. Thus, design adjustment at a later visit targets the population represented at enrolment, carried forward under the hypothetical absence of attrition, rather than necessarily representing the population present at that calendar time. Interpreting the estimates as contemporaneous population quantities therefore requires additional stability or transportability assumptions. The present framework does not accommodate changes in population membership or network structure after recruitment. Therefore, we need to assume that the population at recruitment remains stable over follow-up.

Uncertainty quantification also requires further investigation. In the simulations, the neighbourhood bootstrap tended to underestimate variability, whereas the tree bootstrap produced larger standard errors and was frequently conservative. One important caveat is that our simulation studies used a relatively small sample size compared to previous studies. Therefore, our findings regarding bootstrap uncertainty estimation are compatible with those of previous authors \citep{yauck2021general, yauck2022neighborhood}. A similar situation was observed in the Engage analysis, where substantive conclusions concerning the prevalence differences changed with the uncertainty procedure. Future work should investigate alternative resampling methods that preserve both dependence along the RDS recruitment structure and repeated observations within participants. Potential approaches include network model-assisted RDS resampling procedures \citep{gile2015network}.

Finally, the present analysis treats visits as discrete assessment occasions and does not explicitly account for variation in their timing. This is particularly limiting for survival and recurrent-event outcomes, for which both the timing of events and the observation process may be informative. An important extension would jointly model the visit and outcome processes in continuous time, while simultaneously accounting for informative censoring and baseline RDS recruitment. Approaches based on inverse-intensity weighting and joint modelling may provide useful starting points. Such developments would extend the proposed framework beyond visit-specific analyses while retaining its central objective: accounting simultaneously for selection at recruitment and selection during follow-up in RDS longitudinal studies.

\section*{Acknowledgements}

The Engage Study is led by the following principal investigators (in alphabetical order): Joseph Cox, Daniel Grace, Trevor Hart, Jody Jollimore, Nathan Lachowsky, Gilles Lambert, David Moore and Darrell HS Tan. The authors thank Engage Study participants, research staff and community
engagement committee members, as well as their community partner agencies.

VM holds a postdoctoral fellowship from the Natural Sciences and Engineering Research Council of Canada (NSERC). EEMM holds a Canada Research Chair (Tier 1) from the Canadian Institutes of Health Research in Statistical Methods for Precision Medicine. 

The authors declare no conflicts of interest.

\FloatBarrier

\bibliographystyle{plainnat}
\bibliography{main}

\section*{Supporting Information}
The Web Appendices, tables, and figures referenced in Sections 1–4 are included below.

\clearpage
\appendix
\setcounter{figure}{0}
\setcounter{table}{0}
\setcounter{equation}{0}
\renewcommand{\theHfigure}{web.\thefigure}
\renewcommand{\theHtable}{web.\thetable}
\renewcommand{\theHequation}{web.\theequation}

\newcounter{webappendix}

\renewcommand{\thewebappendix}{\Alph{webappendix}}
\renewcommand{\thefigure}{\thewebappendix\arabic{figure}}
\renewcommand{\thetable}{\thewebappendix\arabic{table}}

% Unique hyperlink identifiers when counters restart
\renewcommand{\theHfigure}
  {\arabic{webappendix}.\arabic{figure}}
\renewcommand{\theHtable}
  {\arabic{webappendix}.\arabic{table}}

\newcommand{\webappendix}[2]{%
    \FloatBarrier
    \refstepcounter{webappendix}%
    \setcounter{figure}{0}%
    \setcounter{table}{0}%
    \section*{Web Appendix \thewebappendix: #1}%
    \label{#2}%
}

\usetikzlibrary{positioning,shapes.geometric,arrows.meta}

%\newtheorem{proposition}{Proposition}
%\newcommand{\ind}{\perp\!\!\!\!\perp}

% Biometrics asks supplementary components to be named Web Appendices,
% Web Tables, and Web Figures.
\renewcommand{\figurename}{Web Figure}
\renewcommand{\tablename}{Web Table}
\setcounter{figure}{0}
\setcounter{table}{0}

\label{firstpage}

% The legacy biom class distributed with some projects does not provide a
% reliable \maketitle command for a stand-alone supplement.  Typeset the
% supplementary title page explicitly instead.
\begin{center}
{\Large\bfseries Supplementary Material for ``Accounting for Attrition in
Respondent-Driven Sampling Cohort Studies''\par}
\vspace{1.25em}
{\normalsize Vanessa McNealis$^{1,*}$, Erica E. M. Moodie$^{2}$,
Alain Fourmigue$^{3}$, Milada Dvorakova$^{3}$, and Joseph Cox$^{2}$\par}
\vspace{0.75em}
{\small
$^{1}$School of Mathematics and Statistics, University of Glasgow,
Glasgow, United Kingdom\\
$^{2}$Department of Epidemiology and Biostatistics, McGill University,
Montr\'eal, Canada\\
$^{3}$Research Institute of the McGill University Health Centre,
Montr\'eal, Canada\\[0.5em]
$^{*}$Corresponding author: \texttt{vanessa.mcnealis@glasgow.ac.uk}\par}
\end{center}
%\vspace{2em}

% -------------------------------------------------------------------------
% Rename, add, or remove Web Appendices as needed. When moving material from
% the main paper, retain useful equation labels but check for duplicates.
% -------------------------------------------------------------------------

\webappendix{Related work on combining design, propensity score and non-response weights}{app:relatedwork}
% Introductory paragraph
The analysis of longitudinal survey data often requires combining methods for complex survey data with methods for longitudinal data analysis \citep{thompson2015using}. In longitudinal settings, weighting methods may be used to explore causal hypotheses, for example through inverse probability of treatment (IPT) weighting, or to correct for attrition and intermittent observation, for example through inverse probability of observation (IPO) weighting. A longstanding methodological question is whether the models used to construct these weights should account for the original survey design. Here, we briefly review two related literatures that have addressed this question: propensity score methods for complex survey data, and non-response adjustment in surveys.

Several authors have studied the use of propensity score methods with complex survey data in causal analyses. \cite{brumback2010testing} discuss the estimation of marginal structural model parameters using inverse weights formed as the product of a survey weight, which adjusts the sample to represent the unstandardized population, and an IPT weight, which standardizes the population to adjust for confounding. They argue that the IPT weights should be estimated using the survey weights, because the treatment model is intended to describe the target population rather than the realised sample. By contrast, \cite{zanutto2006comparison} and \cite{dugoff2014generalizing} argue that the propensity score need not be estimated using survey weights, because they view propensity scores as an in-sample balancing device rather than a population model. In response to these conflicting recommendations, \cite{ridgeway2015propensity} show through analytical derivations and simulation that, unless all features used to construct the sampling weights are available for propensity score estimation, consistent estimation of population parameters generally requires estimating the IPT weights using the survey weights.

Parallel debates have arisen in the survey non-response literature. \cite{little2003weighting} and \cite{chen2015incorporating} provide theoretical and simulation-based evidence that simply including the design weight in a logistic non-response model is not necessarily beneficial, and recommend instead incorporating design information through model covariates or calibration cells. Similarly, \cite{grau2006nonresponse} found in an applied setting that weighted and unweighted logistic response-propensity adjustments produced comparable results, providing little empirical support for routinely weighting the non-response model. In contrast, \cite{kott2012one} argues that, under certain response models, using design weights in response-propensity adjustment or calibration can yield more efficient and nearly unbiased estimates. Overall, this literature cautions against the routine inclusion of sampling weights in non-response models, while recognizing that weighting may be advantageous under some assumptions.

Together, these literatures suggest that whether and how survey weights should enter the estimation of IPT or IPO weights depends on the target estimand, the interpretation of the weighting model, and the extent to which the design information encoded in the sampling weights is otherwise represented in the model. These issues are especially relevant for RDS studies. RDS is not just another survey design: inclusion probabilities are closely tied to degree/network position, and those same features may plausibly predict attrition. Therefore, we make the case that RDS design information should be included in the non-response model.

%\newpage
\webappendix{Review of RDS methodology}{app:review}

\section*{Sampling design and assumptions}\label{subsec21}
In this section, we review the assumptions underlying the RDS design. Suppose that a target population consists of $N$ individuals (e.g., GBM in Montreal). We assume that the individuals in the population are connected by social ties. We define this as the population network and state the following:

\begin{assumption}
\label{rds:network}
The social network connecting the $N$ individuals in the population exists and can be represented as the undirected graph $G=(\mathcal{N}, \mathbb{E})$ devoid of parallel edges or self-loops.
\end{assumption}

 The elements in $\mathcal{N} = \{1, 2, \ldots, N\}$ are the $N$ individuals in the population and $\mathcal{E}$ is a set of edges with generic element $\{i,j\} = \{j,i\}$ denoting the presence of an edge between nodes $i$ and $j$. We assume that $G$ has no parallel edges, meaning that each pair of vertices in $G$ is connected by at most one edge, and no self-loops, meaning that no vertex is connected to itself. We represent the network $G$ by the sociomatrix $\mathbf{B}$, which has entries $B_{ij} = 1$ if $\{i,j\} \in \mathbb{E}$, $B_{ij} = 0$ otherwise, and $B_{ii} = 0$. We define the neighbourhood $\mathcal{N}_i$ of individual $i$, $i=1,2,\ldots, N$, as the set of nodes $j$ which satisfy $B_{ij} = 1$. We define the cardinality of $\mathcal{N}_i$, given by $d_i = |\mathcal{N}_i|$, as the degree of individual $i$.

 Unlike traditional surveys, RDS does not require the specification of a sampling frame, which is why it has emerged as a convenient method for sampling hard-to-reach populations. The typical RDS process proceeds as follows. First, in wave zero of recruitment, a group of $s$ seed participants are recruited to form a convenience sample, which aims to be representative of the target population. Seed participants are then given $c$ uniquely identified coupons and asked to distribute them among peers who are also members of the population of interest. Each seed recruits additional participants up to the number of coupons received $c$. It is assumed that each individual recruits only individuals with whom they share social ties. This process continues through a number of successive waves until the target sample size $n$ is achieved. 

We formalize this description by stating the following assumptions regarding the graphical structure of the RDS process on the population network $G$:

\begin{assumption}
\label{rds:undirected}
The RDS recruitment process takes place across edges of the undirected graph $G$.
\end{assumption}

\begin{assumption}
\label{rds:wor}
No individual can be recruited more than once into the study.
\end{assumption}

Assumption \ref{rds:wor} implies that successive samples are obtained by sampling from the remaining unsampled neighbours of sampled nodes. We let $G_T = (\mathcal{N}_T, \mathcal{E}_T)$ denote the resulting directed recruitment forest, where $\mathcal{N}_T \subseteq \mathcal{N}$ is the sampled vertex set with $|\mathcal{N}_T| = n$, and $\mathcal{E}_T \subseteq \mathbb{E}$ contains the observed recruiter-to-recruit edges. 

\section*{Parameter inference under RDS}\label{subsec22}

Assume that for nodes in $\mathcal{N}$, we observe the population data $\mathcal{DP}= \{O_i = (Y_i, \bm{L}_i) ; i = 1, \ldots N\}$, where each pair $O_i = (Y_i, \bm{L}_i)$ consists of an outcome $Y_i$ and a vector of covariates $\bm{L}_i$ of dimension $p$ associated to node $i$. Similar to \cite{yauck2021general}, we assume that $\mathcal{DP}$ is a random realization of a random sample from a distribution $\mathcal{F}_0$ with mean $\mu_0$ and variance $\sigma^2_0$. Then, following the RDS sampling process, we observe the tree data $\mathcal{DT}= \{O_i = (Y_i, \bm{L}_i) ; i \in \mathcal{N}_T\}$, which is a subset of $\mathcal{DP}$.

\subsection*{Population prevalences and means}

We first discuss inference for finite population parameters, which pertain to $\mathcal{DP}$. A common target of inference in RDS studies is the mean in the finite population of nodes $\mathcal{N}$, i.e.,  $\mu = N^{-1} \sum_{i \in \mathcal{N}} Y_i$. If $Y_i$ is a two-valued variable indicating the presence or absence of an attribute, such as the presence or absence of bSTI, then $\mu$ is a population prevalence. Given a sample of $n$ participants formed through an RDS strategy from $\mathcal{N}$, then the Horvitz-Thompson estimator for $\mu$ is given by 
$$ \hat{\mu}_{IPW} = \frac{1}{N}\sum_{i \in \mathcal{N}} \frac{Y_i S_i}{{\pi}_i},$$
where $S_i$ is a sampling indicator, with $S_i = 1$ if a unit is included in the sample, and $S_i = 0$ otherwise, $\pi_i =\mathbb{E}[S_i|\mathcal{S}]$, and $\mathcal{S}$ is the true sampling design.

Unlike traditional surveys, the sampling probabilities $\pi_i$ in an RDS process are unknown to the study investigators. Under Assumptions \ref{rds:network}--\ref{rds:wor} and additional irreducibility conditions documented in \cite{gile20107}, the RDS process can be approximated by a Markov process on $G$ with unique stationary distribution where draw-wise probabilities $p_i$ for a single sample selection step are proportional to degree, i.e., $p_i = \alpha d_i$ for some constant $\alpha$. Building upon the form of the Hansen-Hurwitz estimator for proportional-to-size sampling designs, \cite{volz2008probability} used this property to construct the following RDS-II estimator of $\mu$:
$$ \hat{\mu}_{VH} = \frac{\sum_{i \in \mathcal{N}} S_i \frac{Y_i}{{\hat{\pi}^{VH}_{i}}}}{\sum_{i\in \mathcal{N}} S_i\frac{1}{\hat{\pi}_i^{VH}}},$$
where $\hat{\pi}_i^{VH}=  n d_i/\sum_{i \in \mathcal{N}_T} d_i$ is the Volz-Heckathorn (VH) estimator of the inclusion probability of unit $i$ in the RDS sample. These estimated estimated stationary draw probabilities reflect the Markov approximation of the sampling process, in that they are proportional to an individual's degree or `popularity'. Also note that they do not require knowledge of the population size $N$. From this point on, we will refer to the corresponding inverse probability weights, $w_i^{VH} = 1/\hat{\pi}_i^{VH}$ as the RDS-II weights. 

The sampling weights of the VH estimator are derived from a single non-branching, with-replacement random walk process at equilibrium, which is a simplification of the true RDS process. The validity of the VH estimator hinges upon a large network size with respect to the sample size (i.e., a small sampling fraction). Further, many waves of sampling are required to justify its reliance on a stationary distribution. The successive sampling approximation, introduced by \cite{gile2011improved}, assumes a self-avoiding, without-replacement random walk process in which nodes with higher degree tend to be sampled in earlier recruitment waves. \cite{gile2011improved} proposed the successive sampling (SS) estimator given by
$$ \hat{\mu}_{SS} = \frac{\sum_{i \in \mathcal{N}} S_i \frac{Y_i}{{\hat{\pi}^{SS}_{i}}}}{\sum_{i\in \mathcal{N}} S_i\frac{1}{\hat{\pi}_i^{SS}}},$$
where the estimated inclusion probabilities $\hat{\pi}_i^{SS} \equiv \hat{\pi}_i^{SS}(d_i)$ are computed by simulating successive samples under a proportional to size without replacement design, where the size variable is a node's degree \citep{gile2018methods}. As opposed to VH, the SS estimator accounts for finite population effects, which can be important for large sampling fractions \citep{gile2018methods, yauck2021general}. However, knowledge of the population size $N$ or an estimate of the latter $\hat{N}$ is required for the computation of SS weights, given by $w_i^{SS} = 1/\hat{\pi}^{SS}_{i}$.

\subsection*{Regression models}
It might also be of interest to RDS investigators to estimate a regression model from $\mathcal{DT}$ by relating the outcome $Y_i$ to the covariates $\bm{L}_i$ such as
$$g(\mathbb{E}[Y_i | \bm{L}_i]) = \bm{\beta}\bm{L}_i,$$
where $g(\cdot)$ is a suitable link function and the target of inference $\bm{\beta}$ is a vector of regression coefficients of dimension $p$. For instance, we might want to understand the influence of certain sexual practices on the risk of being diagnosed with a bSTI. While multivariate modelling using respondent-driven sampling data also presents challenges, recent research efforts have focused on studying properties of regression estimators under different strategies and providing guidance \citep{spiller2009regression, gile2018methods, avery2019unweighted, yauck2021general}. In particular, generalized linear mixed model extensions were investigated to account for clustering by a shared recruiter or seed, as well as autoregressive correlation along recruitment chains \citep{spiller2009regression, avery2019unweighted, yauck2021general}. Letting $j$ denote the seed or shared recruiter index and $i$ denote the recruitee index within the cluster defined by seed or shared recruiter, we can rewrite the observed sample data $\mathcal{DT}$ as $\mathcal{DT}= \{O_{ij} = (Y_{ij}, \bm{L}_{ij}) ; j = 1, \ldots, m; i = 1, \ldots, n_j\}$, where $m$ is the total number of distinct seeds or shared recruiters, and $n_j$ denotes the total number of recruitees for the $j$th shared seed or recruiter. Using this data representation, it is common to consider the following generalized linear mixed model
$$g(\mu_{ij}) = \bm{\beta}\bm{L}_{ij} + b_j + \varepsilon_{ij}, $$ 
where $\mu_{ij} = \mathbb{E}[Y_{ij} | \bm{L}_{ij}, \gamma_{j}]$, $b_j$ is a shared seed-level or shared recruiter-level random effect and $\varepsilon_{ij}$ is a measurement error term for the $i$th individual with seed or shared recruiter $j$. We further assume that $\bm{\varepsilon}_j = (\varepsilon_{1j}, \ldots, \varepsilon_{n_j j}) \sim N(\bm{0}, \bm{\Sigma}_j)$. The additional error term $\varepsilon_{ij}$ departs from the usual GLMM formulation and is used to account for potential autocorrelation along recruitment chains \citep{yauck2021general}. For example, an analyst can impose a conditional autoregressive or a simultaneous autoregressive (SAR) structure for the measurement error vector $\bm{\varepsilon}_j$. The RDS design is then accounted for by incorporating RDS-II or SS weights in the model likelihood. 

% Introduce GEEs here

A second common approach to modelling longitudinal survey data involves a variant of the generalized estimating equation (GEE) approach described
by \cite{LiangZeger1986}. GEE models extend generalized linear models to accommodate correlated or clustered response data, such as repeated measurements on the same unit \citep{LiangZeger1986}. Rather than specifying a full joint distribution for the responses, GEEs model the marginal mean while accounting for within-cluster correlation through a working covariance structure.
Let $\tilde{\mu}_{ij} = \mathbb{E}[Y_{ij}]$ denote the marginal mean for subject $i$ with seed or shared recruiter $j$. The marginal mean is related to the set of covariates $\bm{L}_{ij}$ through a link function $g(\cdot)$,
\[
g(\tilde{\mu}_{ij}) = \bm{l}_{ij}^\top \boldsymbol{\gamma},
\]
where $\boldsymbol{\gamma}$ is a vector of regression coefficients representing population-averaged effects and $\bm{l}_{ij} = (l_{1,ij}, \ldots l_{p,ij})^{\top}$ denotes the covariate vector for individual $i$ in cluster $j$. We distinguish them from the parameter vector $\boldsymbol{\beta}$, which have been used to represent the mixed model conditional parameters. %The marginal variance of the outcomes is modeled as $\mathrm{Var}(\mathbf{Y}) = \mathbf{V}$, where $ \mathbf{V} = \mathbf{V}_i (\bm{\gamma}, \bm{\alpha})$ and $\bm{\alpha}$ represent additional parameters.

If weights are incorporated in the model, the GEE estimator $\hat{\boldsymbol{\gamma}}$ is obtained by solving the estimating equation
\begin{align*}
\sum_{j=1}^{m} \mathbf{D}_j^\top \mathbf{\Omega}_j \mathbf{V}_j^{-1}(\mathbf{Y}_j - \boldsymbol{\mu}_j) = \mathbf{0},
\label{eqn:gee}
\end{align*}
where $\bm{Y}_j = (Y_{1j}, \ldots, Y_{n_j,j})^\top$, $\boldsymbol{\mu}_j = (\mu_{1j}, \ldots, \mu_{n_j,j})^\top$, $\mathbf{D}_j = \partial \boldsymbol{\mu}_j / \partial \boldsymbol{\gamma}^\top$, $\mathbf{V}_j = \mathbf{V}_j(\bm{\gamma}, \hat{\bm{\alpha}})$ denotes the working covariance matrix of $\mathbf{Y}_j$, $\hat{\bm{\alpha}}$ is a consistent estimator of $\bm{\alpha}$, and $\mathbf{\Omega}_j = \mathrm{diag}(\Omega_{1j}, \ldots, \Omega_{n_j, j})$ is a diagonal matrix containing weights, such as RDS weights. 

An interesting property of GEEs is that even if the working correlation structure is misspecified, the resulting estimator remains consistent for $\boldsymbol{\gamma}$ provided the marginal mean model is correctly specified, with inference typically based on a robust sandwich variance estimator.

%It might also be of interest to RDS investigators to estimate a regression model from $\mathcal{DT}$ by relating the outcome $Y_i$ to the covariates $\bm{L}_i$ such as
%$g(\mathbb{E}[Y_i | \bm{L}_i]) = \bm{\beta}\bm{L}_i,$
%where $g(\cdot)$ is a suitable link function and the target of inference $\bm{\beta}$ is a vector of regression coefficients of dimension $p$. For instance, we might want to understand the influence of certain sexual practices on the risk of being diagnosed with a bSTI. Common strategies include generalized linear mixed models to account for clustering by recruiter or seed \citep{spiller2009regression, yauck2021general}, as well as generalized estimating equations \citep{LiangZeger1986, avery2019unweighted}. A brief review of multivariate modelling using respondent-driven sampling data is presented in Web Appendix A. 

\subsection*{Uncertainty estimation}\label{subsec23}

Two main bootstrap methods have been proposed for uncertainty estimation in RDS: (i) the tree bootstrap \citep{baraff2016estimating} and (ii) the neighbourhood bootstrap \citep{yauck2022neighborhood}. The tree bootstrap proposed by \cite{baraff2016estimating} resamples the observed recruitment tree while preserving its hierarchical structure. Seeds are resampled with (or without) replacement first, followed by their recruits, then subsequent waves, until no further recruits are available. By reproducing key features of the recruitment process such as chains, waves, and the number of seeds, this method accounts for the network structure underlying RDS. Although it has been shown to outperform earlier bootstrap approaches, it tends to overestimate uncertainty \citep{gile2018methods, yauck2022neighborhood}. The neighbourhood bootstrap proposed by \citep{yauck2022neighborhood} resamples individuals together with their local connections in the recruitment tree. The first step selects a sample of $n_r$ recruiters in the RDS sample, where $n_r$ is the number of recruiters, after which all their recruits are included in the bootstrap sample. The algorithm can be visualised in Figure 1 of \cite{yauck2022neighborhood}. This approach emphasises local network structure rather than the global branching pattern of the tree. It has been shown to be consistent and, in small samples, generally yields lower coverage probabilities and interval widths compared to the tree bootstrap \citep{yauck2022neighborhood}.

Uncertainty estimation for regression parameters can also be based on a robust sandwich variance estimator, also known as the Huber-White estimator. In the context of weighted generalized estimating equations, the variance of $\widehat{\boldsymbol{\gamma}}$, denoted $\mathbf{V}_{\mathrm{sand}}(\widehat{\boldsymbol{\gamma}})$, takes the sandwich form
\begin{align*}
\small
\left(
\sum_{j=1}^{m} \mathbf{D}_j^\top \mathbf{\Omega}_j \mathbf{V}_j^{-1} \mathbf{D}_j
\right)^{-1}\left(
\sum_{j=1}^{m} \mathbf{D}_j^\top \mathbf{\Omega}_j \mathbf{V}_j^{-1}
\mathrm{Cov}(\bm{Y}_j)
\mathbf{V}_j^{-1} \mathbf{\Omega}_j \mathbf{D}_j
\right)
\left(
\sum_{j=1}^{m} \mathbf{D}_j^\top \mathbf{\Omega}_j\mathbf{V}_j^{-1} \mathbf{D}_j
\right)^{-1}.
\end{align*}

In practice, an estimator of $\mathbf{V}_{\mathrm{sand}}(\hat{\gamma})$ is found by substituting $\mathrm{Cov}(\bm{Y}_j)$ with an empirical estimator. This estimator is consistent for the standard error of $\hat{\boldsymbol{\gamma}}$ under mild regularity conditions, even if the working covariance structure $\mathbf{V}_i(\boldsymbol{\gamma},\boldsymbol{\alpha})$ is misspecified.

Given a variance estimator $\hat{V}_{\theta}$ of a scalar parameter $\theta$, obtained via the sandwich formula for an instance, an approximate $(1-\alpha)$ confidence interval can then be constructed using the usual Wald form,
\[
\hat{\theta} \pm z_{1-\alpha/2} \sqrt{\hat{V}_{\theta}},
\]
where $z_{1-\alpha/2}$ denotes the $(1-\alpha/2)$ quantile of the standard normal distribution.

\newpage
\webappendix{Proof of Proposition 1}{webapp:proof}

\label{app1}
\vspace*{12pt}

\begin{proof}
For this proof, we consider the estimation of a population mean or prevalence and assume a point-treatment study, implying that there is only one follow-up visit ($K=1$) and $D_{i, k-1} =1$ by convention. In this simplified setting, dropout and visit attendance are indistinguishable, and the non-response indicator $\delta$ alone suffices. We first consider the RDS-II estimator using weights based population-standardized response mechanism, $p_i^{O*} = P(\delta_{i} = 1 \ | \ \mathcal{H}_i)$, as opposed to the sample-conditional response probability $p_i^O = P(\delta_{i} = 1 \ | \ S_i = 1, \mathcal{H}_i)$. For the numerator, assuming missingness at random, i.e.,  $Y_i \ind \delta_i \ | \ \mathcal{H}_i$, we have
\begin{align*}
& \sum_{i=1}^N \frac{\delta_i S_i Y_i^{\text{obs}}}{p_i^{O^*} \pi_i} \\
&\rightarrow n \int \int \frac{y}{P(\delta=1 \ | \ \bm{h}) P(S=1 \ | \ y, \bm{h}, \delta=1)} f(y, \bm{h} \ | \ \delta=1, S=1) dy d\bm{h}\\
&= n \int \int y \frac{P(\delta=1 \ | \ y, \bm{h}) f(y, \bm{h}) f(y, \bm{h}\ | \ \delta=1, S=1)}{P(\delta=1 \ | \ \bm{h}) f(y, \bm{h} \ | \ \delta=1, S=1)P(\delta = 1, S=1)} dy d\bm{h}\\
&= \frac{n}{P(\delta=1, S=1)} \int \int y \frac{f(\delta=1 \ | \ \bm{h}) f(y, \bm{h}) }{P(\delta=1 \ | \ \bm{h}) } dy d\bm{h}\\
&= \frac{n}{P(\delta=1, S=1)} \int \int y f(y, \bm{h}) dy d\bm{h}.\\
\end{align*}
Bringing in the denominator normalizing term, we have that
$$\tilde{\mu} \rightarrow \int \int y  f(y, \bm{h}) dy d\bm{h} = \mathbb{E}[Y].$$

Now, following the recommendation of \cite{dugoff2014generalizing}, let us instead re-weight using $p_i^O = P(\delta_i=1 \ | \ S_i=1, \mathcal{H}_i)$ rather than $P(\delta_i=1 |  \mathcal{H}_i)$. We have
\begin{align*}
&\sum_{i=1}^N \frac{\delta_i S_i Y_i}{p^O_{i} \pi_i} \\
&\rightarrow n \int \int \frac{y}{P(\delta=1  \ | \ \bm{h}, S=1) P(S=1 \ | \ y, \bm{h}, \delta=1)} f(y, \bm{h} \ | \ \delta=1, S=1) dy d\bm{h}\\
&= \frac{n}{f(\delta=1, S=1)} \int \int y \frac{P(\delta=1 \ | \ y, \bm{h}, S=1) P(S=1 \ | \ y, \bm{h})}{P(\delta=1 \ | \ \bm{h}, S=1) P(S=1| y, \bm{h}, \delta=1)}f(y, \bm{h}) dy d\bm{h}\\
&= \frac{n}{P(\delta=1, S=1)} \int \int y \frac{P(\delta=1 \ |\ \bm{h}, S=1) P(S=1\ |\ \bm{h}, y)}{P(\delta=1 \ | \ \bm{h}, S=1) f(S=1 \ | \ y, \bm{h}, \delta=1)}f(y, \bm{h}) dy d\bm{h}\\
&= \frac{n}{P(\delta=1, S=1)} \int \int y \frac{ P(S=1 \ | \ \bm{h}, y)}{P(S=1 \ | \ y, \bm{h}, \delta=1)}f(y, \bm{h}) dy d\bm{h}.
\end{align*} 
\end{proof}

We see that we do not achieve consistency unless we make additional assumptions. For this estimator to converge in probability to $\mu$, we need to assume that $S_i \ind \delta_i \ | \ Y_i, \mathcal{H}_i$. This assumption could be violated if there is a variable $W$ that is used in the construction of the sampling weights that is not used or unavailable for the estimation of the non-response propensity. In general survey designs, examples of these include the fielding method, geography, and household, all of which could be related to outcomes of interest \citep{ridgeway2015propensity}. In the context of RDS under population-level MAR, i.e., $Y_i \ind \delta_i \ |  \ \mathcal{H}_i$, this could manifest as a latent variable influencing both the degree and non-response, which would imply
$$P(\delta_i \ | \mathcal{H}_i) \neq P(\delta_i \ | \ \mathcal{H}_i, S_i),$$
requiring design-weighting of the non-response model. An example of such a scenario is represented in Figure \ref{fig:dags}(b) and explored further in numerical studies of Section 4 of the main text.

\newpage
\webappendix{Delta-method variance for the standardized prevalence difference}{webapp:deltamethod}
In the simulation scenario with residual dependence between outcomes, we use the robust sandwich covariance matrix from the regression fit and apply the delta-method to obtain a standard error for the standardized prevalence difference estimator. Let
$$ m_i(e;\bm{\beta}) = \mathrm{expit}\left(\boldsymbol{x}_i(e)^{\mathsf{T}}\boldsymbol{\beta}\right) $$
denote the model-based probability under exposure level $e$ where $\boldsymbol{x}_i(e)$ is the model covariate vector obtained by setting $E_i=e$ and retaining the observed $Z_i$ and/or $W_i$. Thus, $\boldsymbol{x}_i(1)$ and $\boldsymbol{x}_i(0)$ represent the two counterfactual covariate vectors for individual $i$.

The estimator $\widehat{\Delta}_{\omega}$ can then be written as
$$\hat{\Delta}_{\omega} = g\left(\hat{\bm{\beta}}\right) = \sum_{i \in \mathcal{N}_T} a_i \left\{ m_i(1;\hat{\bm{\beta}}) - m_i(0;\hat{\bm{\beta}}) \right\},$$
where $a_i = r_i/\sum_{i \in \mathcal{N}_T} r_i $ and $r_i = \omega_i \delta_i$. For a logistic outcome model, its gradient is given by 
$$G(\bm{\beta}) =  \nabla g(\bm{\beta}) = \sum_{i \in \mathcal{N}_T} a_i\left\{ {m}_i(1)[1- {m}_i(1)]\bm{E}_i(1) - {m}_i(0)[1- {m}_i(0)]\bm{E}_i(0)\right\}. $$
Therefore, a sandwich estimator of $\mathrm{Var}(\hat{\Delta}_{\omega})$ is given by
\begin{equation}
\widehat{\mathrm{Var}}(\hat{\Delta}_{\omega}) = {{G}(\hat{\bm{\beta}})}^{\top}
\widehat{\boldsymbol{V}}_{\mathrm{sand}}({\hat{\bm{\beta}}})
{{G}(\hat{\bm{\beta}})},
\end{equation}
where $\widehat{\boldsymbol{V}}_{\mathrm{sand}}({\hat{\bm{\beta}}})$ is the sandwich variance estimator of the variance of $\hat{\bm{\beta}}$. An approximate $100(1-\alpha) $ confidence interval is then constructed as
\begin{equation}
\widehat{\Delta}{\omega}^{(\mathrm{M},\mathrm{R})}
\pm
z_{1-\alpha/2}
\sqrt{\widehat{\mathrm{Var}}(\hat{\Delta}_{\omega})}.
\label{eq}
\end{equation}
This variance estimator propagates the uncertainty in the estimated outcome-model parameters through the standardization procedure. The use of the robust covariance matrix allows for residual within-cluster correlation in the outcomes, provided that the clusters are independent. However, the calculation is conditional on the estimated weights $\omega_i$, the response indicators $\delta_i$, and the observed covariate distribution. It therefore does not account for additional uncertainty arising from the estimation of the response probabilities or the RDS sampling weights.

\newpage
\webappendix{Simulation design details}
{webapp:simulation-design}

\section*{Population network and baseline covariates generation}

For each individual in the population, we generated a binary baseline covariate denoted $Z_i$ representing the $i$-th individual's age group where $Z_i=0$ corresponds to a person aged less than 30 years old, and $Z_i=1$ corresponds to a person aged 30 years old or more. In all scenarios $C_1$--$C_3$, we set the proportion of individuals aged 30 years old or more as $P(Z_i=1) = 0.67$ to match the overall characteristics of the Engage case study, illustrated in Table E1 in Web Appendix E. For each individual in the population, we also generated a standard normal variable, $W_i \sim \mathrm{Normal}(0,1)$.

The population network was generated in two stages. First, target degrees $d^*$ were drawn from a negative binomial distribution, allowing for overdispersion in social connectedness as is common to see in practice. The individual target degrees were generated as $d_i^* \sim  \mathrm{NegativeBinomial}(\psi_i, \phi)$. The mean parameter was given by $$\psi_i \equiv \psi_i(Z_i, W_i) = \xi_i\exp(\kappa_i W_i - 0.5\kappa_i^2),$$ where $\xi_i \equiv \xi(Z_i) = \xi_1 Z_i + \xi_0(1-Z_i)$, $\xi_1 = 60$, $\xi_0 = 30$, and $\kappa_i \equiv \kappa_i(Z_i) = 0.35 + 0.70Z_i$, allowing target degrees to depend on $Z$ and $W$ jointly. The dispersion parameter was set to $\phi=1.2$ for all individuals. 

The target degree sequence was then bounded to lie in the interval $[1,150]$. The lower bound excludes isolated vertices, which cannot participate in RDS recruitment chains, while the upper bound prevents a small number of unrealistically large hubs from dominating the sampling process or causing instability in graph generation. Finally, the generated degree sequence was examined to ensure the sum of degrees was an even number. Otherwise, an increment or decrement of $1$ was applied to an individual's degree selected at random to satisfy this property of undirected graphs.

\begin{figure}
\centering
\includegraphics[scale=0.24]{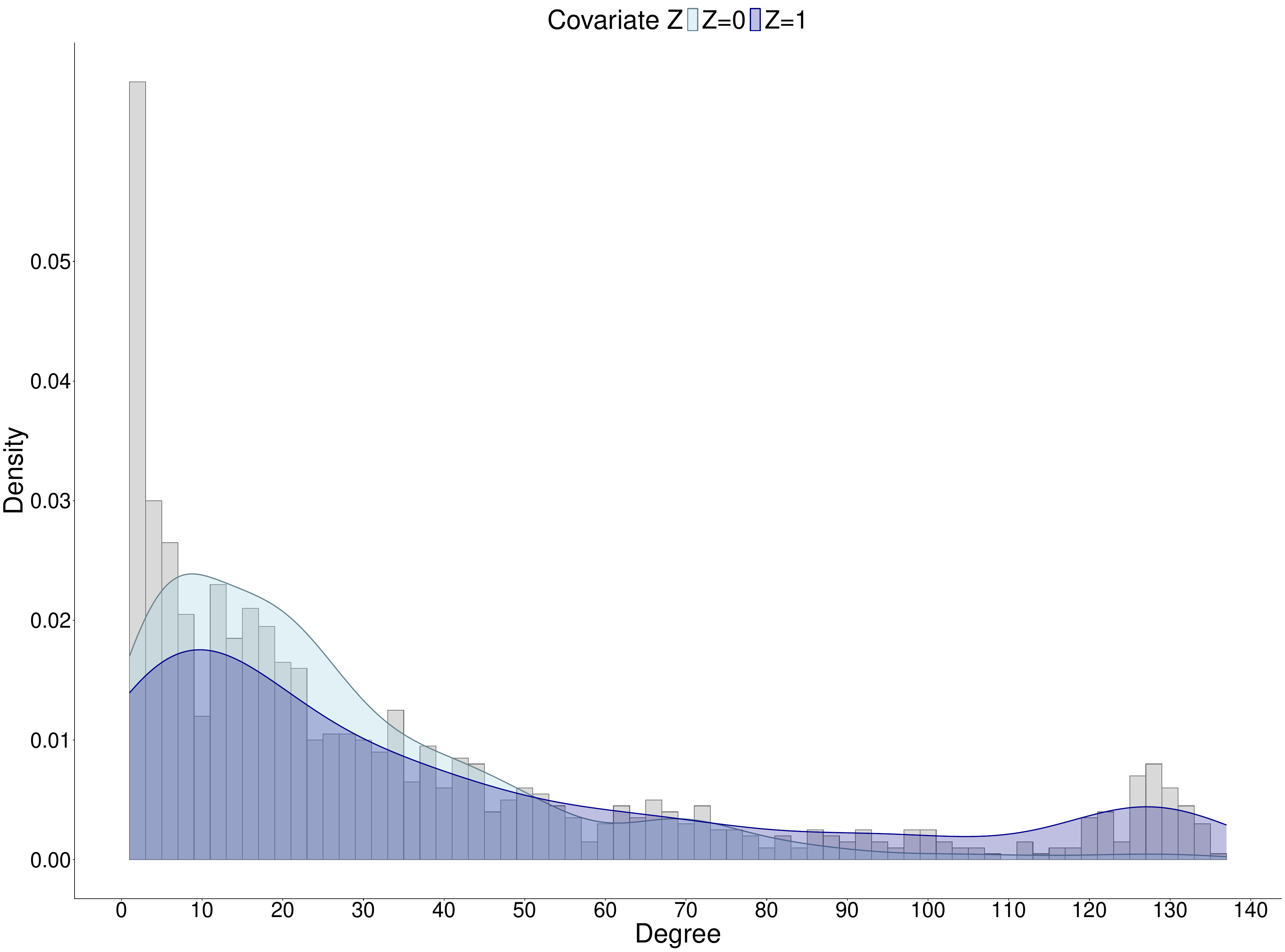}
\caption{Overall and $Z$-specific distributions of network degree in the simulated population. Grey bars represent the density-scaled histogram of the overall degree distribution, while the light- and dark-blue curves represent kernel density estimates of degree conditional on (Z=0) and (Z=1), respectively.}
\label{fig:degreedist}
\end{figure}

Second, a random graph was generated from this target degree sequence using the configuration model \citep{newman20218configuration}, a random graph model of a network with a specified degree sequence. Since the configuration model may produce self-loops and multiple edges, the resulting multigraph was simplified by deleting self-loops and replacing multiple edges by a single edge. Participants with realised degree zero after simplification were removed so that all remaining individuals were eligible for RDS recruitment through at least one social tie. We let $G=(\mathcal{N},\mathcal{E})$ denote the resulting population network, with adjacency matrix $\mathbf{B}$ and degree $d_i = \sum_{j \in \mathcal{N}} B_{ij}$. Figure \ref{fig:degreedist} displays the overall observed degree distribution in $G$, as well as the degree distributions stratified by $Z$. As expected, individuals with $Z=1$ tend to have higher network degree on average.

\section*{Respondent-driven sampling procedure}
From each simulated population, we drew an RDS sample using the population network $G$. Seeds were sampled with probability proportional to degree. Unless otherwise stated, each simulated RDS study used $s=10$ seeds and each participant received up to $c=3$ coupons. The number of successful recruits generated by each participant was sampled uniformly from $\{0,1,\ldots,c\}$, subject to the availability of unrecruited neighbours and the remaining target sample size. Sampling proceeded without replacement along edges of the population network until the target sample size was reached or no further recruitment was possible. We considered sampling fractions $f \in \{ 0.10, 0.20, 0.30\}$, corresponding to target sample sizes $\lfloor fN \rfloor$.

\section*{Generation of exposure, outcome and response indicator at follow-up}

In addition to the baseline covariates $Z$ and $W$, we generated a binary exposure denoted $E$ representing reported crystal use in the past 6 months. In scenario $C_1$, the proportion of reported recent crystal use was dependent on $W$ alone and was set to $\mathrm{logit}[P(E_i =1 \  \rvert \ W_i)]=0.47 + 0.50 W_i$, whereas in scenario $C_2$, $E$ had no causal parents with $P(E_i = 1) = 0.47$. In scenario $C_3$, $E_i$ depended on both $Z_i$ and $W_i$ through $\mathrm{logit} [P(E_i = 1 \  \rvert \ Z_i, W_i)] = 0.56 - 0.13Z_i + 0.50 W_i$.

We considered a point-treatment study with one follow-up visit, with the possibility of terminal dropout or attrition at follow-up. This scenario is similar to the one considered in the mathematical argument of Appendix \ref{app1}. In scenarios $C_1$ and $C_3$, the binary outcome at follow-up of individual $i$ $Y_i$, representing syphilis infection, was generated as a function of $E_i$, $Z_i$, and $W_i$:
\begin{equation}
    \mathrm{logit}[P(Y_i=1 \ \rvert \ E_i, Z_i, W_i]) = \beta_0 + \beta_1 E_i + \beta_2 Z_i + \beta_3 W_i + \beta_{13} E_iW_i + \beta_{23} Z_iW_i, \tag{M3} \label{eq:M3}
\end{equation}  
with coefficients $\beta_0 = -5$, $\beta_1 = 2$, and $\beta_2 = \beta_3 = \beta_{13} = \beta_{23} = 1$. In scenario $C_2$, the outcome was generated according to the model
\begin{equation} 
\mathrm{logit}[P(Y_i=1 \ \rvert \ E_i, Z_i]) = \eta_0 + \eta_1 E_i + \eta_2 Z_i + \eta_{12} E_i Z_i,  \tag{M2} \label{eqn:M2}
\end{equation}
with coefficients $\eta_0 =-5$, $\eta_1 = 1.5$, $\eta_2 = 2$, and $\eta_{12} = 0.6$. 

Finally, we generated a binary indicator of response at follow-up, denoted $\delta_i$, where $\delta=1$ indicates that individual $i$ had a record at follow-up. In scenario $C_1$, non-response depended on both $E_i$ and $W_i$:
\begin{equation} \mathrm{logit}[P(\delta_i=0 \ \rvert \ E_i, W_i )] = \gamma_0 + \gamma_1 E_i +  \gamma_2 W_i + \gamma_{12} E_iW_i, \tag{R1} \label{eqn:R1} 
\end{equation}
with $\gamma_0 = -1.5$, $\gamma_1=-0.5$, $\gamma_2 = 2$, $\gamma_{12} = 1$.
In scenarios $C_2$ and $C_3$, the response indicator was generated as
\begin{equation}
    \mathrm{logit}[P(\delta_i=0 \ \rvert \ E_i, Z_i, W_i )] = \alpha_0 + \alpha_1 E_i + \alpha_2 Z_i + \alpha_3 W_i + \alpha_{23} Z_i W_i,\tag{R3} \label{eqn:R3}
\end{equation}
with coefficients $\alpha_0 = -3$, $\alpha_1 =  0.5$, $\alpha_2 = \alpha_3=  2$, and $\alpha_{23} = 1$. 

The data-generating scenarios summarised in Figure \ref{fig:dags}, and their respective aims, can be understood as follows. Across all scenarios, the outcome $Y$ is associated with the realised degree $d$, and incidentally with the probability of inclusion in the RDS sample. This association is driven primarily by $Z$, which is latent or unmeasured in scenario $C_1$, but observed and available for modelling in scenarios $C_2$ and $C_3$. In some scenarios, degree is also associated with baseline covariates. This reflects the practical reality that participants' network degrees are often related to their characteristics and creates settings in which RDS weighting becomes important, as supported by previous empirical studies \citep{avery2019unweighted, yauck2021general}.

The role of $W$ is of particular importance in this numerical study. In scenarios $C_1$ and $C_3$, $W$ confounds the association between $E$ and $Y$ and influences both the probability of inclusion in the sample and the probability of non-response. In scenario $C_1$, $W$ is observed and available to the analyst. Conditional on $W$ and the other observed covariates, the outcome is therefore independent of non-response, corresponding to a MAR mechanism and satisfying the NUC assumption. In scenario $C_3$, by contrast, $W$ is latent or unmeasured, giving rise to a boundary case involving both missing-not-at-random non-response and unmeasured confounding.

Scenario $C_2$ is most directly relevant to the methodological contribution of this work. Although $W$ is latent, it is unrelated to the outcome $Y$, so that both the MAR and NUC assumptions remain satisfied. However, $W$ affects both degree and non-response, thereby inducing an association between inclusion in the RDS sample and subsequent non-response. This corresponds precisely to the setting considered in Appendix \ref{app1}, in which incorporating the sampling design into the non-response model is necessary for valid inference.

As a sensitivity analysis, we also considered a setting denoted $C_4$ based on the $C_2$ DAG in Figure \ref{fig:dags}(b) but with residual network dependence in the outcome by adding a simultaneous autoregressive (SAR) effect in the outcome model. In this case, the outcome was simulated as 
$$\mathrm{logit}[P(Y_i=1 \ \rvert \ E_i, Z_i]) = \eta_0 + \eta_1 E_i + \eta_2 Z_i + \eta_{12} E_i Z_i + \varepsilon_{i}, $$
where $\varepsilon_{i}$ is a random effect for the $i$-th participant. Recall that $\mathbf{B}$ denote the sociomatrix of the population network. The vector of random effects $\bm{\varepsilon}$ was generated according to a SAR structure, i.e.,
$$\bm{\varepsilon} = \tau \mathbf{B} \bm{\varepsilon} + \bm{u}, \quad \bm{u} \sim N(\bm{0}, \sigma^2 \mathbf{I}_{N}).$$ Provided the existence of $\{\mathbf{I}_{N} - \tau  \mathbf{B} \}^{-1}$, the covariance matrix of $\bm{\varepsilon}$, $\bm{\Sigma}$, was expressed as 
$$\bm{\Sigma} = \sigma^2 \left[\{\mathbf{I}_{N} - \tau  \mathbf{B} \}^{-1}\right]^2, $$
with $\sigma^2 = 1$ and $\tau = 0.2$. This allowed us to assess whether conclusions were robust to correlation between connected individuals beyond that induced by the measured covariates and the RDS sampling process.

\begin{center}
\begin{minipage}{\textwidth}
\centering

%----------------------------- C1 ---------------------------------
\begin{minipage}[t]{0.31\linewidth}
\centering
\begin{tikzpicture}[
    x=1.15cm,
    y=1.15cm,
    >=Stealth,
    thick,
    var/.style={
        draw,
        circle,
        minimum size=6mm,
        inner sep=0pt
    },
    latent/.style={
        draw,
        circle,
        dashed,
        fill=gray!15,
        minimum size=6mm,
        inner sep=0pt
    },
    selected/.style={
        draw,
        circle,
        fill=red!20,
        minimum size=6mm,
        inner sep=0pt
    }
]
\node[var]      (d)     at (0,2) {$d$};
\node[var]      (W)     at (1,1) {$W$};
\node[latent]   (Z)     at (1,2) {$Z$};
\node[selected] (S)     at (0,1) {$S$};
\node[var]      (Y)     at (2,2) {$Y$};
\node[var]      (delta) at (0,0) {$\delta$};
\node[var]      (E)     at (2,0) {$E$};

\draw[->] (Z) -- (Y);
\draw[->] (W) -- (Y);
\draw[->] (W) -- (delta);
\draw[->] (W) -- (E);
\draw[->] (W) -- (d);
\draw[->] (Z) -- (d);
\draw[->] (E) -- (Y);
\draw[->] (E) -- (delta);
\draw[->] (d) -- (S);
\draw[dotted, bend right=40] (d) to (delta);
\draw[dotted, bend left=40]  (d) to (Y);
\end{tikzpicture}

\par\smallskip
(a) Scenario $C_1$.
\end{minipage}
\hfill
%----------------------------- C2 ---------------------------------
\begin{minipage}[t]{0.31\linewidth}
\centering
\begin{tikzpicture}[
    x=1.15cm,
    y=1.15cm,
    >=Stealth,
    thick,
    var/.style={
        draw,
        circle,
        minimum size=6mm,
        inner sep=0pt
    },
    latent/.style={
        draw,
        circle,
        dashed,
        fill=gray!15,
        minimum size=6mm,
        inner sep=0pt
    },
    selected/.style={
        draw,
        circle,
        fill=red!20,
        minimum size=6mm,
        inner sep=0pt
    }
]
\node[var]      (d)     at (0,2)   {$d$};
\node[latent]   (W)     at (1.2,1) {$W$};
\node[var]      (Z)     at (1,2)   {$Z$};
\node[selected] (S)     at (0,1)   {$S$};
\node[var]      (Y)     at (2,2)   {$Y$};
\node[var]      (delta) at (0,0)   {$\delta$};
\node[var]      (E)     at (2,0)   {$E$};

\draw[->] (Z) -- (Y);
\draw[->] (W) -- (delta);
\draw[->] (Z) -- (d);
\draw[->] (W) -- (d);
\draw[->] (Z) -- (delta);
\draw[->] (E) -- (Y);
\draw[->] (E) -- (delta);
\draw[->] (d) -- (S);
\draw[dotted, bend right=40] (d) to (delta);
\draw[dotted, bend left=40]  (d) to (Y);
\end{tikzpicture}

\par\smallskip
(b) Scenario $C_2$.
\end{minipage}
\hfill
%----------------------------- C3 ---------------------------------
\begin{minipage}[t]{0.31\linewidth}
\centering
\begin{tikzpicture}[
    x=1.15cm,
    y=1.15cm,
    >=Stealth,
    thick,
    var/.style={
        draw,
        circle,
        minimum size=6mm,
        inner sep=0pt
    },
    latent/.style={
        draw,
        circle,
        dashed,
        fill=gray!15,
        minimum size=6mm,
        inner sep=0pt
    },
    selected/.style={
        draw,
        circle,
        fill=red!20,
        minimum size=6mm,
        inner sep=0pt
    }
]
\node[var]      (d)     at (0,2) {$d$};
\node[latent]   (W)     at (1,1) {$W$};
\node[var]      (Z)     at (1,2) {$Z$};
\node[selected] (S)     at (0,1) {$S$};
\node[var]      (Y)     at (2,2) {$Y$};
\node[var]      (delta) at (0,0) {$\delta$};
\node[var]      (E)     at (2,0) {$E$};

\draw[->] (Z) -- (Y);
\draw[->] (W) -- (Y);
\draw[->] (W) -- (delta);
\draw[->] (W) -- (d);
\draw[->] (W) -- (E);
\draw[->] (Z) -- (d);
\draw[->, bend left=5]  (Z) to (E);
\draw[->, bend right=5] (Z) to (delta);
\draw[->] (E) -- (Y);
\draw[->] (E) -- (delta);
\draw[->] (d) -- (S);
\draw[dotted, bend right=40] (d) to (delta);
\draw[dotted, bend left=40]  (d) to (Y);
\end{tikzpicture}

\par\smallskip
(c) Scenario $C_3$.
\end{minipage}

\captionof{figure}{Directed acyclic graph representations for the
data-generating mechanisms considered in the simulation study. Nodes
represent degree $d$, sample inclusion $S$, covariates $Z$ and $W$,
exposure $E$, outcome $Y$, and response indicator $\delta$.
Scenario $C_4$ also follows graph~(b).}
\label{fig:dags}

\end{minipage}
\end{center}

\section*{Fitting procedures and working models for the outcome and non-response}

The main estimators compared for the crude prevalence estimand and marginal prevalence difference estimand are described in the main text. 

In scenario $C_1$, the correctly specified working model (\ref{eqn:R1}) was used to estimate the non-response probability. For the outcome, we considered the following working model
\begin{equation} 
\mathrm{logit}[P(Y_i=1 \ \rvert \ E_i, W_i]) = \nu_0 + \nu_1 E_i + \nu_2 W_i + \nu_{12} W_i E_i,  \tag{M1} \label{eq:M1}
\end{equation}
which contains sufficient information for the MAR assumption to hold. 

Scenario $C_3$ corresponds to an MNAR mechanism because $W$ is unmeasured. In this setting, we compare the reduced non-response model 
\begin{equation}
    \mathrm{logit}[P(\delta_i=0 \ \rvert \ E_i, Z_i )] = \omega_0 + \omega_1 E_i + \omega_2 Z_i + \omega_{12} E_i Z_i, \tag{R2} \label{eqn:R2}
\end{equation} which excludes $W$, with the full model in (\ref{eqn:R3}), which represents the hypothetical and practically unattainable situation in which $W$ had been measured and was available for modelling. For the outcome, we compared the reduced model (\ref{eqn:M2}), under which missingness remains nonignorable because $W$ is unavailable, to the full, correctly specified model (\ref{eq:M3}). 

\newpage
\webappendix{Additional simulation results}
{webapp:simulation-results}

\section*{Scenario \texorpdfstring{$C_1$}{C1}: observed common
causes of degree, response, and outcome}
\label{sec:simulation-results-c1}

Under Scenario $C_1$, represented in Figure~\ref{fig:dags}(a), $W$ was an observed common cause of degree, non-response, exposure, and outcome, whereas $Z$ affected degree and outcome only. The working response model (\ref{eqn:R1}) contained the variables required to adjust for differential response. Although $Z$ was omitted from the working outcome model (\ref{eq:M1}), it did not confound the association between $E$ and $Y$, since $E$ was generated as a function of $W$ alone.

\begin{figure}[H]
\centering
\includegraphics[scale=0.3]{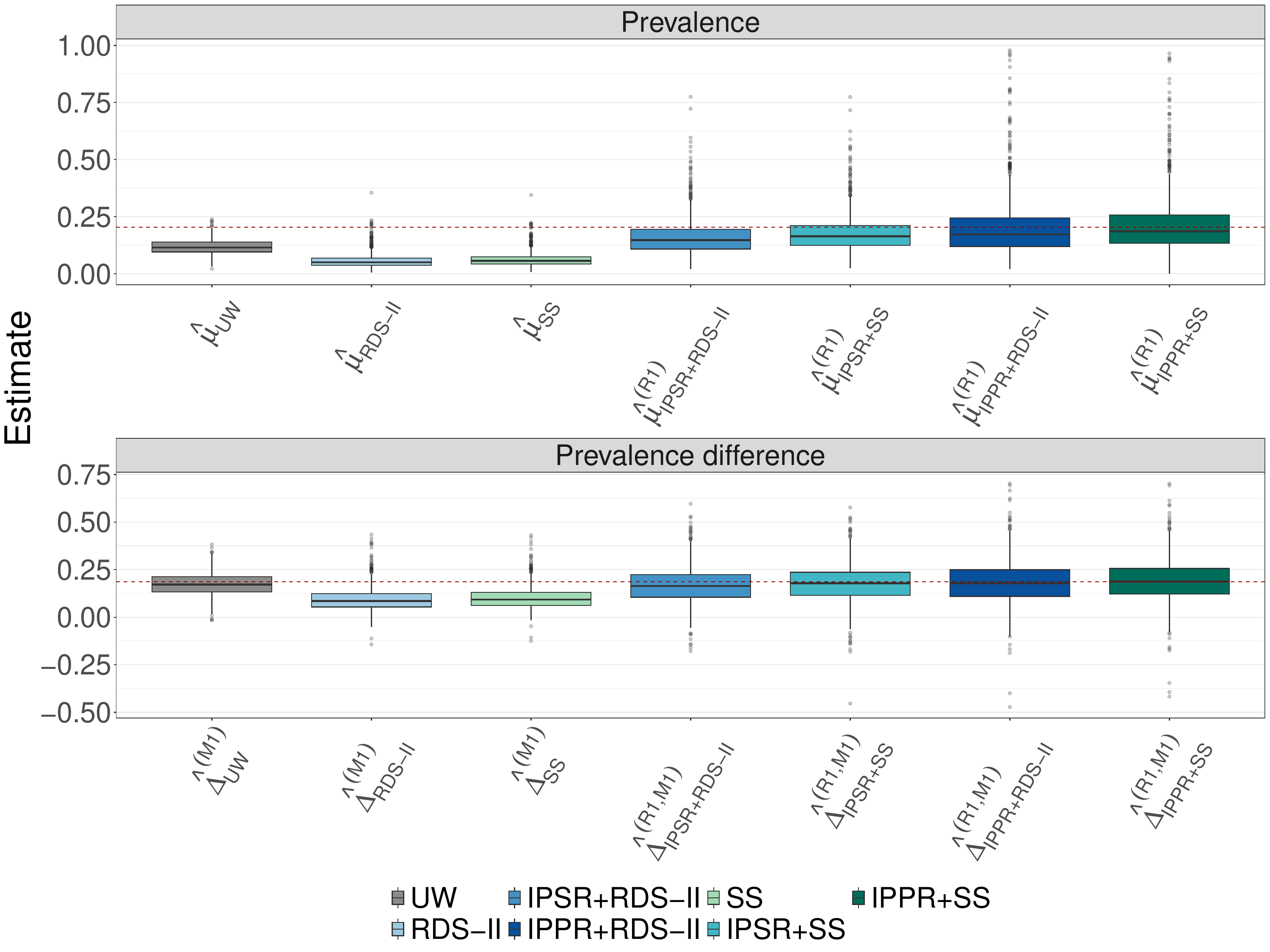}
\caption{Boxplots show the estimated outcome prevalence (top panel) and prevalence difference (bottom panel) under Scenario $C_1$ across 1,000 simulation replicates for the unweighted (UW), RDS, and IPO adjusted estimators (IPSR/IPPR).  Adjusted estimators are combined with either RDS-II or SS sampling weights. The sampling fraction displayed is 20\%. Superscripts R1 and M1 denote response-model specification (\ref{eqn:R1}) and marginal standardization (\ref{eq:M1}), respectively. The red dashed line indicates the mean population-specific true value of the corresponding estimand.}
\label{fig:boxplotsC1}
\end{figure}

Figure~\ref{fig:boxplotsC1} and Table~\ref{tab:simulation-c1} show substantial downward bias in the unweighted prevalence estimator. For an average true prevalence of $\mu=0.204$, the relative bias of $\widehat{\mu}_{\mathrm{UW}}$ ranged from $-38.9\%$ at a $10\%$ sampling fraction to $-44.5\%$ at a $30\%$ sampling fraction. Adjusting only for the RDS design increased the bias, with relative biases of approximately $-69\%$ to $-72\%$ for the RDS-II and SS estimators. As suggested by Figure~\ref{fig:dags}(a), RDS weighting corrects the distribution of degree among respondents but cannot restore individuals systematically absent at follow-up. Because degree, response, and outcome share common causes, design weighting alone can therefore amplify selection due to non-response.

IPO weighting substantially reduced the prevalence bias. When the response model was estimated without baseline RDS weights, relative bias ranged from approximately $-5\%$ to $-17\%$. Estimating the response model in the RDS-reweighted pseudo-population improved bias further, although some overcorrection occurred at the smallest sampling fraction: IPPR relative bias was approximately $10\%$ at $10\%$ sampling, compared with approximately $-5\%$ to $6\%$ at the larger sampling fractions. This improvement came at the cost of additional variability, visible in Figure~\ref{fig:boxplotsC1} through the wider and more right-skewed distributions of the IPSR and IPPR estimators.

The prevalence difference results followed a similar pattern. The unweighted estimator based on outcome model (\ref{eq:M1}) was only moderately biased, with relative bias ranging from $-2.5\%$ to $-10.1\%$, reflecting adjustment for $W$ and hence control of the confounding pathway $E \leftarrow W \rightarrow Y$. In contrast, adding RDS weights without correcting for non-response produced relative biases of approximately $-43\%$ to $-50\%$.

Including IPO weights substantially reduced this bias. In particular, the IPSR estimator combined with SS weights had relative bias close to zero at all three sampling fractions. The IPPR estimators also had relatively small bias at the larger sampling fractions, although they tended to overestimate the prevalence difference at $10\%$ sampling. Differences between RDS-II and SS weighting became more apparent as the sampling fraction increased: at $20\%$ and $30\%$, the IPSR estimator using SS weights remained approximately unbiased, whereas the corresponding RDS-II estimator had relative biases of $-6.4\%$ and $-10.9\%$. This is consistent with SS weights accounting for without-replacement and finite-population effects ignored by the RDS-II approximation, though this comes with the added requirement of knowing $N$.

\begin{table}[p]
\caption{Simulation performance for estimation of population prevalence across scenarios and sampling fractions under Scenario $C_1$.}
\label{tab:simulation-c1}
\scriptsize
\centering
 \begin{threeparttable}
 \small
\setlength{\tabcolsep}{3pt}
\renewcommand{\arraystretch}{0.9}

\begin{adjustbox}{
  max width=\textheight,
  max totalheight=1.3\textwidth,
  keepaspectratio
}
\begin{tabular}{lllrrrrrrrr} 
  \toprule
 &  &  &  &  &  &  & NBoot &  NBoot & TBoot & TBoot \\ 
Scenario & $f$ & Estimator & Av. Est. & Rel. bias (\%) & SD & RRMSE (\%) & Av. SE & Cov. (\%) & Av. SE & Cov. (\%) \\ 
  \midrule
 C1 & 10\% & $\widehat{\mu}_{\mathrm{UW}}$ & 0.125 & -38.857 & 0.049 & 45.551 & 0.040 & 48.785 & 0.064 & 80.364 \\ 
   &  & $\widehat{\mu}_{\mathrm{RDS-II}}$ & 0.062 & -69.586 & 0.045 & 73.001 & 0.029 & 15.182 & 0.047 & 32.591 \\ 
   &  & $\widehat{\mu}_{\mathrm{SS}}$ & 0.064 & -68.422 & 0.044 & 71.807 & 0.030 & 15.283 & 0.047 & 33.401 \\ 
   &  & $\widehat{\mu}_{\mathrm{IPSR+RDS-II}}^{(\mathrm{R1})}$ & 0.168 & -17.541 & 0.108 & 55.564 & 0.070 & 72.065 & 0.099 & 84.818 \\ 
   &  & $\widehat{\mu}_{\mathrm{IPSR+SS}}^{(\mathrm{R1})}$ & 0.176 & -13.688 & 0.109 & 55.019 & 0.071 & 75.000 & 0.100 & 86.336 \\ 
   &  & $\widehat{\mu}_{\mathrm{IPPR+RDS-II}}^{(\mathrm{R1})}$ & 0.224 & 10.003 & 0.173 & 85.318 & 0.117 & 84.818 & 0.168 & 95.142 \\ 
   &  & $\widehat{\mu}_{\mathrm{IPPR+SS}}^{(\mathrm{R1})}$ & 0.225 & 10.647 & 0.168 & 83.004 & 0.112 & 84.514 & 0.162 & 95.040 \\ 
   \addlinespace
 &  & $\widehat{\Delta}_{\mathrm{UW}}^{(\mathrm{M1})}$ & 0.182 & -2.496 & 0.095 & 50.758 & 0.079 & 83.907 & 0.131 & 96.154 \\ 
   &  & $\widehat{\Delta}_{\mathrm{RDS-II}}^{(\mathrm{M1})}$ & 0.102 & -45.369 & 0.085 & 64.387 & 0.063 & 60.121 & 0.105 & 81.579 \\ 
   &  & $\widehat{\Delta}_{\mathrm{SS}}^{(\mathrm{M1})}$ & 0.104 & -44.227 & 0.086 & 64.028 & 0.064 & 61.943 & 0.106 & 82.389 \\ 
   &  & $\widehat{\Delta}_{\mathrm{IPSR+RDS-II}}^{(\mathrm{R1},\mathrm{M1})}$ & 0.181 & -3.135 & 0.136 & 72.692 & 0.100 & 83.704 & 0.151 & 95.648 \\ 
   &  & $\widehat{\Delta}_{\mathrm{IPSR+SS}}^{(\mathrm{R1},\mathrm{M1})}$ & 0.187 & -0.065 & 0.142 & 76.016 & 0.104 & 84.818 & 0.156 & 96.457 \\ 
   &  & $\widehat{\Delta}_{\mathrm{IPPR+RDS-II}}^{(\mathrm{R1},\mathrm{M1})}$ & 0.204 & 9.276 & 0.167 & 90.124 & 0.134 & 86.842 & 0.203 & 97.470 \\ 
   &  & $\widehat{\Delta}_{\mathrm{IPPR+SS}}^{(\mathrm{R1},\mathrm{M1})}$ & 0.206 & 10.121 & 0.169 & 91.159 & 0.134 & 87.551 & 0.203 & 97.672 \\ 
   \midrule
     & 20\% & $\widehat{\mu}_{\mathrm{UW}}$ & 0.119 & -41.532 & 0.032 & 44.243 & 0.027 & 23.647 & 0.051 & 68.938 \\ 
   &  & $\widehat{\mu}_{\mathrm{RDS-II}}$ & 0.060 & -70.396 & 0.033 & 72.248 & 0.022 & 6.413 & 0.041 & 22.946 \\ 
   &  & $\widehat{\mu}_{\mathrm{SS}}$ & 0.065 & -68.004 & 0.032 & 69.779 & 0.022 & 6.413 & 0.041 & 23.347 \\ 
   &  & $\widehat{\mu}_{\mathrm{IPSR+RDS-II}}^{(\mathrm{R1})}$ & 0.168 & -17.281 & 0.087 & 45.655 & 0.054 & 69.840 & 0.093 & 90.681 \\ 
   &  & $\widehat{\mu}_{\mathrm{IPSR+SS}}^{(\mathrm{R1})}$ & 0.185 & -8.916 & 0.088 & 43.720 & 0.056 & 79.359 & 0.096 & 93.186 \\ 
   &  & $\widehat{\mu}_{\mathrm{IPPR+RDS-II}}^{(\mathrm{R1})}$ & 0.204 & 0.195 & 0.128 & 62.530 & 0.085 & 84.970 & 0.153 & 98.196 \\ 
   &  & $\widehat{\mu}_{\mathrm{IPPR+SS}}^{(\mathrm{R1})}$ & 0.215 & 5.539 & 0.123 & 60.105 & 0.078 & 86.273 & 0.143 & 98.497 \\ 
   \addlinespace
  &  & $\widehat{\Delta}_{\mathrm{UW}}^{(\mathrm{M1})}$ & 0.171 & -8.384 & 0.059 & 32.614 & 0.051 & 86.573 & 0.098 & 99.399 \\ 
   &  & $\widehat{\Delta}_{\mathrm{RDS-II}}^{(\mathrm{M1})}$ & 0.096 & -48.633 & 0.062 & 58.803 & 0.044 & 47.896 & 0.083 & 79.359 \\ 
   &  & $\widehat{\Delta}_{\mathrm{SS}}^{(\mathrm{M1})}$ & 0.103 & -45.034 & 0.059 & 54.968 & 0.044 & 52.204 & 0.083 & 82.766 \\ 
   &  & $\widehat{\Delta}_{\mathrm{IPSR+RDS-II}}^{(\mathrm{R1},\mathrm{M1})}$ & 0.173 & -7.276 & 0.095 & 51.557 & 0.071 & 84.669 & 0.127 & 98.597 \\ 
   &  & $\widehat{\Delta}_{\mathrm{IPSR+SS}}^{(\mathrm{R1},\mathrm{M1})}$ & 0.184 & -1.578 & 0.101 & 54.322 & 0.073 & 86.774 & 0.130 & 99.399 \\ 
   &  & $\widehat{\Delta}_{\mathrm{IPPR+RDS-II}}^{(\mathrm{R1},\mathrm{M1})}$ & 0.188 & 0.674 & 0.112 & 60.144 & 0.087 & 86.874 & 0.164 & 99.499 \\ 
   &  & $\widehat{\Delta}_{\mathrm{IPPR+SS}}^{(\mathrm{R1},\mathrm{M1})}$ & 0.197 & 5.781 & 0.110 & 59.114 & 0.084 & 86.373 & 0.160 & 99.699 \\ 
\midrule
 & 30\% & $\widehat{\mu}_{\mathrm{UW}}$ & 0.113 & -44.459 & 0.026 & 46.073 & 0.021 & 8.108 & 0.043 & 51.451 \\ 
   &  & $\widehat{\mu}_{\mathrm{RDS-II}}$ & 0.058 & -71.727 & 0.029 & 73.111 & 0.018 & 4.304 & 0.037 & 15.916 \\ 
   &  & $\widehat{\mu}_{\mathrm{SS}}$ & 0.065 & -67.951 & 0.027 & 69.222 & 0.018 & 4.004 & 0.036 & 15.015 \\ 
   &  & $\widehat{\mu}_{\mathrm{IPSR+RDS-II}}^{(\mathrm{R1})}$ & 0.165 & -18.867 & 0.077 & 42.147 & 0.047 & 65.666 & 0.088 & 90.591 \\ 
   &  & $\widehat{\mu}_{\mathrm{IPSR+SS}}^{(\mathrm{R1})}$ & 0.193 & -5.063 & 0.080 & 39.274 & 0.050 & 83.984 & 0.091 & 96.797 \\ 
   &  & $\widehat{\mu}_{\mathrm{IPPR+RDS-II}}^{(\mathrm{R1})}$ & 0.192 & -5.662 & 0.106 & 52.218 & 0.072 & 82.783 & 0.146 & 99.099 \\ 
   &  & $\widehat{\mu}_{\mathrm{IPPR+SS}}^{(\mathrm{R1})}$ & 0.214 & 4.990 & 0.103 & 50.710 & 0.064 & 85.385 & 0.131 & 99.399 \\ 
   \addlinespace
     & & $\widehat{\Delta}_{\mathrm{UW}}^{(\mathrm{M1})}$ & 0.168 & -10.137 & 0.048 & 27.390 & 0.040 & 85.285 & 0.081 & 99.499 \\ 
   &  & $\widehat{\Delta}_{\mathrm{RDS-II}}^{(\mathrm{M1})}$ & 0.095 & -49.288 & 0.052 & 56.673 & 0.036 & 35.536 & 0.071 & 78.178 \\ 
   &  & $\widehat{\Delta}_{\mathrm{SS}}^{(\mathrm{M1})}$ & 0.104 & -44.074 & 0.049 & 51.160 & 0.036 & 45.946 & 0.071 & 82.482 \\ 
   &  & $\widehat{\Delta}_{\mathrm{IPSR+RDS-II}}^{(\mathrm{R1},\mathrm{M1})}$ & 0.173 & -7.226 & 0.082 & 44.661 & 0.059 & 86.086 & 0.112 & 99.399 \\ 
   &  & $\widehat{\Delta}_{\mathrm{IPSR+SS}}^{(\mathrm{R1},\mathrm{M1})}$ & 0.192 & 2.802 & 0.080 & 43.135 & 0.060 & 84.484 & 0.115 & 99.800 \\ 
   &  & $\widehat{\Delta}_{\mathrm{IPPR+RDS-II}}^{(\mathrm{R1},\mathrm{M1})}$ & 0.182 & -2.615 & 0.092 & 49.543 & 0.069 & 88.388 & 0.144 & 99.800 \\ 
   &  & $\widehat{\Delta}_{\mathrm{IPPR+SS}}^{(\mathrm{R1},\mathrm{M1})}$ & 0.198 & 6.018 & 0.089 & 48.054 & 0.066 & 84.585 & 0.136 & 99.800 \\ 
\bottomrule
%\label{tab:simulation-prevalence}
\end{tabular}
\end{adjustbox}
\begin{tablenotes}
      \small
     \item[] The average true prevalence is $\mu = 0.204$, and the true 
prevalence difference is $\Delta = 0.187$. Abbreviations: Av. Est., 
average estimate; Rel. Bias, relative bias; SD, Monte Carlo standard 
deviation; RRMSE, relative root mean squared error; Av. Boot. SE, average 
bootstrap standard error; Cov., empirical coverage probability; NBoot, neighbourhood bootstrap; TBoot, tree bootstrap.
    \end{tablenotes}
\end{threeparttable}
\end{table}

\section*{Scenario \texorpdfstring{$C_2$}{C2}: an unmeasured
determinant of degree and response}
\label{sec:simulation-results-c2}

Scenario $C_2$, represented in Figure~\ref{fig:dags}(b), was designed to isolate the main methodological question considered in this paper. The latent variable $W$ affected degree and non-response, inducing an association between sample inclusion and response through $W \rightarrow d \rightarrow S$ and $W \rightarrow \delta$, but did not affect the outcome. Consequently, the MAR and NUC assumptions continued to hold after conditioning on the observed variables in models (\ref{eqn:M2}) and (\ref{eqn:R2}). The difficulty was therefore not an unmeasured predictor of the outcome, but the discrepancy between the response process in the realised RDS sample and that in the target population.

Table~\ref{tab:simulation-c2} clearly illustrates this distinction. For an average true prevalence of $\mu=0.196$, the unweighted estimator had approximately $-20\%$ relative bias across sampling fractions, while RDS-II and successive-sampling weighting alone reduced this only to approximately $-13\%$ to $-16\%$. Combining RDS weights with response probabilities estimated in the unweighted RDS sample reversed rather than eliminated the bias, with IPO relative biases of $11\%$ to $16\%$.

In contrast, estimating the response model using baseline RDS weights produced nearly unbiased prevalence estimates, with IPPR relative bias ranging from approximately $-2.4\%$ to $2.4\%$. This directly supports Proposition 1: %~\ref{prop:response_weighting}: 
because $W$ affected both degree and response but was omitted from the working response model, $P(\delta=1\mid S=1,E,Z)$ did not generally equal $P(\delta=1\mid E,Z)$. Fitting the response model in the RDS-reweighted pseudo-population instead targeted the population response probability despite $W$ being unavailable to the analyst.

The prevalence difference results followed the same pattern. For an average true value of $\Delta=0.345$, the unweighted and RDS-only estimators had relative biases of approximately $-12\%$ to $-15\%$, whereas the IPO estimators overestimated it by approximately $7\%$ to $12\%$. The IPPR estimators had relative biases between approximately $-2.6\%$ and $-0.4\%$. Thus, reflecting the RDS design when estimating response probabilities was important for both descriptive population quantities and standardized association estimands; using RDS weights only in the final outcome analysis was insufficient.

This reduction in bias came at the cost of increased variance. At the $10\%$ sampling fraction, the IPPR prevalence estimators had Monte Carlo standard deviations of approximately $0.100$, compared with $0.055$ for the unweighted estimator, and therefore did not always have the smallest RRMSE. Both standard deviations and RRMSE generally declined as the sampling fraction increased, making the benefit of correcting systematic bias more apparent. RDS-II and successive-sampling estimators performed similarly, although the latter tended to have slightly smaller variability at larger sampling fractions.

\begin{table}[H]
\caption{Simulation performance for estimation of population prevalence across scenarios and sampling fractions under Scenario $C_2$.}
\label{tab:simulation-c2}
\scriptsize
\centering
 \begin{threeparttable}

 \small
\setlength{\tabcolsep}{3pt}
\renewcommand{\arraystretch}{0.9}

\begin{adjustbox}{
  max width=\textheight,
  max totalheight=1.3\textwidth,
  keepaspectratio
}

\begin{tabular}{lllrrrrrrrr} 
  \toprule
 &  &  &  &  &  &  & NBoot &  NBoot & TBoot & TBoot \\ 
Scenario & $f$ & Estimator & Av. Est. & Rel. bias (\%) & SD & RRMSE (\%) & Av. SE & Cov. (\%) & Av. SE & Cov. (\%) \\ 
  \midrule
  C2 & 10\% & $\widehat{\mu}_{\mathrm{UW}}$ & 0.157 & -19.823 & 0.055 & 33.664 & 0.045 & 76.406 & 0.074 & 95.281 \\ 
   &  & $\widehat{\mu}_{\mathrm{RDS-II}}$ & 0.170 & -13.304 & 0.109 & 56.533 & 0.067 & 71.787 & 0.105 & 87.751 \\ 
   &  & $\widehat{\mu}_{\mathrm{SS}}$ & 0.170 & -13.493 & 0.107 & 55.448 & 0.066 & 72.490 & 0.103 & 88.052 \\ 
   &  & $\widehat{\mu}_{\mathrm{IPSR+RDS-II}}^{(\mathrm{R2})}$ & 0.228 & 15.920 & 0.132 & 68.749 & 0.082 & 82.129 & 0.126 & 95.080 \\ 
   &  & $\widehat{\mu}_{\mathrm{IPSR+SS}}^{(\mathrm{R2})}$ & 0.227 & 15.799 & 0.130 & 67.433 & 0.080 & 82.329 & 0.124 & 95.783 \\ 
   &  & $\widehat{\mu}_{\mathrm{IPPR+RDS-II}}^{(\mathrm{R2})}$ & 0.201 & 2.112 & 0.106 & 53.231 & 0.069 & 82.430 & 0.108 & 95.181 \\ 
   &  & $\widehat{\mu}_{\mathrm{IPPR+SS}}^{(\mathrm{R2})}$ & 0.201 & 2.419 & 0.104 & 52.211 & 0.068 & 82.530 & 0.107 & 95.582 \\ 
   \addlinespace
  &  & $\widehat{\Delta}_{\mathrm{UW}}^{(\mathrm{M2})}$ & 0.295 & -14.341 & 0.110 & 34.856 & 0.089 & 81.727 & 0.145 & 97.289 \\ 
   &  & $\widehat{\Delta}_{\mathrm{RDS-II}}^{(\mathrm{M2})}$ & 0.304 & -11.769 & 0.183 & 54.423 & 0.117 & 79.317 & 0.181 & 94.779 \\ 
   &  & $\widehat{\Delta}_{\mathrm{SS}}^{(\mathrm{M2})}$ & 0.301 & -12.766 & 0.194 & 57.528 & 0.121 & 80.522 & 0.184 & 95.382 \\ 
   &  & $\widehat{\Delta}_{\mathrm{IPSR+RDS-II}}^{(\mathrm{R2},\mathrm{M2})}$ & 0.380 & 10.247 & 0.224 & 65.791 & 0.141 & 82.028 & 0.212 & 95.080 \\ 
   &  & $\widehat{\Delta}_{\mathrm{IPSR+SS}}^{(\mathrm{R2},\mathrm{M2})}$ & 0.386 & 11.942 & 0.241 & 70.994 & 0.162 & 83.936 & 0.231 & 95.281 \\ 
   &  & $\widehat{\Delta}_{\mathrm{IPPR+RDS-II}}^{(\mathrm{R2},\mathrm{M2})}$ & 0.344 & -0.361 & 0.190 & 55.165 & 0.125 & 81.627 & 0.192 & 95.783 \\ 
   &  & $\widehat{\Delta}_{\mathrm{IPPR+SS}}^{(\mathrm{R2},\mathrm{M2})}$ & 0.342 & -0.830 & 0.210 & 60.913 & 0.139 & 83.735 & 0.206 & 95.984 \\ 
   \midrule
   & 20\% & $\widehat{\mu}_{\mathrm{UW}}$ & 0.156 & -20.750 & 0.037 & 27.687 & 0.031 & 69.439 & 0.058 & 96.493 \\ 
   &  & $\widehat{\mu}_{\mathrm{RDS-II}}$ & 0.166 & -15.522 & 0.075 & 40.947 & 0.051 & 72.745 & 0.093 & 92.485 \\ 
   &  & $\widehat{\mu}_{\mathrm{SS}}$ & 0.165 & -15.872 & 0.071 & 39.197 & 0.048 & 72.345 & 0.089 & 93.086 \\ 
   &  & $\widehat{\mu}_{\mathrm{IPSR+RDS-II}}^{(\mathrm{R2})}$ & 0.221 & 12.660 & 0.094 & 48.927 & 0.063 & 85.170 & 0.112 & 98.297 \\ 
   &  & $\widehat{\mu}_{\mathrm{IPSR+SS}}^{(\mathrm{R2})}$ & 0.221 & 12.395 & 0.089 & 46.442 & 0.059 & 86.673 & 0.108 & 98.497 \\ 
   &  & $\widehat{\mu}_{\mathrm{IPPR+RDS-II}}^{(\mathrm{R2})}$ & 0.192 & -2.051 & 0.075 & 37.691 & 0.052 & 83.267 & 0.094 & 97.695 \\ 
   &  & $\widehat{\mu}_{\mathrm{IPPR+SS}}^{(\mathrm{R2})}$ & 0.194 & -0.968 & 0.071 & 35.897 & 0.049 & 84.870 & 0.090 & 98.096 \\ 
   \addlinespace
   &  & $\widehat{\Delta}_{\mathrm{UW}}^{(\mathrm{M2})}$ & 0.297 & -13.958 & 0.071 & 24.843 & 0.060 & 79.559 & 0.114 & 98.898 \\ 
   &  & $\widehat{\Delta}_{\mathrm{RDS-II}}^{(\mathrm{M2})}$ & 0.302 & -12.447 & 0.132 & 40.298 & 0.090 & 80.261 & 0.159 & 97.495 \\ 
   &  & $\widehat{\Delta}_{\mathrm{SS}}^{(\mathrm{M2})}$ & 0.302 & -12.515 & 0.128 & 39.197 & 0.086 & 80.160 & 0.156 & 97.796 \\ 
   &  & $\widehat{\Delta}_{\mathrm{IPSR+RDS-II}}^{(\mathrm{R2},\mathrm{M2})}$ & 0.376 & 8.916 & 0.176 & 51.720 & 0.111 & 83.768 & 0.189 & 98.397 \\ 
   &  & $\widehat{\Delta}_{\mathrm{IPSR+SS}}^{(\mathrm{R2},\mathrm{M2})}$ & 0.378 & 9.673 & 0.174 & 51.276 & 0.111 & 84.970 & 0.194 & 98.697 \\ 
   &  & $\widehat{\Delta}_{\mathrm{IPPR+RDS-II}}^{(\mathrm{R2},\mathrm{M2})}$ & 0.338 & -2.052 & 0.142 & 41.214 & 0.097 & 85.471 & 0.169 & 98.397 \\ 
   &  & $\widehat{\Delta}_{\mathrm{IPPR+SS}}^{(\mathrm{R2},\mathrm{M2})}$ & 0.340 & -1.360 & 0.153 & 44.435 & 0.097 & 86.373 & 0.172 & 98.597 \\ 
   \midrule 
   & 30\% & $\widehat{\mu}_{\mathrm{UW}}$ & 0.157 & -20.039 & 0.030 & 24.723 & 0.025 & 59.200 & 0.050 & 96.600 \\ 
   &  & $\widehat{\mu}_{\mathrm{RDS-II}}$ & 0.166 & -15.274 & 0.060 & 34.121 & 0.043 & 72.000 & 0.085 & 94.600 \\ 
   &  & $\widehat{\mu}_{\mathrm{SS}}$ & 0.165 & -15.825 & 0.055 & 31.978 & 0.038 & 71.400 & 0.078 & 94.400 \\ 
   &  & $\widehat{\mu}_{\mathrm{IPSR+RDS-II}}^{(\mathrm{R2})}$ & 0.219 & 11.544 & 0.075 & 39.507 & 0.053 & 88.700 & 0.102 & 99.100 \\ 
   &  & $\widehat{\mu}_{\mathrm{IPSR+SS}}^{(\mathrm{R2})}$ & 0.218 & 11.079 & 0.068 & 36.024 & 0.047 & 88.500 & 0.095 & 99.200 \\ 
   &  & $\widehat{\mu}_{\mathrm{IPPR+RDS-II}}^{(\mathrm{R2})}$ & 0.192 & -2.353 & 0.061 & 30.884 & 0.044 & 85.500 & 0.085 & 98.700 \\ 
   &  & $\widehat{\mu}_{\mathrm{IPPR+SS}}^{(\mathrm{R2})}$ & 0.195 & -0.575 & 0.057 & 28.292 & 0.039 & 87.400 & 0.079 & 99.300 \\ 
   \addlinespace
    &  & $\widehat{\Delta}_{\mathrm{UW}}^{(\mathrm{M2})}$ & 0.295 & -14.562 & 0.057 & 22.033 & 0.048 & 73.000 & 0.097 & 99.000 \\ 
   &  & $\widehat{\Delta}_{\mathrm{RDS-II}}^{(\mathrm{M2})}$ & 0.302 & -12.403 & 0.109 & 33.916 & 0.076 & 81.100 & 0.144 & 98.500 \\ 
   &  & $\widehat{\Delta}_{\mathrm{SS}}^{(\mathrm{M2})}$ & 0.302 & -12.520 & 0.101 & 31.886 & 0.068 & 80.600 & 0.136 & 98.600 \\ 
   &  & $\widehat{\Delta}_{\mathrm{IPSR+RDS-II}}^{(\mathrm{R2},\mathrm{M2})}$ & 0.373 & 8.242 & 0.133 & 39.529 & 0.094 & 84.600 & 0.171 & 99.400 \\ 
   &  & $\widehat{\Delta}_{\mathrm{IPSR+SS}}^{(\mathrm{R2},\mathrm{M2})}$ & 0.369 & 6.977 & 0.138 & 40.629 & 0.084 & 84.400 & 0.165 & 99.400 \\ 
   &  & $\widehat{\Delta}_{\mathrm{IPPR+RDS-II}}^{(\mathrm{R2},\mathrm{M2})}$ & 0.336 & -2.577 & 0.115 & 33.295 & 0.082 & 85.600 & 0.153 & 99.200 \\ 
   &  & $\widehat{\Delta}_{\mathrm{IPPR+SS}}^{(\mathrm{R2},\mathrm{M2})}$ & 0.341 & -1.198 & 0.118 & 34.081 & 0.076 & 85.600 & 0.148 & 99.200 \\ 
\bottomrule
\end{tabular}
\end{adjustbox}
\begin{tablenotes}
      \small
     \item[] The average true prevalence is $\mu = 0.196$, and the true 
prevalence difference is $\Delta = 0.345$. Abbreviations: Av. Est., 
average estimate; Rel. Bias, relative bias; SD, Monte Carlo standard 
deviation; RRMSE, relative root mean squared error; Av. Boot. SE, average 
bootstrap standard error; Cov., empirical coverage probability; NBoot, neighbourhood bootstrap; TBoot, tree bootstrap.
    \end{tablenotes}
\end{threeparttable}
\end{table}

\section*{Scenario \texorpdfstring{$C_3$}{C3}: latent common
causes of exposure, outcome, degree, and response}
\label{sec:simulation-results-c3}

Scenario $C_3$, represented in Figure~\ref{fig:dags}(c), considered an unobserved $W$ affecting degree, response, exposure, and outcome. Omitting $W$ therefore induced both nonignorable non-response and unmeasured confounding of the association between $E$ and $Y$. Unlike Scenario $C_2$, design-weighting the response model cannot recover the target estimands because the latent variable is also directly associated with the outcome. For conciseness, only estimators using RDS-II weights are shown.

\begin{figure}
\centering
\includegraphics[scale=0.3]{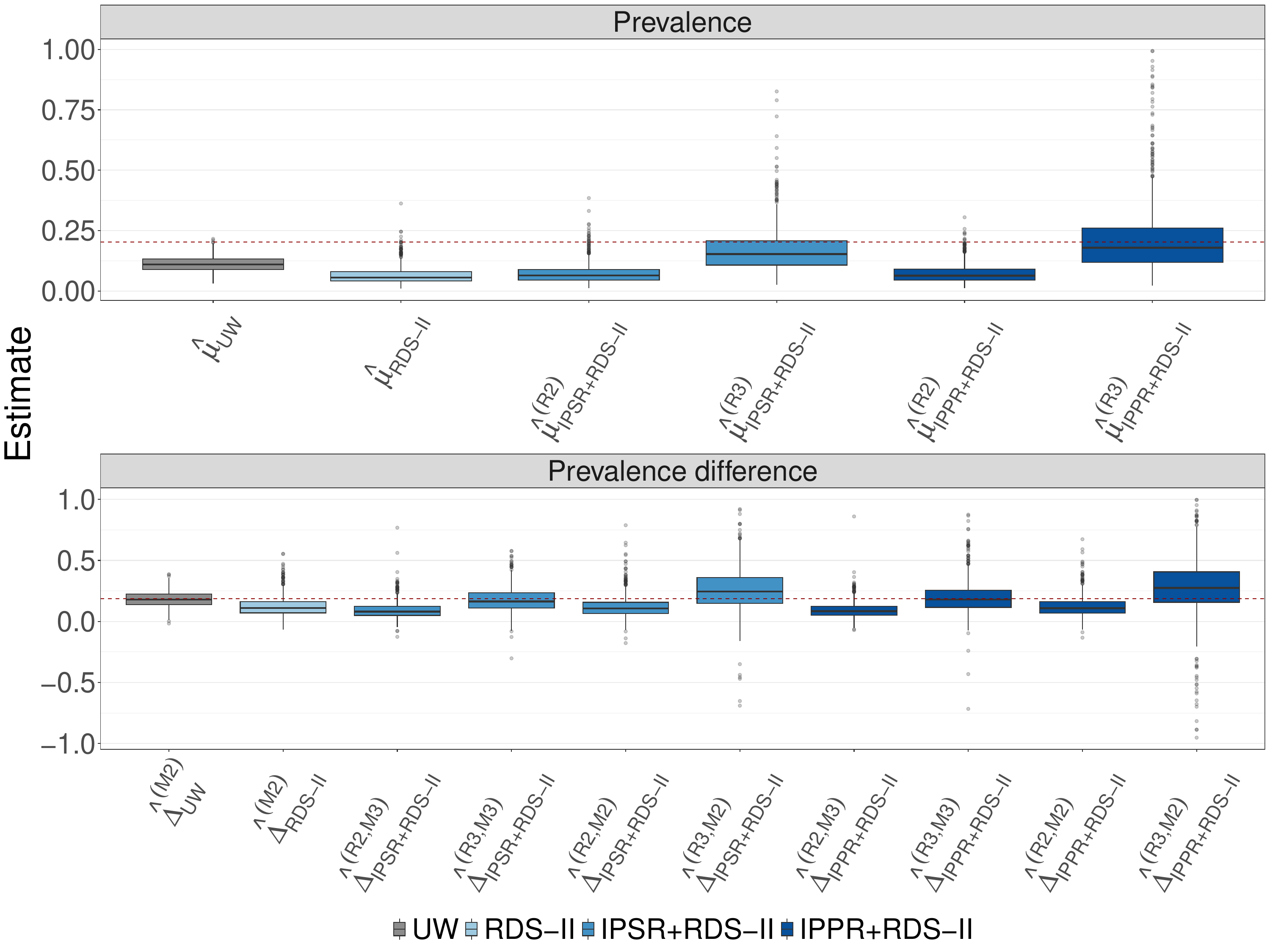}
\caption{Boxplots show the estimated outcome prevalence (top panel) and prevalence difference (bottom panel) under Scenario $C_3$ across 1,000 simulation replicates for the unweighted (UW), RD, and IPO adjusted estimators (IPSR/IPPR).  Adjusted estimators are combined with RDS-II sampling weights. The sampling fraction displayed is 20\%. Superscripts R2 and R3 denote response-model specification (\ref{eqn:R2}) and (\ref{eqn:R3}), respectively, whereas superscripts M2 and M3 denote marginal standardization models (\ref{eqn:M2}) and (\ref{eq:M3}), respectively. The red dashed line indicates the mean population-specific true value of the corresponding estimand.}
\label{fig:boxplotsC3}
\end{figure}

Figure~\ref{fig:boxplotsC3} shows that prevalence estimators based on the reduced response model (\ref{eqn:R2}) remained substantially below the true value, regardless of whether the response model was fitted using RDS weights. Using the oracle response model (\ref{eqn:R3}), which included $W$, moved estimates considerably closer to the truth. The IPPR estimator based on (\ref{eqn:R3}) was the most closely centred, although it also exhibited considerable dispersion and a long upper tail.

For the prevalence difference, both the response and outcome models needed to include $W$. Combining the correctly specified outcome model (\ref{eq:M3}) with the reduced response model (\ref{eqn:R2}) left residual non-response bias, while combining the oracle response model (\ref{eqn:R3}) with the reduced outcome model (\ref{eqn:M2}) left residual bias due to unmeasured confounding. Estimates were most closely centred around the true prevalence difference when both (\ref{eq:M3}) and (\ref{eqn:R3}) were used. Some misspecified combinations appeared less biased, but this could reflect cancellation between non-response and confounding bias rather than identification of the target estimand.

Scenario $C_3$ therefore illustrates the limits of the proposed weighting procedure: design-weighting cannot restore identification when variables required for both the MAR and NUC assumptions are unavailable. Valid estimation hinges on a sufficiently rich observed history for these assumptions to be scientifically plausible.

\section*{Scenario \texorpdfstring{$C_4$}{C4}: an unmeasured
determinant of degree and response with network autocorrelation}

\begin{figure}
\centering
\includegraphics[scale=0.3]{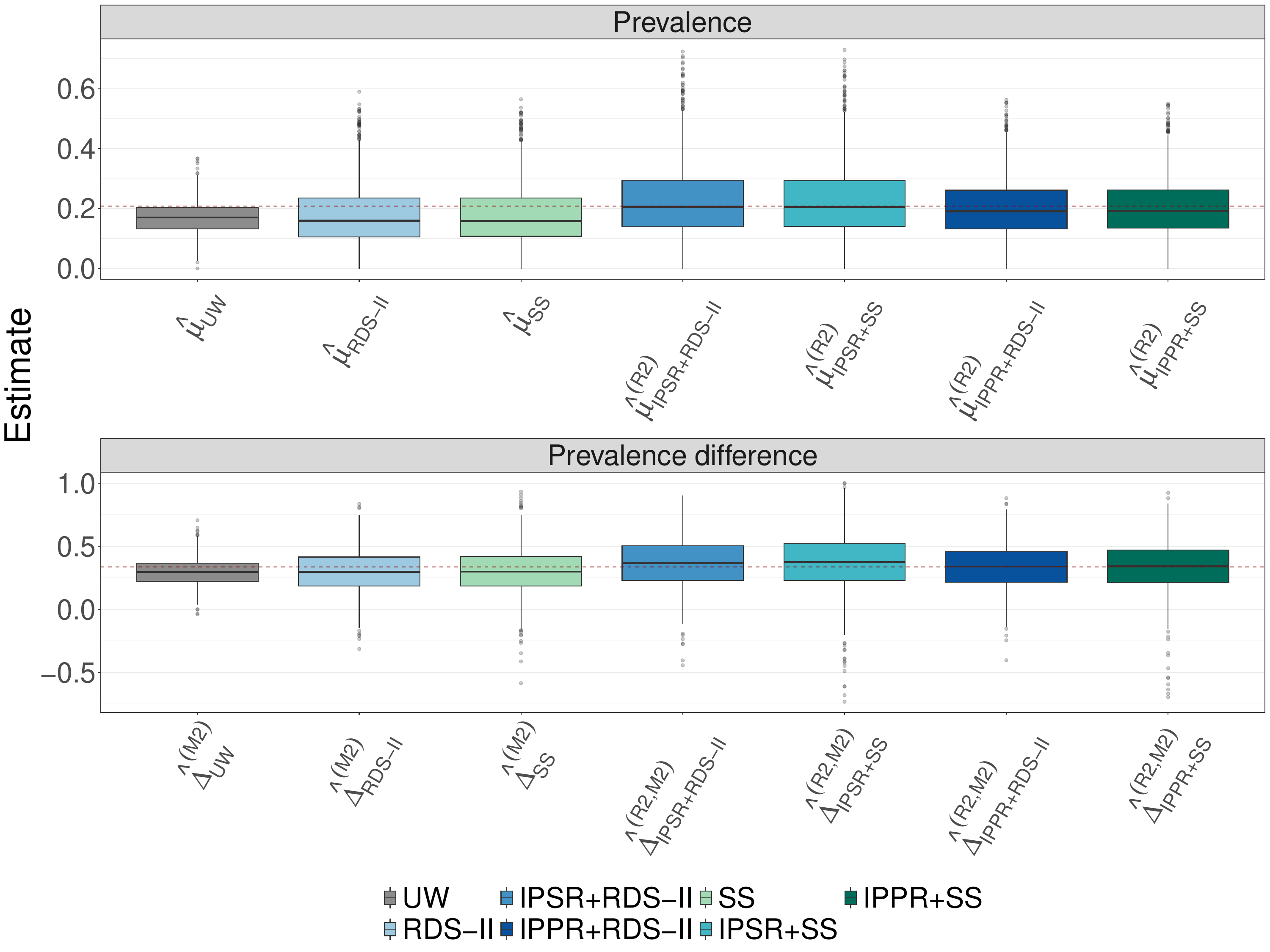}
\caption{Boxplots show the estimated outcome prevalence (top panel) and prevalence difference (bottom panel) under Scenario $C_4$ across 1,000 simulation replicates for the unweighted (UW), RDS-II, successive-sampling (SS), and IPO adjusted estimators (IPSR/IPPR). Adjusted estimators are combined with either RDS-II or SS sampling weights. Superscripts R2 and M1 denote response-model specification 2 and marginal standardization model 1, respectively. The red dashed line indicates the mean population-specific true value of the corresponding estimand.}
\label{fig:boxplotsC4}
\end{figure}

Figure \ref{fig:boxplotsC4} and Table \ref{tab:simulation-inference-C4} present simulation results under the sensitivity scenario $C_4$. The presence of network autocorrelation in the outcome model does not change previous conclusions drawn for $C_2$.

\clearpage
% latex table generated in R 4.4.3 by xtable 1.8-4 package
% Thu Aug  6 20:37:59 2026

\begin{table}
\centering
\caption{Simulation performance under sandwich, neighbourhood-bootstrap, and tree-bootstrap inference under Scenario $C_4$.} 
\label{tab:simulation-inference-C4}

\small
\setlength{\tabcolsep}{3pt}
\renewcommand{\arraystretch}{0.9}

\begin{adjustbox}{
  max width=\textheight,
  max totalheight=1.0\textwidth,
  keepaspectratio
}

\begin{tabular}{llllrrrrrr}
  \toprule
Scenario & $f$ & Estimator & Method & Av. Est. & Rel. Bias (\%) & SD & RRMSE (\%) & Av. SE & Cov. (\%) \\ 
  \midrule
 $C_4$& 10\% & $\widehat{\mu}_{\mathrm{UW}}$ & Neighbour bootstrap & 0.171 & -17.981 & 0.055 & 31.607 & 0.047 & 77.186 \\ 
   &  &  & Tree bootstrap &  &  &  &  & 0.076 & 95.377 \\ 
   &  & $\widehat{\mu}_{\mathrm{RDS-II}}$ & Neighbour bootstrap & 0.181 & -12.976 & 0.106 & 52.182 & 0.069 & 74.171 \\ 
   &  &  & Tree bootstrap &  &  &  &  & 0.108 & 87.538 \\ 
   &  & $\widehat{\mu}_{\mathrm{SS}}$ & Neighbour bootstrap & 0.181 & -13.216 & 0.104 & 50.942 & 0.067 & 74.271 \\ 
   &  &  & Tree bootstrap &  &  &  &  & 0.106 & 87.638 \\ 
   &  & $\widehat{\mu}_{\mathrm{IPSR+RDS-II}}^{(\mathrm{R2})}$ & Neighbour bootstrap & 0.231 & 11.156 & 0.129 & 62.412 & 0.081 & 82.412 \\ 
   &  &  & Tree bootstrap &  &  &  &  & 0.125 & 95.980 \\ 
   &  & $\widehat{\mu}_{\mathrm{IPSR+SS}}^{(\mathrm{R2})}$ & Neighbour bootstrap & 0.231 & 10.952 & 0.126 & 60.856 & 0.079 & 82.714 \\ 
   &  &  & Tree bootstrap &  &  &  &  & 0.123 & 96.181 \\ 
   &  & $\widehat{\mu}_{\mathrm{IPPR+RDS-II}}^{(\mathrm{R2})}$ & Neighbour bootstrap & 0.207 & -0.715 & 0.104 & 49.060 & 0.069 & 82.211 \\ 
   &  &  & Tree bootstrap &  &  &  &  & 0.109 & 96.181 \\ 
   &  & $\widehat{\mu}_{\mathrm{IPPR+SS}}^{(\mathrm{R2})}$ & Neighbour bootstrap & 0.207 & -0.502 & 0.101 & 48.057 & 0.068 & 82.714 \\ 
   &  &  & Tree bootstrap &  &  &  &  & 0.108 & 96.683 \\ 
 &  & $\widehat{\Delta}_{\mathrm{UW}}^{(\mathrm{M2})}$ & Sandwich 1 (seed) & 0.295 & -11.907 & 0.109 & 34.642 & 0.101 & 84.429 \\ 
   &  &  & Sandwich 2 (recruiter) &  &  &  &  & 0.105 & 90.332 \\ 
   &  &  & Neighbour bootstrap &  &  &  &  & 0.092 & 83.417 \\ 
   &  &  & Tree bootstrap &  &  &  &  & 0.154 & 98.693 \\ 
   &  & $\widehat{\Delta}_{\mathrm{RDS-II}}^{(\mathrm{M2})}$ & Sandwich 1 (seed) & 0.306 & -8.785 & 0.172 & 51.920 & 0.169 & 81.966 \\ 
   &  &  & Sandwich 2 (recruiter) &  &  &  &  & 0.138 & 79.516 \\ 
   &  &  & Neighbour bootstrap &  &  &  &  & 0.120 & 82.814 \\ 
   &  &  & Tree bootstrap &  &  &  &  & 0.190 & 96.583 \\ 
   &  & $\widehat{\Delta}_{\mathrm{SS}}^{(\mathrm{M2})}$ & Sandwich 1 (seed) & 0.306 & -8.622 & 0.185 & 55.830 & 0.160 & 82.121 \\ 
   &  &  & Sandwich 2 (recruiter) &  &  &  &  & 0.132 & 79.062 \\ 
   &  &  & Neighbour bootstrap &  &  &  &  & 0.124 & 83.417 \\ 
   &  &  & Tree bootstrap &  &  &  &  & 0.193 & 96.985 \\ 
   &  & $\widehat{\Delta}_{\mathrm{IPSR+RDS-II}}^{(\mathrm{R2},\mathrm{M2})}$ & Sandwich 1 (seed) & 0.368 & 9.924 & 0.200 & 60.522 & 0.200 & 82.759 \\ 
   &  &  & Sandwich 2 (recruiter) &  &  &  &  & 0.165 & 82.067 \\ 
   &  &  & Neighbour bootstrap &  &  &  &  & 0.142 & 83.819 \\ 
   &  &  & Tree bootstrap &  &  &  &  & 0.217 & 98.492 \\ 
   &  & $\widehat{\Delta}_{\mathrm{IPSR+SS}}^{(\mathrm{R2},\mathrm{M2})}$ & Sandwich 1 (seed) & 0.369 & 10.187 & 0.234 & 70.603 & 0.199 & 81.928 \\ 
   &  &  & Sandwich 2 (recruiter) &  &  &  &  & 0.161 & 81.271 \\ 
   &  &  & Neighbour bootstrap &  &  &  &  & 0.164 & 85.930 \\ 
   &  &  & Tree bootstrap &  &  &  &  & 0.236 & 98.794 \\ 
   &  & $\widehat{\Delta}_{\mathrm{IPPR+RDS-II}}^{(\mathrm{R2},\mathrm{M2})}$ & Sandwich 1 (seed) & 0.338 & 0.810 & 0.179 & 53.469 & 0.184 & 83.198 \\ 
   &  &  & Sandwich 2 (recruiter) &  &  &  &  & 0.151 & 82.157 \\ 
   &  &  & Neighbour bootstrap &  &  &  &  & 0.128 & 84.422 \\ 
   &  &  & Tree bootstrap &  &  &  &  & 0.200 & 98.191 \\ 
   &  & $\widehat{\Delta}_{\mathrm{IPPR+SS}}^{(\mathrm{R2},\mathrm{M2})}$ & Sandwich 1 (seed) & 0.337 & 0.645 & 0.207 & 61.838 & 0.176 & 82.529 \\ 
   &  &  & Sandwich 2 (recruiter) &  &  &  &  & 0.148 & 81.075 \\ 
   &  &  & Neighbour bootstrap &  &  &  &  & 0.143 & 85.829 \\ 
   &  &  & Tree bootstrap &  &  &  &  & 0.214 & 98.794 \\ 
   \bottomrule
\end{tabular}
\end{adjustbox}
\end{table}

% Move secondary data-generating details, algorithms, and parameter settings
% here. Keep enough information in the main paper to understand the study.

% Summarize what is shown in the supplementary displays and insert any
% explanatory text here.

\newpage
\webappendix{Additional case study details}{webapp:case-study}

\section*{Attrition pattern and descriptive statistics for the Engage-Montréal cohort}

\begin{figure}
\centering
\includegraphics[scale=0.40]{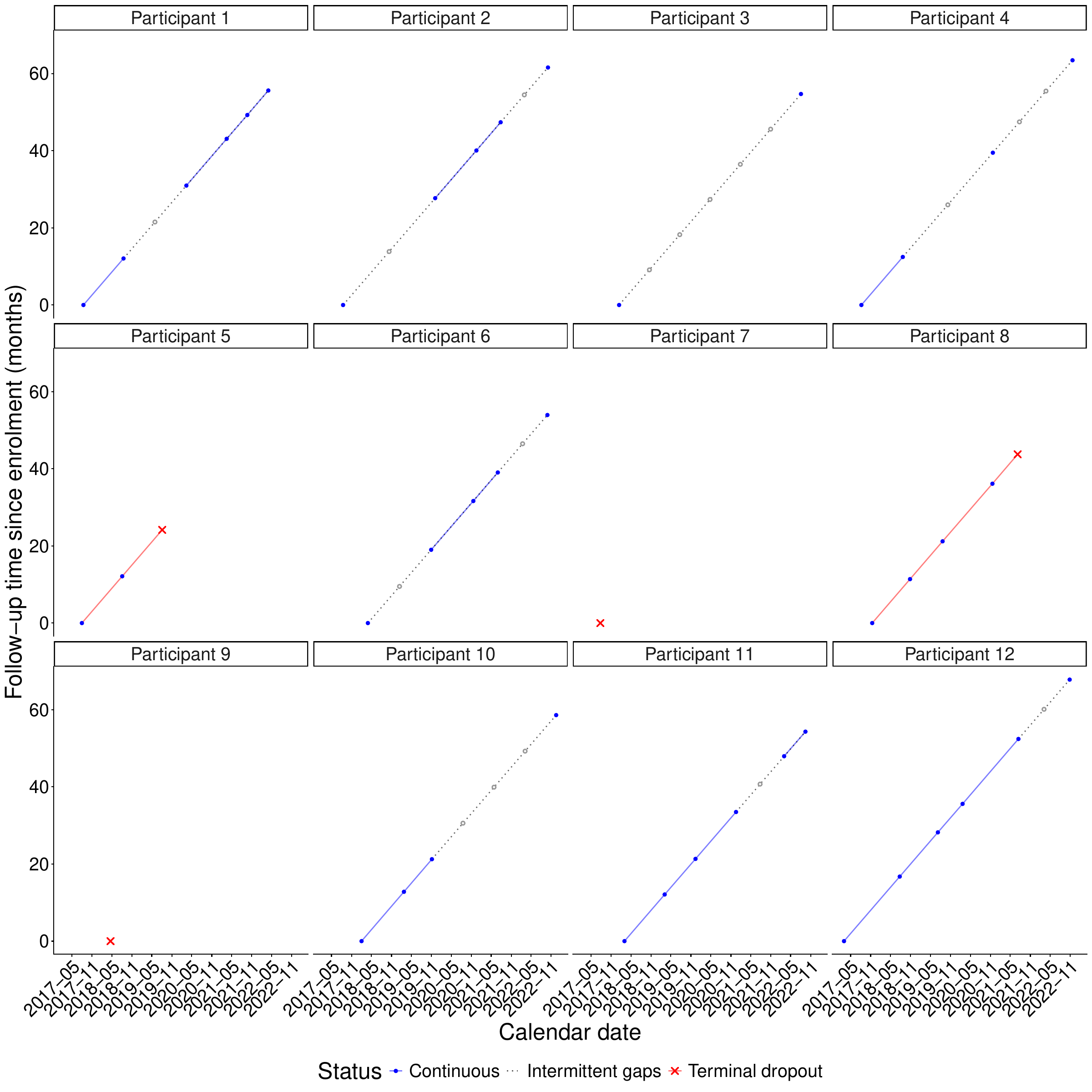}
\caption{Illustration of longitudinal observation patterns for 12 hypothetical participants. Blue points connected by solid blue lines represent consecutive observed visits, grey dotted segments indicate intermittent gaps followed by a return to follow-up, and red crosses indicate terminal dropout, after which no further visits are observed. Follow-up time is measured in months since each participant’s enrolment.}
\label{fig:lexis_ind}
\end{figure}

Figure \ref{fig:lexis_ind} shows longitudinal observation patterns for 12 hypothetical participants and contains examples of either LTFU and INA. A description of the attrition pattern over the course of the study is presented in Table \ref{tab:attrition}.

% latex table generated in R 4.4.3 by xtable 1.8-4 package
% Thu May  7 17:02:51 2026
\begin{table}[ht]
\centering
\caption{Attendance and continued enrolment across the seven planned visits in the Engage Montreal prospective cohort.} 
\label{tab:attrition}
\begin{tabular}{rrrr}
  \hline
Visit & Attended & Missed & Still Enrolled \\ 
  \hline
  Baseline (1) & 1179 (100.0\%) & 0 (0.0\%) & 1179 (100.0\%)  \\ 
    2 & 887 (75.2\%) & 292 (24.8\%) & 1005 (85.2\%)  \\ 
    3 & 850 (72.1\%) & 329 (27.9\%) & 942 (79.9\%)  \\ 
4 & 802 (68.0\%) & 377 (32.0\%) & 896 (76.0\%)  \\ 
    5 & 764 (64.8\%) & 415 (35.2\%) & 847 (71.8\%)  \\ 
    6 & 758 (64.3\%) & 421 (35.7\%) & 808 (68.5\%)  \\ 
    7 & 753 (63.9\%) & 426 (36.1\%) & 753 (63.9\%)  \\ 
   \hline
\end{tabular}
\end{table}

Unweighted baseline characteristics are presented by eventual LTFU status in Table~\ref{tab:baseline_characteristics_combined}, with the corresponding RDS-weighted characteristics presented in Table~\ref{tab:baseline_characteristics_RDSII}. Assuming that GBM represented 6.4\% of the male population of Montreal, the target population size was estimated to be approximately $N=54{,}500$. Given the resulting small sampling fraction ($1179/54{,}500 \approx 2.2\%$), the normalized RDS-II and successive-sampling (SS) weights were nearly identical. We therefore used RDS-II weights for the weighted descriptive analyses.

The unweighted statistics in Table~\ref{tab:baseline_characteristics_combined} indicate that participants classified as having terminal dropout (LTFU) reported a smaller median network size than those without terminal dropout (30 [IQR: 12--50] versus 35 [IQR: 20--100]). Age distributions were broadly similar, although participants with terminal dropout were somewhat more likely to be younger than 30 years. Participants without terminal dropout were more likely to identify as gay, to have completed education beyond high school, and to report neither injection drug use nor a history of sexual abuse. After applying the RDS-II weights (Table~\ref{tab:baseline_characteristics_RDSII}), the imbalance in network size remained and several imbalances by LTFU status became more pronounced, particularly those in sexual identity, education, marital status, chemsex-related substance use, injection drug use, and history of sexual abuse.

\clearpage
\begin{table}[p]
\caption{Baseline characteristics stratified by terminal dropout status.} 
\label{tab:baseline_characteristics_combined}
\centering
\small
\setlength{\tabcolsep}{3pt}
\renewcommand{\arraystretch}{0.9}

\begin{adjustbox}{
  max width=\textheight,
  max totalheight=1.5\textwidth,
  keepaspectratio
}
\begin{tabular}{lcccc}
  \toprule
 & Overall & No dropout & Terminal dropout & SMD \\ 
  \midrule
$n$ &  1179 &   808 &   371 &  \\ 
\addlinespace
  Reported network size (median (IQR)) & 30 [15, 80] & 35 [20, 100] & 30 [10, 50] &  0.254 \\
  \addlinespace
  Age (median (IQR)) & 34.00 [27.00, 49.00] & 34.00 [28.00, 50.00] & 34.00 [27.00, 49.00] &  0.049 \\ 
  \addlinespace
  Age group ($n$ (\%)) &   &    &    &  0.093 \\ 
  \ \ \ \ $<$ 30 y.o.&   384 (32.6)  &   252 (31.2)  &   132 (35.6)  &  \\ 
  \ \ \ \ 30--44 y.o &   420 (35.6)  &   294 (36.4)  &   126 (34.0)  &  \\ 
   \ \ \ \ $\geq$ 45 y.o &   375 (31.8)  &   262 (32.4)  &   113 (30.5)  &  \\ 
   \addlinespace
  Sexual identity ($n$ (\%)) &   &    &    &  0.309 \\ 
  \ \ \ \ Gay &   964 (81.8)  &   691 (85.5)  &   273 (73.6)  &  \\ 
 \ \ \ \ Bisexual &    99 ( 8.4)  &    49 ( 6.1)  &    50 (13.5)  &  \\ 
  \ \ \ \ Other &   116 ( 9.8)  &    68 ( 8.4)  &    48 (12.9)  &  \\ 
  \addlinespace
  Trans participant ($n$ (\%)) &    26 ( 2.2)  &    15 ( 1.9)  &    11 ( 3.0)  &  0.072 \\ 
  \addlinespace
  Education group: Greater than high school ($n$ (\%)) &   936 (79.4)  &   673 (83.3)  &   263 (70.9)  &  0.298 \\ 
  \addlinespace
  Income group &   &    &    &  0.164 \\ 
  \ \ \ \ $<$ 30,000 CAD &   678 (57.5)  &   447 (55.3)  &   231 (62.3)  &  \\ 
  \ \ \ \ 30,000 -- 59,999 CAD &   362 (30.7)  &   255 (31.6)  &   107 (28.8)  &  \\ 
  \ \ \ \ $\geq$ 60,000 CAD &   139 (11.8)  &   106 (13.1)  &    33 ( 8.9)  &  \\ 
  \addlinespace
  BIPOC ($n$ (\%))  &   &    &    &  0.183 \\ 
  \ \ \ \ Yes &   119 (10.1)  &    77 ( 9.5)  &    42 (11.3)  &  \\
   \ \ \ \ No &   894 (75.8)  &   602 (74.5)  &   292 (78.7)  &  \\ 
   \ \ \ \ Other/Unknown &   166 (14.1)  &   129 (16.0)  &    37 (10.0)  &  \\ 
  \addlinespace
  Marital status ($n$ (\%)) &   &    &    &  0.209 \\ 
  \ \ \ \ Married &    42 ( 3.6)  &    32 ( 4.0)  &    10 ( 2.7)  &  \\ 
  \ \ \ \ Common law &   187 (15.9)  &   142 (17.6)  &    45 (12.1)  &  \\ 
  \ \ \ \ Separated &    29 ( 2.5)  &    20 ( 2.5)  &     9 ( 2.4)  &  \\ 
  \ \ \ \ Divorced &    56 ( 4.7)  &    43 ( 5.3)  &    13 ( 3.5)  &  \\ 
 \ \ \ \ Widowed  &    18 ( 1.5)  &    13 ( 1.6)  &     5 ( 1.3)  &  \\ 
  \ \ \ \ Single  &   847 (71.8)  &   558 (69.1)  &   289 (77.9)  &  \\ 
  \addlinespace
  Immigration status ($n$ (\%)) &   &    &    &  0.210 \\ 
  \ \ \ \ Canadian citizen &   958 (81.3)  &   651 (80.6)  &   307 (82.7)  &  \\ 
  \ \ \ \ Permanent resident &    92 ( 7.8)  &    72 ( 8.9)  &    20 ( 5.4)  &  \\ 
  \ \ \ \ Refugee applicant &    14 ( 1.2)  &     7 ( 0.9)  &     7 ( 1.9)  &  \\ 
  \ \ \ \ Temporary work permit &    53 ( 4.5)  &    31 ( 3.8)  &    22 ( 5.9)  &  \\ 
  \ \ \ \ Student visa  &    53 ( 4.5)  &    41 ( 5.1)  &    12 ( 3.2)  &  \\ 
  \ \ \ \ Undocumented &     2 ( 0.2)  &     1 ( 0.1)  &     1 ( 0.3)  &  \\ 
  \ \ \ \ Other &     7 ( 0.6)  &     5 ( 0.6)  &     2 ( 0.5)  &  \\ 
  \addlinespace
  Condomless anal sex within the past six months ($n$ (\%)) &   554 (47.0)  &   388 (48.0)  &   166 (44.7)  &  0.066 \\ 
  \addlinespace
 Number of sex partners ($n$ (\%)) &   &    &    &  0.094 \\ 
  \ \ \ \ 0--5 &   655 (55.6)  &   441 (54.6)  &   214 (57.7)  &  \\ 
  \ \ \ \  6--10 &   231 (19.6)  &   156 (19.3)  &    75 (20.2)  &  \\ 
 \ \ \ \  $\geq$ 11 &   293 (24.9)  &   211 (26.1)  &    82 (22.1)  &  \\ 
  \addlinespace
 Use of any chemsex-related substance ($n$ (\%)) &   &    &    &  0.135 \\ 
  \ \ \ \ Yes &   265 (22.5)  &   178 (22.0)  &    87 (23.5)  &  \\ 
  \ \ \ \  No &   876 (74.3)  &   610 (75.5)  &   266 (71.7)  &  \\ 
  \ \ \ \ Missing/Unknown &    38 ( 3.2)  &    20 ( 2.5)  &    18 ( 4.9)  &  \\ 
  \addlinespace
  Ever been diagnosed with an STI ($n$ (\%)) &   &    &    &  0.181 \\ 
    \ \ \ \ Yes &   750 (63.6)  &   535 (66.2)  &   215 (58.0)  &  \\ 
   \ \ \ \  No &   407 (34.5)  &   257 (31.8)  &   150 (40.4)  &  \\ 
  \ \ \ \ Missing/Unknown &    22 ( 1.9)  &    16 ( 2.0)  &     6 ( 1.6)  &  \\ 
  \addlinespace
  Injection drug use in the past six months ($n$ (\%)) &   &    &    &  0.290 \\ 
  \ \ \ \ No, never &  1059 (89.8)  &   749 (92.7)  &   310 (83.6)  &  \\ 
  \ \ \ \ Yes, in the past six months  &    63 ( 5.3)  &    29 ( 3.6)  &    34 ( 9.2)  &  \\ 
  \ \ \ \ Yes, but more than six months ago  &    49 ( 4.2)  &    25 ( 3.1)  &    24 ( 6.5)  &  \\ 
  \ \ \ \ Missing/Unknown &     8 ( 0.7)  &     5 ( 0.6)  &     3 ( 0.8)  &  \\ 
  \addlinespace
  I believe that I was sexually abused ($n$ (\%))  &   &    &    &  0.300 \\ 
 \ \ \ \  Never &   830 (70.4)  &   579 (71.7)  &   251 (67.7)  &  \\ 
  \ \ \ \ Rarely &    99 ( 8.4)  &    77 ( 9.5)  &    22 ( 5.9)  &  \\ 
  \ \ \ \  Sometimes &    77 ( 6.5)  &    52 ( 6.4)  &    25 ( 6.7)  &  \\ 
  \ \ \ \ Often &    44 ( 3.7)  &    33 ( 4.1)  &    11 ( 3.0)  &  \\ 
  \ \ \ \ Very often &    40 ( 3.4)  &    25 ( 3.1)  &    15 ( 4.0)  &  \\ 
  \ \ \ \ Missing/Unknown &    89 ( 7.6)  &    47 ( 5.2)  &    54 (12.7)  & \\
   \bottomrule
\end{tabular}
\end{adjustbox}
\end{table}

\clearpage

% latex table generated in R 4.4.3 by xtable 1.8-4 package
% Thu May  7 14:45:47 2026
\begin{table}[p]
\caption{RDS-II weighted baseline characteristics, stratified by terminal dropout status.} 
\label{tab:baseline_characteristics_RDSII}
\centering
\small
\setlength{\tabcolsep}{3pt}
\renewcommand{\arraystretch}{0.9}

\begin{adjustbox}{
  max width=\textheight,
  max totalheight=1.5\textwidth,
  keepaspectratio
}
\begin{tabular}{lcccc}
  \toprule
 & Overall & No dropout & Terminal dropout & SMD \\ 
  \midrule
$n$ & 1179.00 & 733.82 & 445.18 &  \\ 
\addlinespace
  Reported network size (median (IQR)) &   10.00 [4.00, 20.00] &  10.00 [4.00, 20.00] &   7.00 [4.00, 15.00] &  0.190 \\ 
  \addlinespace
  Age (median (IQR)) &   33.00 [27.00, 49.00] &  33.00 [27.00, 48.00] &  35.00 [27.00, 49.00] &  0.029 \\ 
   \addlinespace
  Age group ($n$ (\%)) &   &   &   &  0.063 \\ 
  \ \ \ \ $<$ 30 y.o. &   428.9 (36.4)  &  263.0 (35.8)  &  165.9 (37.3)  &  \\ 
  \ \ \ \ 30--44 y.o &   407.5 (34.6)  &  261.8 (35.7)  &  145.7 (32.7)  &  \\ 
   \ \ \ \ $\geq$ 45 y.o  &   342.6 (29.1)  &  209.0 (28.5)  &  133.6 (30.0)  &  \\ 
   \addlinespace
  Sexual identity &   &   &   &  0.475 \\ 
  \ \ \ \ Gay &   897.2 (76.1)  &  614.4 (83.7)  &  282.9 (63.5)  &  \\ 
   \ \ \ \ Bisexual &   150.2 (12.7)  &   68.5 ( 9.3)  &   81.7 (18.3)  &  \\ 
  \ \ \ \ Other  &   131.6 (11.2)  &   50.9 ( 6.9)  &   80.6 (18.1)  &  \\ 
  \addlinespace
  Trans participant ($n$ (\%)) &    40.9 ( 3.5)  &   19.1 ( 2.6)  &   21.8 ( 4.9)  &  0.121 \\ 
  \addlinespace
  Education group: Greater than high school ($n$ (\%)) &   880.9 (74.7)  &  591.1 (80.6)  &  289.8 (65.1)  &  0.353 \\ 
  \addlinespace
  Income group &   &   &   &  0.117 \\ 
  \ \ \ \ $<$ 30,000 CAD &   788.1 (66.8)  &  486.2 (66.3)  &  301.9 (67.8)  &  \\ 
   \ \ \ \ 30,000 -- 59,999 CAD &   298.3 (25.3)  &  181.5 (24.7)  &  116.8 (26.2)  &  \\ 
  \ \ \ \ $\geq$ 60,000 CAD &    92.7 ( 7.9)  &   66.1 ( 9.0)  &   26.5 ( 6.0)  &  \\ 
  \addlinespace
  BIPOC ($n$ (\%)) &   &   &   &  0.345 \\ 
  \ \ \ \ Yes&   145.7 (12.4)  &   94.9 (12.9)  &   50.8 (11.4)  &  \\ 
  \ \ \ \ No &   820.8 (69.6)  &  473.3 (64.5)  &  347.5 (78.1)  &  \\ 
  \ \ \ \ Other/Unknown &   212.5 (18.0)  &  165.7 (22.6)  &   46.8 (10.5)  &  \\ 
  \addlinespace
  Marital status ($n$ (\%))  &   &   &   &  0.378 \\
  \ \ \ \ Married &    45.6 ( 3.9)  &   33.2 ( 4.5)  &   12.5 ( 2.8)  &  \\ 
  \ \ \ \ Common law  &   166.2 (14.1)  &  128.6 (17.5)  &   37.5 ( 8.4)  &  \\ 
  \ \ \ \ Separated &    38.5 ( 3.3)  &   14.2 ( 1.9)  &   24.4 ( 5.5)  &  \\ 
  \ \ \ \ Divorced &    44.5 ( 3.8)  &   34.5 ( 4.7)  &   10.0 ( 2.2)  &  \\ 
  \ \ \ \ Widowed &    10.0 ( 0.8)  &    7.1 ( 1.0)  &    2.9 ( 0.6)  &  \\ 
  \ \ \ \ Single &   874.2 (74.1)  &  516.3 (70.4)  &  357.9 (80.4)  &  \\ 
  \addlinespace
  Immigration status ($n$ (\%))&   &   &   &  0.419 \\ 
  \ \ \ \ Canadian citizen &   875.9 (74.3)  &  516.2 (70.3)  &  359.7 (80.8)  &  \\ 
  \ \ \ \ Permanent resident &   116.6 ( 9.9)  &   96.4 (13.1)  &   20.2 ( 4.5)  &  \\ 
  \ \ \ \ Refugee applicant &    29.5 ( 2.5)  &   21.8 ( 3.0)  &    7.7 ( 1.7)  &  \\ 
  \ \ \ \ Temporary work permit &    64.2 ( 5.4)  &   41.9 ( 5.7)  &   22.3 ( 5.0)  &  \\ 
  \ \ \ \ Student visa &    69.9 ( 5.9)  &   51.9 ( 7.1)  &   18.0 ( 4.0)  &  \\ 
   \ \ \ \ Undocumented  &     8.8 ( 0.7)  &    0.5 ( 0.1)  &    8.3 ( 1.9)  &  \\
   \ \ \ \ Other &  14.1 ( 1.2)  &    5.2 ( 0.7)  &    9.0 ( 2.0)  &  \\ 
  \addlinespace
  Condomless anal sex within the past six months ($n$ (\%)) &   459.8 (39.0)  &  305.0 (41.6)  &  154.9 (34.8)  &  0.140 \\ 
  \addlinespace
 Number of sex partners ($n$ (\%)) &   &   &   &  0.098 \\ 
  \ \ \ \ 0--5 &   783.3 (66.4)  &  485.1 (66.1)  &  298.2 (67.0)  &  \\
  \ \ \ \  6--10 &   223.8 (19.0)  &  133.1 (18.1)  &   90.7 (20.4)  &  \\ 
  \ \ \ \  $\geq$ 11 &   171.9 (14.6)  &  115.7 (15.8)  &   56.2 (12.6)  &  \\ 
  \addlinespace
 Use of chemsex in the past six months ($n$ (\%)) &   &   &   &  0.363 \\ 
    \ \ \ \ Yes &   194.2 (16.5)  &   94.5 (12.9)  &   99.7 (22.4)  &  \\ 
 \ \ \ \  No &   910.4 (77.2)  &  608.6 (82.9)  &  301.7 (67.8)  &  \\ 
 \ \ \ \ Missing/Unknown &    74.5 ( 6.3)  &   30.7 ( 4.2)  &   43.7 ( 9.8)  &  \\ 
   \addlinespace
  Ever been diagnosed with an STI ($n$ (\%)) &   &   &   &  0.078 \\ 
  \ \ \ \   Yes &   607.3 (51.5)  &  377.3 (51.4)  &  230.0 (51.7)  &  \\ 
    \ \ \ \  No  &   528.2 (44.8)  &  325.5 (44.4)  &  202.7 (45.5)  &  \\ 
  \ \ \ \ Missing/Unknown&    43.5 ( 3.7)  &   31.0 ( 4.2)  &   12.5 ( 2.8)  &  \\ 
   \addlinespace
  Injection drug use in the past six months ($n$ (\%)) &   &   &   &  0.381 \\ 
 \ \ \ \  No, never  &  1043.9 (88.5)  &  672.4 (91.6)  &  371.5 (83.4)  &  \\ 
   \ \ \ \  Yes, in the past six months &    64.6 ( 5.5)  &   29.7 ( 4.0)  &   34.9 ( 7.8)  &  \\ 
  \ \ \ \  Yes, but more than six months ago &    46.4 ( 3.9)  &   11.2 ( 1.5)  &   35.2 ( 7.9)  &  \\ 
 \ \ \ \  Missing/Unknown &    24.1 ( 2.0)  &   20.5 ( 2.8)  &    3.6 ( 0.8)  &  \\ 
   \addlinespace
 I believe that I was sexually abused ($n$ (\%)) &   &   &   &  0.403 \\ 
 \ \ \ \  Never &   767.1 (65.1)  &  501.1 (68.3)  &  266.0 (59.8)  &  \\ 
 \ \ \ \ Rarely &   106.9 ( 9.1)  &   67.4 ( 9.2)  &   39.4 ( 8.9)  &  \\ 
 \ \ \ \ Sometimes&    72.5 ( 6.1)  &   31.4 ( 4.3)  &   41.0 ( 9.2)  &  \\ 
 \ \ \ \ Often &    48.1 ( 4.1)  &   41.0 ( 5.6)  &    7.1 ( 1.6)  &  \\ 
 \ \ \ \ Very often &    29.5 ( 2.5)  &    9.7 ( 1.3)  &   19.8 ( 4.5)  &  \\ 
 \ \ \ \ Missing/Unknown &   155.0 (13.2)  &   83.2 ( 11.4)  &   71.8 (16.1)  &  \\ 
   \bottomrule
\end{tabular}
\end{adjustbox}
\end{table}

\FloatBarrier

\section*{Detailed analysis results}

Table \ref{tab:sti_prevalence} displays estimated prevalences for syphilis, CT, and NG for each weighting method and by study visit.  Table \ref{tab:sti_prevalence_diff} shows estimated standardized prevalence differences for the three bSTIs comparing a hypothetical scenario in which all participants used at least one chemsex-related substance than under the scenario in which none did. A detailed discussion of these results can be found in the main manuscript.

\begin{table}
\caption{Estimated prevalence of bSTIs at selected study visits, by weighting method.} 
\label{tab:sti_prevalence}
\centering
\scriptsize
\begin{tabular}{lllcccc}
  \toprule
STI & Study wave & Weighting method & Prevalence ($\hat{\mu}$) & Sandwich CI & NBoot CI & TBoot CI \\ 
  \midrule
\multirow{12}{*}{Syphilis} & \multirow{4}{*}{\shortstack[l]{Baseline\\(Feb 2017--June 2018)}} & Unweighted & 0.031 & [0.018, 0.044] & [0.020, 0.037] & [0.009, 0.061] \\ 
   &  & RDS-II & 0.018 & [0.008, 0.027] & [0.010, 0.023] & [0.005, 0.035] \\ 
   &  & IPSR+RDS-II & 0.020 & [0.012, 0.028] & [0.012, 0.027] & [0.006, 0.040] \\ 
   &  & IPPR+RDS-II & 0.020 & [0.011, 0.028] & [0.012, 0.027] & [0.006, 0.040] \\ 
   \cline{2-7}
 & \multirow{4}{*}{\shortstack[l]{Visit 4\\(Jan 2020--Apr 2021)}} & Unweighted & 0.019 & [0.007, 0.031] & [0.011, 0.027] & [0.004, 0.039] \\ 
   &  & RDS-II & 0.010 & [0.001, 0.019] & [0.005, 0.019] & [0.001, 0.033] \\ 
   &  & IPSR+RDS-II & 0.015 & [-0.003, 0.033] & [0.008, 0.029] & [0.001, 0.044] \\ 
   &  & IPPR+RDS-II & 0.016 & [-0.006, 0.038] & [0.008, 0.034] & [0.001, 0.049] \\ 
   \cline{2-7}
 & \multirow{4}{*}{\shortstack[l]{Visit 6\\(Jul 2021--Aug 2022)}} & Unweighted & 0.016 & [0.006, 0.025] & [0.008, 0.024] & [0.001, 0.038] \\ 
   &  & RDS-II & 0.010 & [0.001, 0.019] & [0.003, 0.017] & [0.000, 0.032] \\ 
   &  & IPSR+RDS-II & 0.016 & [0.005, 0.028] & [0.005, 0.026] & [0.000, 0.050] \\ 
   &  & IPPR+RDS-II & 0.016 & [0.005, 0.027] & [0.005, 0.024] & [0.000, 0.047] \\ 
   \hline
\multirow{12}{*}{Chlamydia} & \multirow{4}{*}{\shortstack[l]{Baseline\\(Feb 2017--June 2018)}} & Unweighted & 0.041 & [0.034, 0.048] & [0.030, 0.049] & [0.020, 0.066] \\ 
   &  & RDS-II & 0.030 & [0.011, 0.049] & [0.018, 0.038] & [0.010, 0.062] \\ 
   &  & IPSR+RDS-II & 0.034 & [0.020, 0.048] & [0.020, 0.042] & [0.012, 0.066] \\ 
   &  & IPPR+RDS-II & 0.034 & [0.019, 0.048] & [0.020, 0.042] & [0.011, 0.066] \\ 
   \cline{2-7}
 & \multirow{4}{*}{\shortstack[l]{Visit 4\\(Jan 2020--Apr 2021)}} & Unweighted & 0.041 & [0.026, 0.056] & [0.031, 0.054] & [0.016, 0.072] \\ 
   &  & RDS-II & 0.056 & [-0.046, 0.157] & [0.037, 0.100] & [0.011, 0.135] \\ 
   &  & IPSR+RDS-II & 0.040 & [0.020, 0.061] & [0.026, 0.058] & [0.012, 0.075] \\ 
   &  & IPPR+RDS-II & 0.039 & [0.018, 0.061] & [0.025, 0.058] & [0.011, 0.075] \\ 
   \cline{2-7}
 & \multirow{4}{*}{\shortstack[l]{Visit 6\\(Jul 2021--Aug 2022)}} & Unweighted & 0.011 & [0.006, 0.015] & [0.005, 0.016] & [0.000, 0.026] \\ 
   &  & RDS-II & 0.012 & [0.005, 0.020] & [0.004, 0.027] & [0.000, 0.038] \\ 
   &  & IPSR+RDS-II & 0.009 & [-0.004, 0.023] & [0.002, 0.022] & [0.000, 0.032] \\ 
   &  & IPPR+RDS-II & 0.009 & [-0.005, 0.023] & [0.002, 0.021] & [0.000, 0.033] \\ 
   \hline
\multirow{12}{*}{Gonorrhea} & \multirow{4}{*}{\shortstack[l]{Baseline\\(Feb 2017--June 2018)}} & Unweighted & 0.068 & [0.034, 0.101] & [0.054, 0.080] & [0.023, 0.117] \\ 
   &  & RDS-II & 0.057 & [0.018, 0.095] & [0.041, 0.074] & [0.014, 0.116] \\ 
   &  & IPSR+RDS-II & 0.060 & [0.027, 0.093] & [0.043, 0.077] & [0.023, 0.117] \\ 
   &  & IPPR+RDS-II & 0.060 & [0.027, 0.093] & [0.044, 0.076] & [0.018, 0.114] \\ 
   \cline{2-7}
 & \multirow{4}{*}{\shortstack[l]{Visit 4\\(Jan 2020--Apr 2021)}} & Unweighted & 0.031 & [0.018, 0.044] & [0.025, 0.046] & [0.007, 0.061] \\ 
   &  & RDS-II & 0.027 & [-0.024, 0.077] & [0.013, 0.057] & [0.002, 0.091] \\ 
   &  & IPSR+RDS-II & 0.025 & [0.012, 0.039] & [0.017, 0.049] & [0.003, 0.060] \\ 
   &  & IPPR+RDS-II & 0.026 & [0.013, 0.040] & [0.017, 0.049] & [0.003, 0.069] \\ 
   \cline{2-7}
 & \multirow{4}{*}{\shortstack[l]{Visit 6\\(Jul 2021--Aug 2022)}} & Unweighted & 0.037 & [0.020, 0.054] & [0.026, 0.050] & [0.011, 0.078] \\ 
   &  & RDS-II & 0.034 & [0.015, 0.053] & [0.018, 0.048] & [0.008, 0.084] \\ 
   &  & IPSR+RDS-II & 0.033 & [0.023, 0.044] & [0.019, 0.048] & [0.006, 0.088] \\ 
   &  & IPPR+RDS-II & 0.032 & [0.023, 0.042] & [0.020, 0.048] & [0.006, 0.082] \\ 
   \bottomrule
\end{tabular}
\end{table}

% latex table generated in R 4.4.3 by xtable 1.8-4 package
% Tue Aug 25 13:58:05 2026
\begin{table}
\caption{Model-standardized bSTI prevalence differences comparing participants reporting chemsex-related substance use with those reporting no such use, standardized to a common covariate distribution, at selected study visits and by weighting method.} 
\label{tab:sti_prevalence_diff}
\centering
\scriptsize
\begin{tabular}{lllcccc}
  \toprule
bSTI & Study wave & Weighting method & Prevalence diff. ($\hat{\Delta}$) & Sandwich CI & NBoot CI & TBoot CI \\ 
  \midrule
\multirow{12}{*}{Syphilis} & \multirow{4}{*}{\shortstack[l]{Baseline\\(Feb 2017--June 2018)}} & Unweighted & 0.022 & [0.009, 0.034] & [0.006, 0.032] & [-0.000, 0.075] \\ 
   &  & RDS-II & 0.022 & [0.001, 0.043] & [0.005, 0.039] & [-0.010, 0.073] \\ 
   &  & IPSR+RDS-II & 0.022 & [-0.009, 0.053] & [0.004, 0.041] & [-0.009, 0.070] \\ 
   &  & IPPR+RDS-II & 0.023 & [-0.008, 0.053] & [0.006, 0.042] & [-0.010, 0.070] \\ 
   \cline{2-7}
 & \multirow{4}{*}{\shortstack[l]{Visit 4\\(Jan 2020--Apr 2021)}} & Unweighted & 0.016 & [0.009, 0.023] & [0.005, 0.027] & [-0.000, 0.046] \\ 
   &  & RDS-II & 0.013 & [0.001, 0.026] & [0.004, 0.032] & [-0.003, 0.034] \\ 
   &  & IPSR+RDS-II & 0.016 & [-0.012, 0.044] & [0.004, 0.044] & [-0.004, 0.047] \\ 
   &  & IPPR+RDS-II & 0.018 & [-0.014, 0.049] & [0.004, 0.049] & [-0.004, 0.068] \\ 
   \cline{2-7}
 & \multirow{4}{*}{\shortstack[l]{Visit 6\\(Jul 2021--Aug 2022)}} & Unweighted & 0.014 & [-0.000, 0.028] & [0.004, 0.027] & [-0.000, 0.047] \\ 
   &  & RDS-II & 0.014 & [0.004, 0.024] & [0.003, 0.031] & [-0.010, 0.067] \\ 
   &  & IPSR+RDS-II & 0.019 & [0.006, 0.031] & [0.003, 0.033] & [-0.012, 0.069] \\ 
   &  & IPPR+RDS-II & 0.019 & [0.008, 0.031] & [0.003, 0.034] & [-0.017, 0.069] \\ 
   \hline
\multirow{12}{*}{Chlamydia} & \multirow{4}{*}{\shortstack[l]{Baseline\\(Feb 2017--June 2018)}} & Unweighted & 0.022 & [0.005, 0.038] & [0.007, 0.032] & [-0.006, 0.058] \\ 
   &  & RDS-II & 0.018 & [-0.011, 0.048] & [0.004, 0.033] & [-0.011, 0.086] \\ 
   &  & IPSR+RDS-II & 0.021 & [-0.010, 0.052] & [0.004, 0.037] & [-0.009, 0.079] \\ 
   &  & IPPR+RDS-II & 0.020 & [-0.011, 0.052] & [0.004, 0.037] & [-0.012, 0.087] \\ 
   \cline{2-7}
 & \multirow{4}{*}{\shortstack[l]{Visit 4\\(Jan 2020--Apr 2021)}} & Unweighted & 0.021 & [0.004, 0.038] & [0.007, 0.031] & [-0.007, 0.063] \\ 
   &  & RDS-II & 0.028 & [-0.004, 0.061] & [0.007, 0.064] & [-0.015, 0.106] \\ 
   &  & IPSR+RDS-II & 0.026 & [-0.004, 0.056] & [0.005, 0.052] & [-0.013, 0.100] \\ 
   &  & IPPR+RDS-II & 0.025 & [-0.004, 0.054] & [0.006, 0.049] & [-0.017, 0.098] \\ 
   \cline{2-7}
 & \multirow{4}{*}{\shortstack[l]{Visit 6\\(Jul 2021--Aug 2022)}} & Unweighted & 0.005 & [0.001, 0.009] & [0.001, 0.010] & [-0.001, 0.021] \\ 
   &  & RDS-II & 0.007 & [-0.001, 0.015] & [0.001, 0.022] & [-0.006, 0.038] \\ 
   &  & IPSR+RDS-II & 0.007 & [0.001, 0.014] & [0.001, 0.020] & [-0.004, 0.037] \\ 
   &  & IPPR+RDS-II & 0.007 & [0.001, 0.013] & [0.001, 0.018] & [-0.004, 0.032] \\ 
   \hline
\multirow{12}{*}{Gonorrhea} & \multirow{4}{*}{\shortstack[l]{Baseline\\(Feb 2017--June 2018)}} & Unweighted & 0.069 & [0.036, 0.102] & [0.049, 0.094] & [0.011, 0.137] \\ 
   &  & RDS-II & 0.043 & [-0.004, 0.090] & [0.027, 0.087] & [-0.011, 0.129] \\ 
   &  & IPSR+RDS-II & 0.055 & [0.007, 0.104] & [0.036, 0.101] & [-0.004, 0.141] \\ 
   &  & IPPR+RDS-II & 0.058 & [0.009, 0.106] & [0.039, 0.105] & [-0.005, 0.144] \\ 
   \cline{2-7}
 & \multirow{4}{*}{\shortstack[l]{Visit 4\\(Jan 2020--Apr 2021)}} & Unweighted & 0.038 & [0.023, 0.052] & [0.028, 0.067] & [0.006, 0.086] \\ 
   &  & RDS-II & 0.025 & [-0.005, 0.055] & [0.015, 0.078] & [-0.008, 0.087] \\ 
   &  & IPSR+RDS-II & 0.028 & [0.009, 0.046] & [0.018, 0.070] & [-0.002, 0.075] \\ 
   &  & IPPR+RDS-II & 0.030 & [0.008, 0.051] & [0.020, 0.076] & [-0.003, 0.087] \\ 
   \cline{2-7}
 & \multirow{4}{*}{\shortstack[l]{Visit 6\\(Jul 2021--Aug 2022)}} & Unweighted & 0.044 & [0.024, 0.064] & [0.029, 0.070] & [0.006, 0.106] \\ 
   &  & RDS-II & 0.031 & [0.011, 0.052] & [0.019, 0.070] & [-0.010, 0.104] \\ 
   &  & IPSR+RDS-II & 0.039 & [0.021, 0.057] & [0.023, 0.073] & [-0.005, 0.094] \\ 
   &  & IPPR+RDS-II & 0.039 & [0.018, 0.061] & [0.024, 0.077] & [-0.006, 0.108] \\
   \bottomrule
\end{tabular}
\end{table}

% Move extended Engage descriptive material and secondary analyses here.

% -------------------------------------------------------------------------
% Examples. Use ordinary figure and table environments with the biom class;
% avoid figure* and table* in referee mode because delayed-float handling in
% the legacy template can fail for starred environments.
% -------------------------------------------------------------------------

% \begin{figure}[p]
%   \centering
%   \includegraphics[width=0.9\textwidth]{pointestimates_C4.pdf}
%   \caption{Additional simulation results under Scenario $C_4$.}
%   \label{webfig:c4}
% \end{figure}

% \begin{table}[p]
%   \centering
%   \caption{Additional simulation results.}
%   \label{webtab:simulation-results}
%   \begin{threeparttable}
%     \begin{tabular}{lrr}
%       \toprule
%       Method & Estimate & Standard error \\
%       \midrule
%       Example method & 0.000 & 0.000 \\
%       \bottomrule
%     \end{tabular}
%     \begin{tablenotes}
%       \small
%       \item[] Note: Replace this example with the relevant table.
%     \end{tablenotes}
%   \end{threeparttable}
% \end{table}

\clearpage

% This uses the same BibTeX database and Biometrics bibliography style as the
% main manuscript. Only works cited in this supplement will be printed.

\end{document}